\documentclass[onecolumn,12pt]{article}
\usepackage{jheppub}
\usepackage{ifpdf}
\usepackage[bb=boondox]{mathalfa}
\usepackage{graphicx,subcaption}
\graphicspath{ {figures/} }
\usepackage{amsfonts}
\usepackage{amsmath,braket}
\usepackage{amssymb}
\usepackage{amsthm}
\usepackage{epsfig}
\usepackage{tensor}
\usepackage{pdfpages}
\usepackage{bbm}
\usepackage{graphicx,epstopdf}
\usepackage[numbers]{natbib} 
\usepackage[makeroom]{cancel}
\usepackage{hyperref}
\usepackage{array}
\usepackage[export]{adjustbox}

\usepackage[normalem]{ulem}
\usepackage{romannum}

\usepackage{ulem}

\usepackage{color}
\usepackage{here}

\usepackage{tikz}
\usepackage{tikz-3dplot}
\usetikzlibrary{calc}

\numberwithin{equation}{section}									

\def\d#1{\,{\rm d}#1}
\usepackage{comment}

\newcommand{\be}{\begin{equation}}
	\newcommand{\ba}{\begin{eqnarray}}
		\newcommand{\ea}{\end{eqnarray}}
	\newcommand{\ee}{\end{equation}}

\newcommand{\ti}{\tilde}

\newcommand{\ddd}{\cdot\cdot\cdot}

\newcommand{\la}{\langle}
\newcommand{\lb}{\rangle}
\newcommand{\bea}{\begin{eqnarray}}
	\newcommand{\eea}{\end{eqnarray}}
\newcommand{\bes}{\begin{equation*}}
	\newcommand{\beas}{\begin{eqnarray*}}
		\newcommand{\eeas}{\end{eqnarray*}}
	\newcommand{\bas}{\begin{array*}}
		\newcommand{\eas}{\end{array*}}
	\newcommand{\ees}{\end{equation*}}

\newcommand{\ov}{\overline}

\newcommand{\Tr}{\textrm{Tr}}

\let\a=\alpha \let\b=\beta  \let\d=\delta        
    \let\x=\xi

\newcommand{\tr}{{\rm tr}}

\newcommand{\arctanh}{\text{arctanh}}
\newcommand{\arccosh}{\text{arccosh}}
\newcommand{\arcsinh}{\text{arcsinh}}

\subheader{\today}
\title{The Entanglement Wedge Polygon}
\author[a]{Kosei Fujiki,}
\author[b,a]{Jonathan Harper,}
\author[a,c]{Tadashi Takayanagi,}
\author[a]{Nicol\`o Zenoni}

\affiliation[a]{Center for Gravitational Physics and Quantum Information, Yukawa Institute for Theoretical Physics, Kyoto University,\\
	Kitashirakawa Oiwakecho, 
    Sakyo-ku, Kyoto 606-8502, Japan}
    \affiliation[b]{Department of Physics,
Case Western Reserve University,\\ 10900 Euclid Ave, Cleveland, OH 44106 USA}
\affiliation[c]{Inamori Research Institute for Science,\\
	620 Suiginya-cho, Shimogyo-ku,Kyoto 600-8411 Japan}

\emailAdd{kosei.fujiki@yukawa.kyoto-u.ac.jp}
\emailAdd{jonathan.harper@case.edu}
\emailAdd{takayana@yukawa.kyoto-u.ac.jp}
\emailAdd{nicolo@yukawa.kyoto-u.ac.jp}

\abstract{In this work we consider a particular codimension-1 region of a holographic spacetime which we call the entanglement wedge polygon (EWP). For a pure state and a partition of the boundary into a number of regions $A_i$ the EWP is defined as the region external to all the individual homology regions $r_{A_i}$ which consists of the intersection of the entanglement wedge EW($A_i$) with the time slice. In vacuum AdS$_3$ and BTZ spacetime, the quantity is topological as a direct consequence of the Gauss-Bonnet theorem. In higher dimensions we make progress by considering a number of concrete examples including vacuum, black brane, and soliton solutions of AdS$_{d+1}$ as well as spacetime geometries with end of the world branes dual to boundary conformal field theories. We provide a suitable generalization to mixed states and comment on possible connections between the EWP and measures of multi-partite entanglement.
}

\begin{document} 
	
	\begin{flushright}
		YITP-26-76
        \\
	\end{flushright}
	\maketitle
	\flushbottom

\section{Introduction}
\label{sec:intro}

In holography \cite{tHooft:1993dmi,Susskind:1994vu,Maldacena:1997re,Gubser:1998bc,Witten:1998qj},
entanglement entropy \cite{Bombelli:1986rw,Srednicki:1993im,Holzhey:1994we,Calabrese:2004eu} has played a key role to connect quantum information to the geometry of gravitational spacetimes \cite{Ryu:2006bv,Ryu:2006ef,Hubeny:2007xt}, 
leading to the idea that the spacetimes may be regarded as tensor networks \cite{Swingle:2009bg,Pastawski:2015qua,Hayden:2016cfa}. The entanglement entropy measures a bi-partite correlation (i.e. quantum entanglement) in quantum many-body systems when the total system is pure. However, it is well-known that quantification of multi-partite quantum entanglement, which is also closely related to quantum entanglement in mixed states, is much more difficult and  has no unified picture, though there have been many progresses and ideas \cite{Linden:1999qve,PhysRevA.62.062314,Verstraete:2002gqj,Walter:2016lgl,RevModPhys.81.865,Ma:2023ecg,Horodecki:2024bgc}. 

In the context of holography, the simplest quantity which measures multi-partite correlations will be multi-partite mutual information \cite{Hayden:2011ag}, which includes not only quantum correlations, but also classical ones. For bi-partite correlations for mixed states the holographic dual of entanglement of purification was proposed in \cite{Takayanagi:2017knl,Nguyen:2017yqw} and was generalized to multi-partite systems in \cite{Umemoto:2018jpc}. The Markov gap \cite{Hayden:2021gno}, based on the reflected entropy \cite{Dutta:2019gen}, is also an intriguing quantity, though it has been noted that the reflected entropy is not always a correlation measure \cite{Hayden:2023yij}.

Moreover, the multi-entropy \cite{Gadde:2022cqi,Penington:2022dhr,Gadde:2023zzj}
provides a natural generalization of entanglement entropy to multi-partite reduced density matrices
and its quantum information theoretic properties have been worked out \cite{Gadde:2023zni,Gadde:2025csh}. 
This quantity can be computable both in field theoretic calculations \cite{Harper:2024ker,Liu:2024ulq,Gadde:2025csh,Harper:2025uui,Fujiki:2026qdt}
and holographic ones \cite{Gadde:2022cqi,Penington:2022dhr,Iizuka:2025ioc,Iizuka:2025bcc,Anegawa:2025prn,Hu:2026bhg,Fujiki:2026qdt} (see also \cite{Bao:2018gck}). The multi-entropy generally includes not only the multi-partite correlations but also bi-partite correlations. Therefore to extract the former we need to subtract the bi-partite entanglement entropy \cite{Harper:2024ker,Iizuka:2025ioc}.

\begin{figure}[htbp]
		\centering
		\includegraphics[scale=0.15]{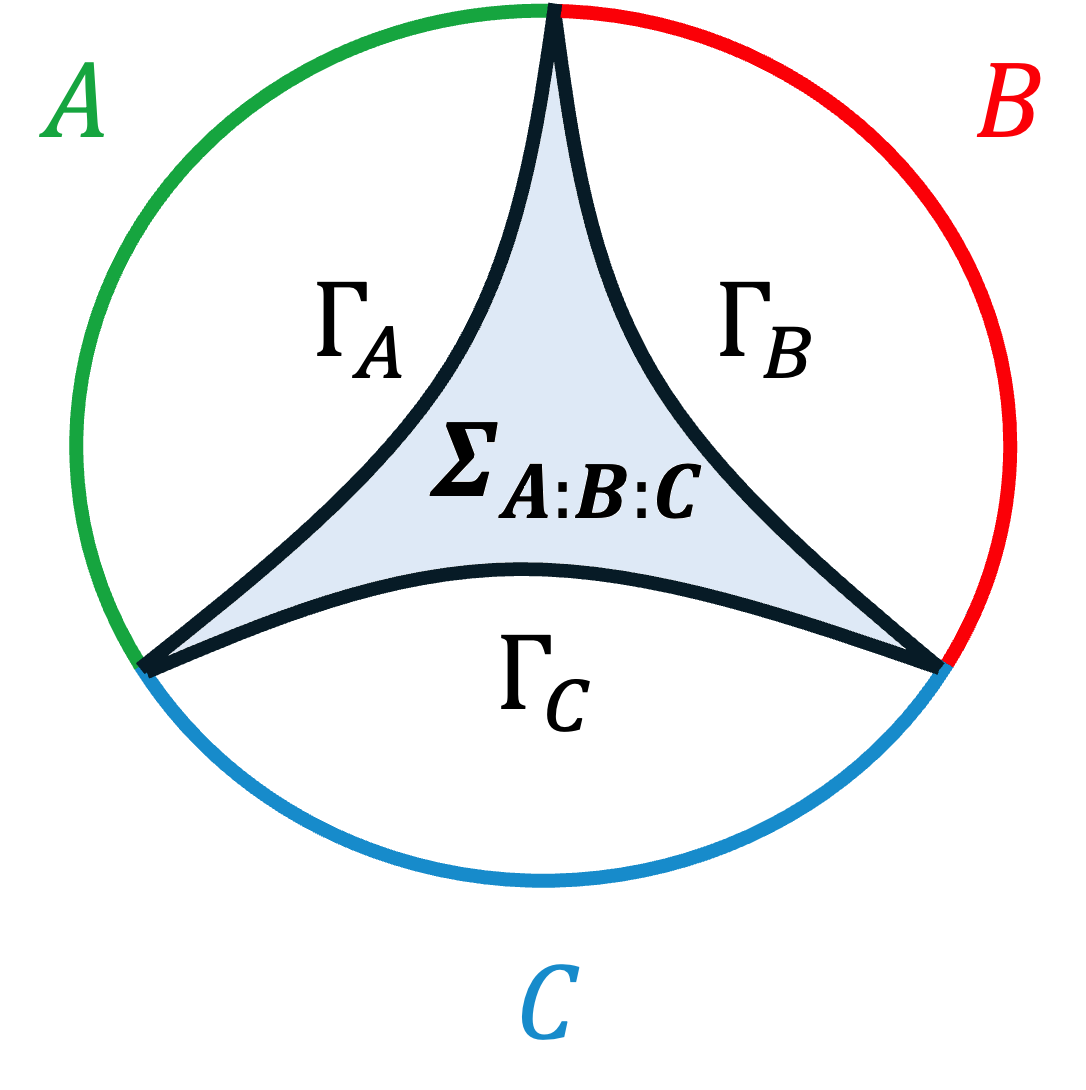}
        		\caption{The entanglement wedge polygon $\Sigma_{A:B:C}$ for a pure state $|\Psi\lb_{ABC}$. $\Sigma_{A:B:C}$ is surrounded by the extremal surfaces 
                $\Gamma_A$, $\Gamma_B$, and $\Gamma_C$.} 
		\label{fig:EWPintro}
\end{figure}

In this paper, we would like to propose and examine a simpler geometric quantity which can possibly measure the amount of multi-partite quantum entanglement, motivated by the ideas of entanglement wedges \cite{Headrick:2014cta,Wall:2012uf,Czech:2012bh} and the tensor network descriptions of holography \cite{Swingle:2009bg,Pastawski:2015qua,Hayden:2016cfa}.
We introduce a codimension-one region which we call the entanglement wedge polygon (EWP), depicted in Fig.~\ref{fig:EWPintro}, and argue that its volume can be considered as a possible measure of multi-partite quantum entanglement. This is highly motivated by the observation that the tensors which are responsible for the multi-partite entanglement are expected to be in this polygon region assuming that the AdS spacetime can be regarded as a tensor network or quantum circuits 
\cite{Swingle:2009bg,Pastawski:2015qua,Hayden:2016cfa,Miyaji:2015yva,Caputa:2017yrh,Takayanagi:2018pml}. 
Notice that the volume of the full time slice or that of an entanglement wedge has already been introduced in the context of holographic computational complexity \cite{Susskind:2014rva,Stanford:2014jda,Alishahiha:2015rta}. Thus, we can also think that our quantity can be a multi-partite generalization of complexity. 
In particular, in the static case the volume of the EWP for a tri-partition $A,B,C$ of a pure state matches the difference between volume subregion complexity of $A \cup B$ and volume subregion complexity of the single $A$ and $B$. This difference has been studied in several backgrounds \cite{Ben-Ami:2016qex,Carmi:2016wjl,Abt:2017pmf,Abt:2018ywl,Agon:2018zso,Auzzi:2019fnp} and defined as (the opposite of) mutual complexity, initially introduced for bi-partitions of pure states \cite{Alishahiha:2018lfv} and later generalized to bi-partitions of mixed states \cite{Caceres:2019pgf}. We stress that in the covariant case, however, the volume of the EWP cannot be obtained by subtraction of volume subregion complexities, thus providing a different measure compared to mutual complexity.
We would also like to note that an example of volume of entanglement wedge square was already discussed in \cite{Pastawski:2015qua} in a connection between the quantum error correction codes and multi-partite entanglement. See also  \cite{Gan:2017qkz} for the connection to subsystem complexity and \cite{Leutheusser:2025zvp} for an algebraic interpretation.
We also note that modifications of the IR geometry inside the region that we here refer to as the EWP and their effect on various measures of (multi-partite) entanglement have been investigated in \cite{Ju:2025eyn}.

For a simple setup, consider a three-partite system $ABC$ for a CFT vacuum, where $\rho_{ABC}=|\Psi\lb_{ABC}\la\Psi|_{ABC}$ is a pure state. On a time slice of its dual AdS geometry, the entanglement wedge (projected to this slice) for the subsystem $A$ is given by the subregion surrounded by $A$ and $\Gamma_A$, where $\Gamma_A$ is the minimal surface which computes the entanglement entropy $S_A$. This is the homology region $r_A$. Similarly we can construct the entanglement wedges for $B$ and $C$. The entanglement wedge triangle $\Sigma_{A:B:C}$ is defined by the triangle region given by the time slice of the AdS with the three homology regions excluded as depicted in Fig.~\ref{fig:EWPintro}.

\begin{figure}[H]
		\centering
		\includegraphics[scale=0.15]{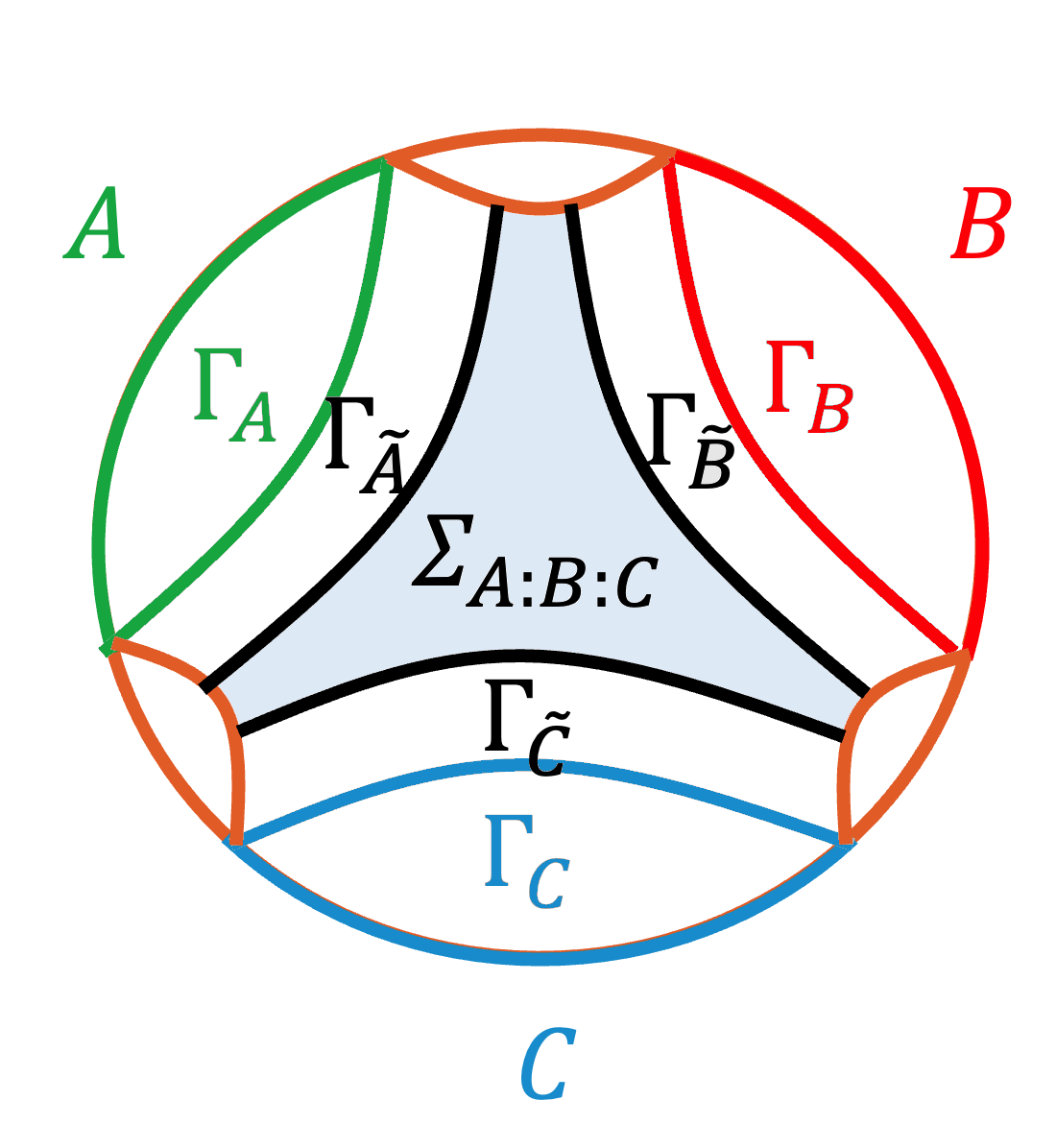}
        		\caption{The entanglement wedge polygon $\Sigma_{A:B:C}$ for a mixed state $\rho_{ABC}$. The extremal surface $\Gamma_{\ti{A}}$ is defined by minimizing the area among the cross sections of the entanglement wedge between $A$ and $BC$. Similarly, we can define the surfaces $\Gamma_{\ti{B}}$ and $\Gamma_{\ti{C}}$.} 
		\label{fig:EWPmix}
\end{figure}

We can generalize this to the case when $A$, $B$, and $C$ are disconnected such that $\rho_{ABC}$ is a mixed state, as depicted in Fig.~\ref{fig:EWPmix}. The entanglement wedge polygon for this mixed state is the region surrounded by the borders of the entanglement wedge for $\rho_{ABC}$ and the three minimal surfaces $\Gamma_{\ti{A}}$, $\Gamma_{\ti{B}}$, and $\Gamma_{\ti{C}}$, which are the entanglement wedge cross sections dual to the entanglement of purification \cite{Takayanagi:2017knl,Nguyen:2017yqw}. Clearly, this construction has straightforward generalizations to higher partite entanglement.

We will examine the volume of the entanglement wedge polygon in various examples in diverse dimensions, paying attention to both general properties and peculiar behaviors which distinguish it from other quantum information theoretic quantities in holography such as the holographic entanglement entropy and the multi-entropy. In pure gravity on AdS$_3$, we will find that the volume is quantized in an interesting way and thus becomes topological, though this property is lost in the presence of an end-of-the-world brane for the AdS/BCFT duality \cite{Takayanagi:2011zk,Fujita:2011fp}. 
This topological nature is carried over to our extension to the mixed states. In higher dimensions or in the presence of matter fields or time-dependence, we will find richer structures including various phase transitions.

This paper is organized as follows: In section~\ref{sec:def-pure}, we present the definition of entanglement wedge polygon for pure states and its basic properties.
In section~\ref{sec:AdS3-pure}, we compute the entanglement wedge polygon in AdS$_3$ for various setups.
In section~\ref{sec:Higher-d}, we analyze the entanglement wedge polygon in higher dimensional AdS setups, namely Poincar\'e AdS, the AdS black brane, the AdS/BCFT, and the AdS soliton geometry dual to a confining gauge theory. In section~\ref{sec:quenches}, we analyze the time evolution of the entanglement wedge polygon for the global quench dual to a one-sided AdS black hole with an end-of-the-world brane behind the horizon, the thermofield double state, and a global quench dual to AdS Vaidya spacetime.
In section~\ref{sec:def-mixed}, we introduce the definition of entanglement wedge polygon for mixed states and related properties.
In section~\ref{sec:AdS3-mixed}, we discuss the volume of the entanglement wedge polygon for mixed states in various three-dimensional holographic setups.
In section~\ref{sec:Higher-d-mixed}, we calculate the volume of the entanglement wedge polygon for mixed states in higher-dimensional Poincar\'e AdS and discuss its monotonicity property.
In section~\ref{sec:conclusions}, we draw conclusions and discuss future problems. We defer details on our computations to the Appendices. 

While in preparation we became aware of related work \cite{Caputa:2026ldd} where a similar quantity is discussed from the perspective of subsystem complexity. We thank Pawel Caputa for sharing an early draft and for discussion and comments.

\section{Entanglement Wedge Polygon for pure states}
\label{sec:def-pure}

In holography, given a boundary pure state $\rho$ with a bulk dual, there is a known correspondence between reduced density matrices and bulk regions.
In particular, all the information on $\rho_A = \tr_{\ov{A}} (\rho)$ is contained into the \textit{entanglement wedge} (EW) associated to the boundary subregion $A$.
In order to build the EW$(A)$, one first identifies the extremal codimension-two surface $\Gamma_A$ homologous to $A$, whose area is proportional to the \textit{entanglement entropy} $S_A = - \tr_A(\rho_A \log \rho_A)$.
Then, denoting by $r_A$ the codimension-one homology surface, $\partial r_A = \Gamma_A \cup A$, the EW is the codimension$-0$ region corresponding to the domain of dependence of $r_A$.  
Heuristically, the EW should take into account the bi-partite entanglement between $A$ and its complement $\ov{A}$.

For a higher number of partitions of the pure state $\rho$, multi-partite entanglement enters into play.
It is interesting to understand which region in the bulk dual accounts for the purely multi-partite entanglement of $\rho_{A_1 \ddd A_q}$, neglecting the bi-partite contributions.
The multi-partite correlations are encoded in the EW$(A_1 \ddd A_q)$, while the bi-partite ones in EW$(A_i)$, as discussed above.
Therefore, a natural candidate is the portion of EW$(A_1 \ddd A_q)$ external to all the EWs of the single boundary subregions $A_i$.
In this work, we restrict to the codimension-one Cauchy slice with maximal volume and refer to the resulting region as the \textit{entanglement wedge polygon} (EWP).
The maximality condition is crucial for some properties of the EWP to hold, as we will see later.
The EWP for a tri-partition of the vacuum state is represented in Fig.~\ref{fig:EWPintro}.
The proposal is that the volume of the EWP is responsible for the purely multi-partite entanglement among the boundary subregions.\footnote{As we will show later, 
the volume of the EWP does not satisfy the monotonicity property with respect to the geometrical size of subsystems. In this sense, we do not argue that the volume measures the amount of multi-partite correlations. Instead we would like to argue that it is related to a certain quantity which characterizes the multi-partite entanglement motivated from the tensor network viewpoint, e.g. the number of tensors which participate in the multi-partite entanglement.}
Below is a formal definition of this quantity.
\\ 

{\bf Definition $P$.}
Let us consider $q$ \textit{disjoint} subregions $A_i$ which can share boundaries, and assume that $\rho_{A_1 \ddd A_q}$ is a \textit{pure} state.
We denote by $\Gamma_{A_i}$ the codimension-two extremal bulk surface homologous to the subregion $A_i$.
We define the \textit{entanglement wedge polygon} (EWP) as the space-like codimension-one surface $\Sigma_{A_1: \ddd: A_q}$ with \textit{maximal} volume satisfying 
\begin{equation}
\label{eq:def-Sigma}
    \partial \Sigma_{A_1: \ddd :A_q} = \bigcup_{i=1}^q \Gamma_{A_i} \, .
\end{equation}
We denote the volume $V(\Sigma_{A_1: \dots: A_q})$ by the shorthand notation $V_{\{ A_i \}}$, where it is subtended that we are considering $q$ subregions.

Note that in the static case, definition~\eqref{eq:def-Sigma} is equivalent to removing the homology surfaces $r_{A_i}$ to the full Cauchy slice 
\begin{align}
    \Sigma_{A_1: \ddd :A_q} &= r_{(A_1 \ddd A_q)} \setminus \bigcup_{i=1}^q r_{A_i}  \\
    &= r_{(A_1 \ddd A_{q-1})} \setminus \bigcup_{i=1}^{q-1} r_{A_i} \, ,
    \label{eq:def-Sigma-static}
\end{align}
where the second inequality directly follows from the purity of $\rho_{A_1 \ddd A_q}$ and $\Gamma_{A_q} = \Gamma_{(A_1 \ddd A_{q-1})}$.
However, definition \eqref{eq:def-Sigma} is more general and applies to the covariant case as well.
We list two properties of the EWP for pure states.
\\

{\bf Property $P1$.}
The volume of the EWP vanishes for bi-partitions of pure states:
\begin{equation}
    V_{A:\ov{A}} =  0 \, .
\end{equation}
\begin{proof}
    From $\Gamma_{\ov{A}} = \Gamma_{A}$, we clearly have $\Sigma_{A:\ov{A}}=\emptyset$.
\end{proof}
Property $P1$ resonates with the fact that the volume of the EWP captures only the multi-partite entanglement among the boundary subregions, neglecting the bi-partite contributions.  
Another property which supports this interpretation is the following.
\\

{\bf Property $P2$.}
The volume of the EWP increases for finer grained partitions of pure states:
\begin{equation}
\label{eq:inequality-pure}
    V_{(AB):C:O} \leq V_{A:B:C:O} \, ,
    \qquad
    O= \ov{ABC} \, ,
\end{equation}
where the equality holds if and only if $\Gamma_{(AB)}=\Gamma_A \cup \Gamma_B$, i.e. there are no correlations between $A$ and $B$, $S_{(AB)}= S_A +S_B$.\footnote{Despite $\Gamma_{(AB)}=\Gamma_A \cup \Gamma_B$ implies $S_{(AB)} =S_A +S_B$, the reverse has some loopholes. Some examples are the phase transition between a connected and a disconnected phase, and the presence of a black hole, in which case $\Gamma_{(AB)}$ may also include the event horizon.}
\begin{proof}
If $\Gamma_{(AB)}$ is in the disconnected phase, i.e. $\Gamma_{(AB)}=\Gamma_A \cup \Gamma_B$. Then, from the definition \eqref{eq:def-Sigma}, we clearly have $\Sigma_{(AB):C:O}=\Sigma_{A:B:C:O}$.

Let us now assume that $A$ and $B$ are correlated. In other words, $\Gamma_{(AB)}\neq\Gamma_A \cup \Gamma_B$.
To prove that the inequality holds, it is enough to observe that $\Sigma_{(AB):C:O} \subset \Sigma_{A:B:C:O}$.

Let us start from the static case.
Denoting by $r_A$ the maximal codimension-one surface bounded by $\Gamma_A \cup A$, we have $r_A \cup r_B \subset r_{(AB)}$ \cite{Wall:2012uf}. 
From this and eq.~\eqref{eq:def-Sigma-static}, it directly follows that
$\Sigma_{(AB):C:O} = r_{(ABC)} \setminus \left( r_{(AB)} \cup r_C \right) \subset r_{(ABC)} \setminus \left( r_A \cup r_B \cup r_C \right) = \Sigma_{A:B:C:O}$.\footnote{Note that $r_{(AB)} \cup r_C \subseteq r_{(ABC)}$, where equality holds when $\Gamma_{(ABC)}=\Gamma_{(AB)} \cup \Gamma_C$, $(AB)$ and $C$ are in the disconnected phase.}

In the covariant case, let us assume that the bulk spacetime satisfies the null curvature condition $R_{\mu \nu} \xi^\mu \xi^\nu \geq 0$ for any null vector $\xi^\mu$, and in the presence of a black hole the singularities are spacelike and the metric close to the singularity is of the Kasner form. Under these assumptions, the extremal codimension-two surface $\Gamma_A$ can be built by the maximin prescription \cite{Wall:2012uf}. By theorem 17 and corollary 17 (h) of \cite{Wall:2012uf}, since $A,B \subset AB \subset ABC$ and $A,B,C$ are disjoint, then $\Gamma_{A}$, $\Gamma_{B}$, $\Gamma_{C}$, $\Gamma_{(AB)}$, and $\Gamma_{O}=\Gamma_{(ABC)}$ are all minimal on the same codimension-one achronal slice $\Sigma$. 
Let us define the sub-slice $\hat{\Sigma}_{A:B:C:O} \subset \Sigma$ satisfying $\partial \hat{\Sigma}_{A:B:C:O} = \Gamma_A \cup \Gamma_B \cup \Gamma_C \cup \Gamma_{(ABC)}$.
Similarly, we define the sub-slice $\hat{\Sigma}_{(AB):C:O} \subset \Sigma$ satisfying $\partial \hat{\Sigma}_{(AB):C:O} = \Gamma_{(AB)} \cup \Gamma_C \cup \Gamma_{(ABC)}$, see Fig.~\ref{fig:V-decompos}.
By applying the same proof as in the static case, we promptly get $\hat{\Sigma}_{(AB):C:O} \subset \hat{\Sigma}_{A:B:C:O}$.
Now, we can modify both $\hat{\Sigma}_{(AB):C:O}$ and $\hat{\Sigma}_{A:B:C:O}$ by keeping their boundaries fixed and deforming the slice along the orthogonal future and past direction. The aim is to obtain the corresponding slices $\Sigma_{(AB):C:O}$ and $\Sigma_{A:B:C:O}$ with maximal volume for the given boundary conditions, and compare their volumes. Let us distinguish the following cases:
\begin{itemize}
    \item If $\Sigma_{(AB):C:O}=\hat{\Sigma}_{(AB):C:O}$ and $\Sigma_{A:B:C:O}=\hat{\Sigma}_{A:B:C:O}$, i.e. both slices have maximal volume, we immediately conclude $V_{(AB):C:O} < V_{A:B:C:O}$.
    \item If $\hat{\Sigma}_{(AB):C:O}$ and $\hat{\Sigma}_{A:B:C:O}$ are not maximal, let us define $\hat{\Sigma}_{\rm comp} = \hat{\Sigma}_{A:B:C:O} \setminus \hat{\Sigma}_{(AB):C:O}$, see Fig.~\ref{fig:V-decompos}.
    In other words, 
    \begin{equation}
    \label{eq:V-decompos}
    V(\hat{\Sigma}_{A:B:C:O}) = V(\hat{\Sigma}_{(AB):C:O}) + V(\hat{\Sigma}_{\rm comp}) \, .
    \end{equation}
    Now, we can locally deform $\hat{\Sigma}_{(AB):C:O}$ by keeping $\partial \hat{\Sigma}_{(AB):C:O}$ fixed to obtain $\hat{\Sigma}_{A:B:C:O} = \hat{\Sigma}_{(AB):C:O} \cup \hat{\Sigma}_{\rm comp} \to \Sigma_{(AB):C:O} \cup \hat{\Sigma}_{\rm comp}$.
    Under this process,
    \begin{align}
        V_{(AB):C:O} = \max_{ \partial \tilde{\Sigma}_{(AB):C:O} = \partial \hat{\Sigma}_{(AB):C:O}} V(\tilde{\Sigma}_{(AB):C:O}) 
        &< \max_{ \partial \tilde{\Sigma}_{(AB):C:O} = \partial \hat{\Sigma}_{(AB):C:O}} V(\tilde{\Sigma}_{A:B:C:O}) \nonumber \\
        &\leq V_{A:B:C:O}
        \, ,
    \end{align}
    where the first inequality follows from generalizing eq.~\eqref{eq:V-decompos} to $V(\tilde{\Sigma}_{A:B:C:O}) = V(\tilde{\Sigma}_{(AB):C:O}) + V(\hat{\Sigma}_{\rm comp})$.
    The second inequality follows from the fact that $V_{A:B:C:O}$ is the maximal volume among all the possible deformations of $\hat{\Sigma}_{A:B:C:O}$ with $\partial \hat{\Sigma}_{A:B:C:O}$ fixed.
    Indeed, note that the first deformation of $\hat{\Sigma}_{(AB):C:O}$ does not modify $\partial \hat{\Sigma}_{A:B:C:O}$.
\end{itemize}
\begin{figure}
    \centering
    \begin{tikzpicture}
    \begin{scope}[yshift=5cm]
      \draw (-0.2,0) -- (6.4,0);
      \draw[thick] (0.6,0) arc (180:0:1.0);     
      \draw[thick] (2.6,0) arc (180:0:0.8);     
      \draw[thick] (4.2,0) arc (180:0:0.7);     
      \draw[thick] (5.6,0) arc (0:180:2.5);     
      \node[below] at (0.2,0) {$O$};
      \node[below] at (1.6,0) {$A$};
      \node[below] at (3.4,0) {$B$};
      \node[below] at (4.9,0) {$C$};
      \node[below] at (6,0) {$O$};
      \fill[cyan!15]
        (0.6,0) arc (180:0:2.5)
        -- (5.6,0) arc (0:180:0.7)
        -- (4.2,0) arc (0:180:1.8)
        -- cycle;
      \draw[dotted] (4.2,0) arc (0:180:1.8);
      \node at (2.6,1.2) {$\hat{\Sigma}_{\rm comp}$};
      \node at (4.3,1.6) {$\hat{\Sigma}_{(AB):C:O}$};
    \end{scope}
  \end{tikzpicture}
    \caption{The codimension-one slice $\Sigma$ contains the sub-slice $\hat{\Sigma}_{A:B:C:O} = \hat{\Sigma}_{(AB):C:O} \cup \hat{\Sigma}_{\rm comp}$.}
    \label{fig:V-decompos}
\end{figure}
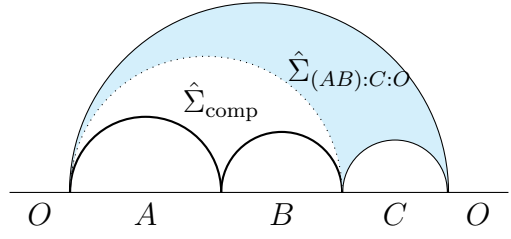
\end{proof}
From Property $P2$, $V_{\{A_i\}}$ can be interpreted as the number of tensors (or gates) which produce $k-$partite entanglement, with $3\leq k\leq q$, when reconstructing the state $\rho_{\{A_i\}}$.

\section{Entanglement Wedge Polygon in AdS$_3$ for pure states}
\label{sec:AdS3-pure}
In three-dimensional spacetime, the volume of the EWP can be computed by the Gauss-Bonnet theorem:\footnote{See \cite{Pastawski:2015qua} for a similar analysis and \cite{Abt:2017pmf} for a related observation regarding the volume of the EW.}
\begin{equation}
\label{eq:GB-theorem}
    -\int_{\Sigma} K \, dA = \int_{\partial\Sigma} k_g \, ds - 2 \pi \,  \chi(\Sigma) \, .
\end{equation}
Here $K$ is the Gaussian curvature of the two-dimensional spacelike slice $\Sigma$, $k_g$ is the geodesic curvature of the boundary $\partial\Sigma$, and $\chi(\Sigma)$ is the Euler characteristic of $\Sigma$.
The Gaussian curvature can be expressed in terms of the intrinsic curvature of $\Sigma$ as $K = R^{(2)}/2$.
In the case of the hyperbolic plane $R^{(2)}=-2/L^2$,
the Gaussian curvature of $\Sigma$ is constant, so eq.~\eqref{eq:GB-theorem} directly gives the volume $V(\Sigma)$.
Regarding the boundary integral, since $\partial\Sigma$ is piece-wise smooth, we have to sum over all the smooth pieces and add the defect angle $\delta_i$ of every smooth piece with respect to the previous one. 
More details can be found in Appendix~\ref{sec:GB-details}.
The Gauss-Bonnet theorem then gives
\begin{equation}
\label{eq:GB-theorem-general}
   \frac{V(\Sigma)}{L^2} = \sum_i \int_{\partial \Sigma_i} k_g \, ds + \sum_i \delta_i - 2 \pi \, \chi(\Sigma) \, .
\end{equation}
When the EWP is bounded by geodesics, for which the geodesic curvature $k_g$ vanishes, its volume depends only on the number of connected boundary subregions and on the topology of the EWP. 
Since geodesics end normally to the AdS boundary \cite{Rangamani:2016dms}, the deficit angles where two geodesics meet at the AdS boundary is $\delta_i=\pi$.
We then get the general result
\begin{equation}
\label{eq:V-topological}
    V_{\{A_i\}} = (n-2 \chi(\Sigma_{\{A_i\}})) \pi L^2 \, ,
\end{equation}
with $n$ the number of \textit{vertices} of the EWP $\Sigma_{\{A_i\}}$.
For $q+1$ \textit{connected} partitions of a pure state $\rho_{A_1 \ddd A_q O}$, we have $n \leq q+1$.

\subsection{Vacuum state in AdS$_3$}
\label{subsec:pure_AdS3}
\begin{figure}
    \centering
    \begin{tikzpicture}
    \begin{scope}[yshift=5cm]
      \draw[->] (-0.2,0) -- (6.0,0) node[right] {$x$};
      \draw[->] (0,-0.2) -- (0,3.0) node[above] {$z$};
      \draw[thick] (0.6,0) arc (180:0:0.6);     
      \draw[thick] (1.8,0) arc (180:0:0.4);     
      \draw[thick] (2.6,0) arc (180:0:0.8);     
      \draw[thick] (4.2,0) arc (180:0:0.7);     
      \draw[thick] (5.6,0) arc (0:180:2.5);     
      \node[below] at (0.3,0) {$O$};
      \node[below] at (1.2,0) {$A_1$};
      \node[below] at (2.2,0) {$A_2$};
      \node[below] at (3.4,0) {$A_3$};
      \node[below] at (4.9,0) {$A_4$};
      \node[below] at (5.8,0) {$O$};
      \fill[cyan!15]
        (0.6,0) arc (180:0:2.5)
        -- (5.6,0) arc (0:180:0.7)
        -- (4.2,0) arc (0:180:0.8)
        -- (2.6,0) arc (0:180:0.4)
        -- (1.8,0) arc (0:180:0.6)
        -- cycle;
      \draw[dotted] (2.6,0) arc (0:180:1.0);
      \draw[dotted] (4.2,0) arc (0:180:1.8);
      \node at (1.8,0.7) {$\pi L^2$};
      \node at (2.6,1.4) {$\pi L^2$};
      \node at (4.3,1.6) {$\pi L^2$};
    \end{scope}
  \end{tikzpicture}
  \qquad \qquad
    \includegraphics[scale=0.6]{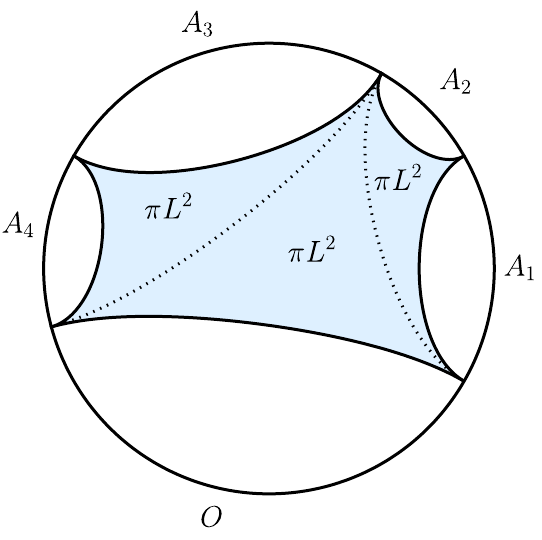}
    \caption{Ideal polygon with $n=5$ on the hyperbolic plane (left) and disk (right).}
    \label{fig:ideal-polygon}
\end{figure}
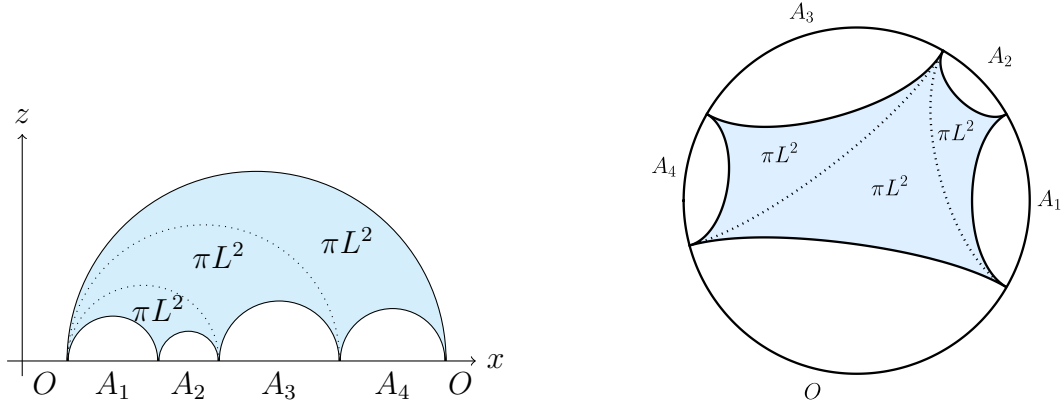

An ideal polygon on the hyperbolic plane has edges which are geodesics ending at the boundary. As such, all the vertices have zero-degree interior angles. Any ideal polygon with $n$ sides can be subdivided into $n-2$ ideal triangles, as in Fig.~\ref{fig:ideal-polygon}. From the known fact that any ideal triangle on the hyperbolic plane has area $\pi L^2$, where $L$ is the radius of the hyperbolic plane, we conclude that any ideal polygon with $n$ sides has area $(n-2) \pi L^2$. This agrees with eq.~\eqref{eq:V-topological}, by taking into account that an ideal polygon has the topology of a disk, $\chi=1$.

\begin{figure}[t]
  \centering
  \begin{tikzpicture}
    \begin{scope}[scale=0.9]
      \draw[->] (-0.2,0) -- (6.0,0) node[right] {$x$};
      \draw[->] (0,-0.2) -- (0,3.0) node[above] {$z$};
      \draw[dotted] (0.6,0) arc (180:0:0.6);     
      \draw[very thick] (1.8,0) arc (180:0:0.3);     
      \draw[dotted] (2.4,0) arc (180:0:0.6);     
      \draw[very thick] (3.6,0) arc (180:0:0.3);     
      \draw[dotted] (4.2,0) arc (180:0:0.6);     
      \node[below] at (0.3,0) {$O$};
      \node[below] at (1.2,0) {$A_1$};
      \node[below] at (2.1,0) {$O$};
      \node[below] at (3.0,0) {$A_2$};
      \node[below] at (3.9,0) {$O$};
      \node[below] at (4.8,0) {$A_3$};
      \node[below] at (5.6,0) {$O$};
      \draw[very thick] (0.6,0) arc (180:0:2.4);
      \fill[cyan!15]
        (0.6,0) arc (180:0:2.4)
        -- (5.4,0)
        arc (0:180:0.6)
        -- (4.2,0)
        arc (0:180:0.3)
        -- (3.6,0)
        arc (0:180:0.6)
        -- (2.4,0)
        arc (0:180:0.3)
        -- (1.8,0)
        arc (0:180:0.6)
        -- cycle;
      \node at (3.0,1.4) {$\Sigma_{A_1:A_2:A_3:O}$};
    \node at (12.0,1.5) {\includegraphics[scale=0.5]{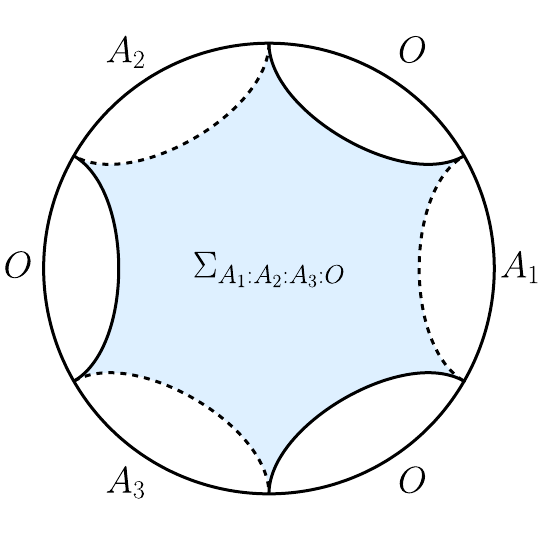}};
    \end{scope} 
  \end{tikzpicture}
  \caption{Fully connected phase in Poincar\'e AdS$_3$ (on the left) and global AdS$_3$ (on the right) for $q=3$. The bold curves represent $\Gamma_O$ and the dotted ones $\Gamma_{A_i}$.}
  \label{fig:AdS3-Phases}
\end{figure}
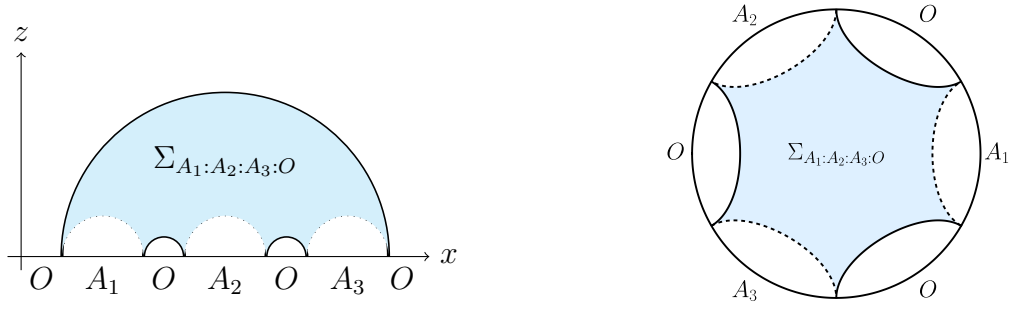

By using this result, it is straightforward to study the EWP for a $(q+1)-$partite vacuum state of a CFT$_2$ on both non-compact and compact space, $\rho_{A_1 \ddd A_q O}= \rho_{\rm vac}$.
Let us consider $q$ connected boundary subregions $A_1, \dots, A_q$ in Poincar\'e or global AdS$_3$ spacetime, and let us denote by $O$ the purifier.
If all the $A_i$ are adjacent and have finite size, 
$O$ is the union of two semi-infinite segments in Poincar\'e AdS$_3$, but a connected subregion with finite size in global AdS$_3$. Regardless of the sizes of the $A_i$, the volume is 
\begin{equation}
\label{eq:pure-AdS3-adjacent}
V_{\{A_i\}:O}= (q-1)\pi L^2 \, , 
\end{equation}
as it is clear from Fig.~\ref{fig:ideal-polygon}.
On the contrary, if the $A_i$ are not adjacent,
the volume depends on their sizes and separations.
The highest volume $V_{\{A_i\}}$ is obtained when all the subregions $A_i$ are non-adjacent, so that $O$ is the union of $q$ (or $q+1$) connected components in global (or Poincar\'e) AdS$_3$. 
If $\Gamma_O$ is fully connected, the EWP is an ideal polygon with $2q$ sides, so $V_{\{A_i\}:O} = 2(q-1) \pi L^2$.
An example of this configuration is shown in Fig.~\ref{fig:AdS3-Phases}.
By decreasing the size of one of the $A_i$, eventually $\Gamma_O$ gets partially disconnected, so the number of vertices of the EWP decreases and/or its Euler characteristic increases. 
As a result, the volume $V_{\{A_i\}:O}$ decreases of an integer multiple of $\pi L^2$.
The process can be iterated until $\Gamma_O$ becomes fully disconnected and $V_{\{A_i\}:O}=0$.
So, the volume of the EWP is quantized in units of $\pi L^2$ and
\begin{equation}
    0 \leq V_{\{A_i\}:O} \leq 2(q-1) \pi L^2 \, ,  \label{eq:rangepure}
\end{equation}
where the mimimal and maximal value are achieved when $\Gamma_O$ is fully disconnected and fully connected, respectively.

As a last example, let us assume that in Poincar\'e AdS$_3$ the subregion $A_1$ is a semi-infinite segment. The EWP is not an ideal polygon. If all the $A_i$ are adjacent, the purifier $O$ is also a semi-infinite segment.
The volume reads
\begin{equation}
\label{eq:pure-AdS3-adjacent-infinite}
V_{\{A_i\}:O}=(q-1)\pi L^2 \, , 
\end{equation}
as it can be promptly computed by the Gauss-Bonnet theorem.
Details are deferred to Appendix~\ref{subsec:GB-infinite}.

\subsection{Pure state in BTZ}
\label{subsec:pure_BTZ}
The next example is a pure state in BTZ spacetime. We discuss two different pure states at Lorentzian time $t=0$: the thermofield double state and a highly excited holographic state.
\\

\subsubsection{Two-sided BTZ}
\label{subsubsec:two-sided_BTZ}
We start from a thermofield double state (TFD) of the product of two identical copies of CFT$_2$,
$\ket{TFD} = \sum_n e^{-\beta E_n/2} \ket{E_n}_L\ket{E_n}_R$ up to a normalization.
The bulk dual is an eternal two-sided BTZ spacetime
\begin{equation}
\label{eq:BTZ-metric}
    ds^2 = -\left( \frac{r^2 - r_h^2}{L^2} \right) dt^2 + \left( \frac{r^2 - r_h^2}{L^2} \right)^{-1} dr^2 + r^2 d\varphi^2 \, ,
\end{equation}
of which geometry we consider the time-reflection invariant slice $t=0$.
The inverse temperature is $\b = 2 \pi L^2/r_h$. 
In the following, we assume that the purifier $O$ contains one full boundary, $L \subset O$.
In general, the purifier can also have a component $O_R$ on the other boundary, where also all the $q$ subregions $A_i$ are located.
The intrinsic curvature of the $t=0$ bulk time-slice of BTZ spacetime is still $R^{(2)}=-2/L^2$, so the result in eq.~\eqref{eq:V-topological} applies to this case as well.
Let us distinguish between the case of a black brane, where $\varphi = x/L$  with $x$ non-compact and ranging in $-\infty < x < +\infty$, and a black hole, where $\varphi$ is compact and $\varphi \sim \varphi + 2\pi$. 
\\

{\bf Planar BTZ.}
For a black brane, if we consider $O=L \cup O_R$ with $O_R$ the union of two disconnected semi-infinite segments, $\Gamma_O$ joins the endpoints of the two segments $O_R$ in the right exterior of the black brane.
In this case, the $q$ partitions $A_i$ of the pure state $\rho_{A_1 \ddd A_q O} = \ket{TFD}\!\bra{TFD}$ are adjacent and have finite size. The discussion runs parallel to that in Poincar\'e AdS$_3$, see the left panel of Fig.~\ref{fig:ideal-polygon}.
Analog conclusions can be drawn if $O_R$ additionally includes disconnected components of finite size, similarly to the left panel of Fig.~\ref{fig:AdS3-Phases}. 
\\

{\bf Global BTZ.}
For a black hole, the homology condition enforces a phase transition for the minimal $\Gamma_A$. 
Let us denote by $2\alpha_A$ the opening angle of a boundary subregion $A$.
For a boundary subregion with $\alpha_A < \alpha_{\rm t}$ the shortest geodesic homologous to $A$ is connected, whereas for $\alpha_A > \alpha_{\rm t}$  it is given by the union of the shortest geodesic for the complementary subregion on the same boundary and the black hole horizon.\footnote{The value of the critical opening angle is given by $\alpha_{\rm t} = \frac{L}{r_h} \coth^{-1} \left( 2 \coth \left( \frac{\pi r_h}{L} \right) -1 \right) >\frac{\pi}{2}$ \cite{Hubeny:2013gta}.} 
Consequently, the topology of $\Sigma_{\{A_i\}:O}$ is modified and a phase transition also occurs for $V_{\{A_i\}:O}$.
We are interested in the $(q+1)-$partite state $\rho_{A_1 \ddd A_q O} = \ket{TFD}\! \bra{TFD}$, where the purifier $O=L \cup O_R$ is the union of the full left boundary and a subregion of the right one.

\begin{figure}[t]
  \centering
  \includegraphics[scale=0.5]{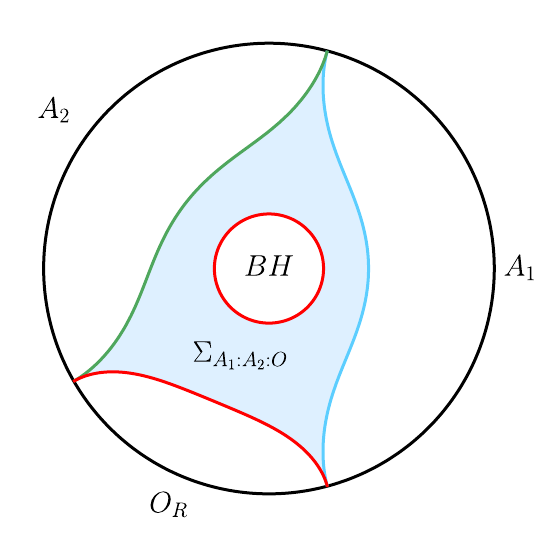} 
  \qquad \qquad
  \includegraphics[scale=0.085]{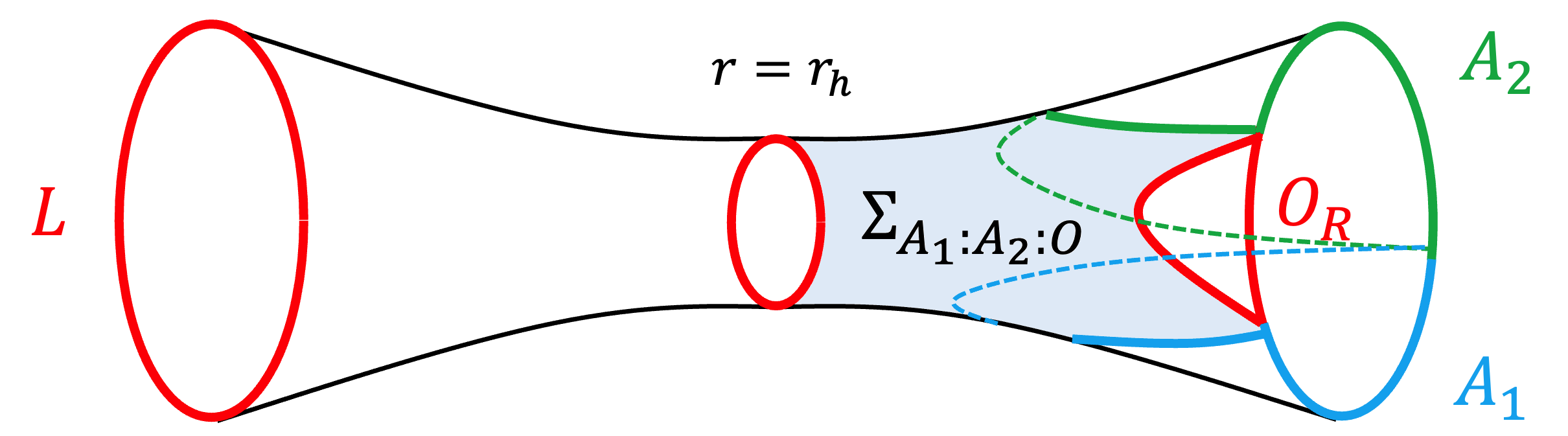} 
  \qquad \qquad
  \includegraphics[scale=0.5]{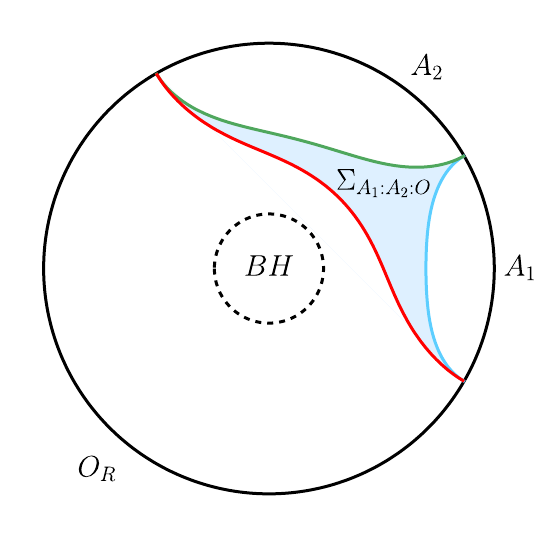}
  \qquad \qquad
  \includegraphics[scale=0.085]{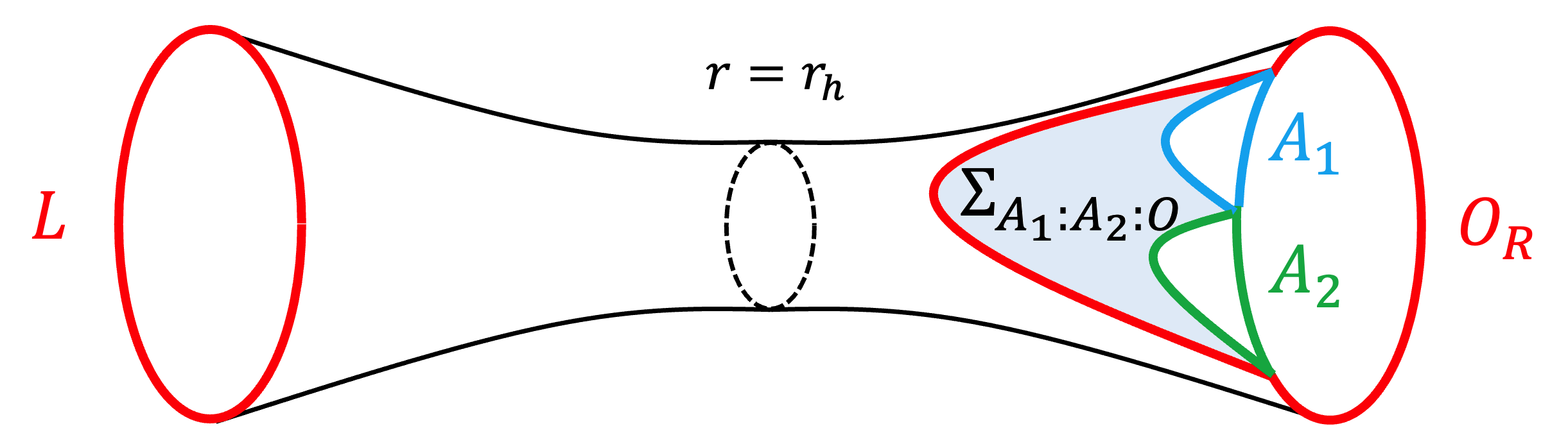} 
  \qquad \qquad
  \includegraphics[scale=0.5]{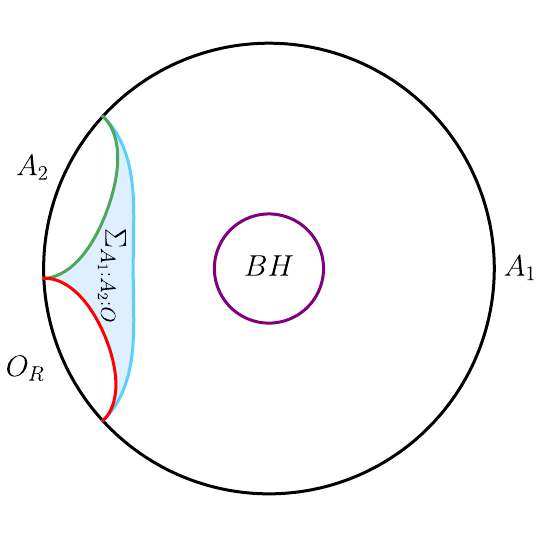}
  \qquad \qquad
  \includegraphics[scale=0.085]{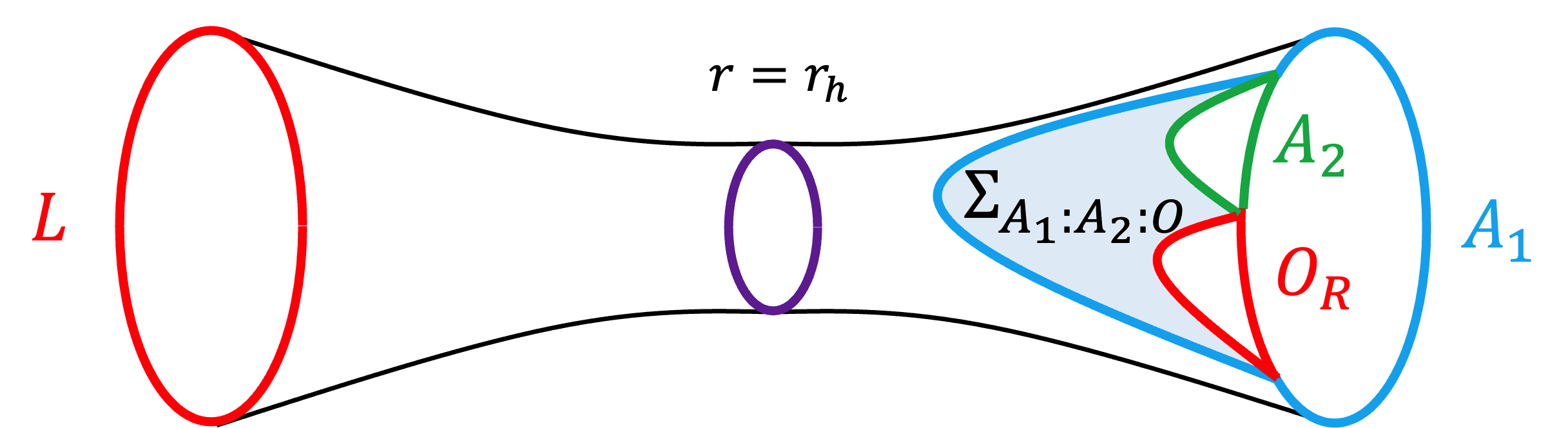} 
  \caption{Allowed configurations of the EWP for the TFD state in compact space, $\rho_{A_1A_2A_3O}=\ket{TFD}\!\bra{TFD}$. 
  The purifier $O=L \cup O_R$ is the union of the full left boundary $L$ and a connected subregion $O_R$ of the right one.
  Top, case (a): $\Gamma_O$ is disconnected and includes the horizon.
  Middle, case (b): $\Gamma_O$ is connected and the horizon is not involved.
  Bottom, case (c): both $\Gamma_O$ and $\Gamma_{A_1}$ are disconnected and include the horizon.
  For pictorial convenience, we have set $r_h=0.4$ and $L=1$.}
  \label{fig:BTZ-global-phases}
\end{figure}

For a connected $O_R$, we have $q$ adjacent subregions $A_1, \dots, A_q$. We can distinguish three cases, displayed in Fig.~\ref{fig:BTZ-global-phases}:
\begin{itemize}
    \item[(a)] $\alpha_i < \alpha_{\rm t}$ but $\sum_{i=1}^q \alpha_i > \alpha_{\rm t}$: $\Sigma_{\{A_i\}:O}$ has the topology of an annulus, $\chi=0$.
    \item[(b)] $\alpha_i < \alpha_{\rm t}$ and $\sum_{i=1}^q \alpha_i < \alpha_{\rm t}$: $\Sigma_{\{A_i\}:O}$ has the topology of a disk, $\chi=1$.
    \item[(c)] One of the $\alpha_i > \alpha_{\rm t}$: $\Sigma_{\{A_i\}:O}$ has again the topology of a disk.
\end{itemize}
Therefore, from eq.~\eqref{eq:V-topological} the volume of the EWP is $V_{\{A_i\}:O} = (q+1)\pi L^2$ in case (a) and $V_{\{A_i\}:O}= (q-1) \pi L^2$ in cases (b),(c).\footnote{When $\cup_{i=1}^q A_i$ is the full right boundary, the purifier $O=L$ has no components on the right boundary, $O_R = \emptyset$ (in this case $\Gamma_O$ is the black hole horizon). Consequently, the EWP has $n=q$ vertices. The volume in the two phases is then $V_{\{A_i\}:L} = q \pi L^2$ and $V_{\{A_i\}:L}= (q-2) \pi L^2$.
}
Note that, contrary to the entanglement entropy, at the phase transition triggered by the change of size of one subregion, $V_{\{A_i\}:O}$ is discontinuous, with a finite jump of $2 \pi L^2$.

For a non-connected $O_R$,
we have $q$ subregions $A_1, \dots, A_q$ with non-vanishing separations.  
Eq.~\eqref{eq:V-topological} implies that the configuration with greatest volume is realized by maximizing the number of adjacent sides of the EWP and by minimizing its Euler characteristic. 
Assuming that the $A_i$ are connected,
we obtain
 \begin{equation}
    0 \leq V_{\{A_i\}:O} \leq 2q \pi L^2 \, ,
 \end{equation}
 where the volume is quantized in units of $\pi L^2$. 
 The upper bound is achieved when $\Gamma_O$ is fully connected and the EWP has the annulus topology, namely all the connected components have size $\alpha_i, \alpha_{O_{R,i}} < \alpha_{\rm t}$.
 For $q=2$ and equal-sized subregions $\a_i=\a$, phase diagrams for the shortest $\Gamma_O$ has been studied in \cite{Ben-Ami:2014gsa}. It has been found that the fully connected configuration can only be realized for small temperature, whereas for  large temperature the dominant configuration is the fully disconnected one, corresponding to $V_{\{A_i\}:O}=0$. 
 \\

\subsubsection{One-sided BTZ}
\label{subsubsec:one-sided_BTZ}
In the case of one-sided BTZ, we can place an end-of-the-world (EOW) brane
inside the black hole horizon \cite{Hartman:2013qma,Takayanagi:2011zk}. In this case the state in a single CFT (for instance the right CFT) is pure, $\ket{\Psi}_R$. We can regard this as a high energy excitation by a primary state in AdS$_3/$CFT$_2$ with energy higher than the black hole threshold.
At initial time $t=0$, the EOW brane coincides with the event horizon $r=r_h$. Therefore, the geometry of the bulk slice $t=0$ looks exactly like half the TFD state case, for instance the right half in the right panels of Fig.~\ref{fig:BTZ-global-phases}, with the event horizon replaced by the EOW brane.

We consider a $(q+1)-$partition of the boundary pure state, $\rho_{A_1 \ddd A_q O} = \ket{\Psi}_R \! \bra{\Psi}_R$.
The EOW brane is homologically trivial, meaning that the minimal geodesics $\Gamma_{A_i}$ can have one end on the brane and, similarly, the EWP can end on the brane.
This brings more possible configurations for the EWP compared to the two-sided case. 
As for the two-sided case, let us distinguish between a black brane with non-compact boundary and a black hole with compact boundary.
\\

{\bf Planar boundary $\mathbb{R}$.}
Let us start from a $(q+1)-$partition of the excited state $\ket{\Psi}_R$ of a CFT$_2$ on non-compact space, $\rho_{A_1 \ddd A_q O}= \ket{\Psi}_R\!\bra{\Psi}_R$.
In other words, we consider $q$ connected and adjacent subregions $A_i$ whose union is the full boundary of a one-sided black brane.
This means that $A_1$ and $O$ are semi-infinite segments.
Clearly, $\Gamma_{A_1}$ and $\Gamma_{O}$ are $x$-constant geodesics with one end at the boundary and the other end at the EOW brane.
The configuration of the remaining $A_i$ instead depends on their size. 
As studied in \cite{Takayanagi:2017knl}, if $l_i < \beta \log(\sqrt{2}+1)/\pi$, with $\beta$ the inverse temperature, the corresponding $\Gamma_{A_i}$ connects the left and right ends of $A_i$.
If $l_i > \beta \log(\sqrt{2}+1)/\pi$, the corresponding $\Gamma_{A_i}$ is disconnected and each component connects the boundary to the EOW brane.
The largest volume is obtained when $l_i < \beta \log(\sqrt{2}+1)/\pi$ for $2\leq i \leq q$, see the left panel of Fig.~\ref{fig:BTZ-planar-pure}.
By the Gauss-Bonnet theorem, the value is
$V_{\{A_i\}:O}= (q-1) \pi L^2$,
see Appendix~\ref{subsec:GB-EOW} for details.
When one of the $l_i > \beta \log(\sqrt{2}+1)/\pi$, $\Sigma_{\{A_i\}:O}$ gets disconnected into two components and the volume decreases of a $\pi L^2$ contribution.
Therefore, the volume is a quantized multiple of $\pi L^2$, with 
\begin{equation}
  0 \leq V_{\{A_i\}:O} \leq (q-1) \pi L^2 \, .
\end{equation}
The lower bound is attained for $l_i > \beta \log(\sqrt{2}+1)/\pi$ for every $i=2, \dots, q$. Note that the upper bound agrees with the Poincar\'e AdS$_3$ result in eq.~\eqref{eq:pure-AdS3-adjacent-infinite}.
\begin{figure}
    \centering
    \includegraphics[scale=0.09]{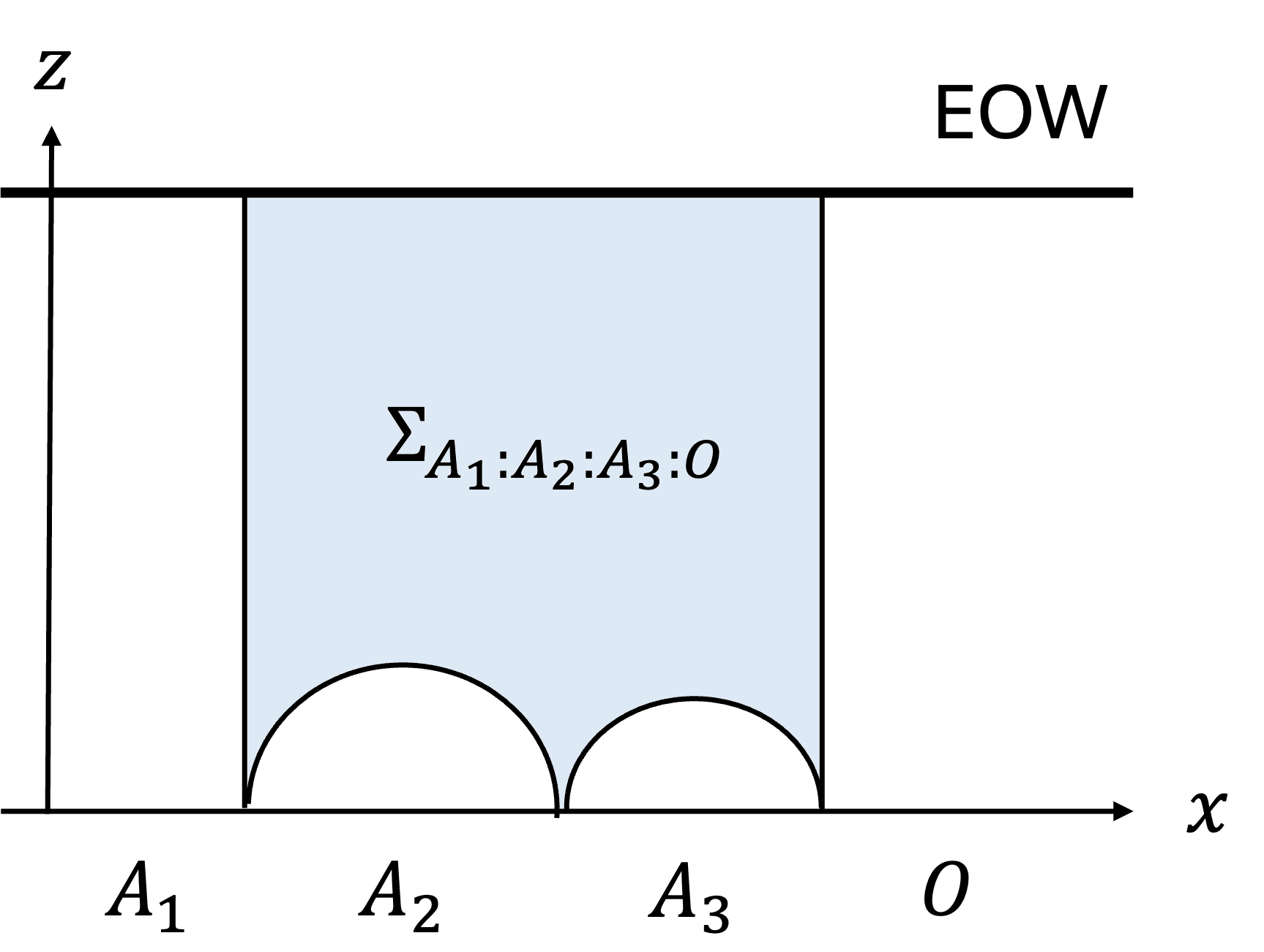}
    \qquad
    \includegraphics[scale=0.09]{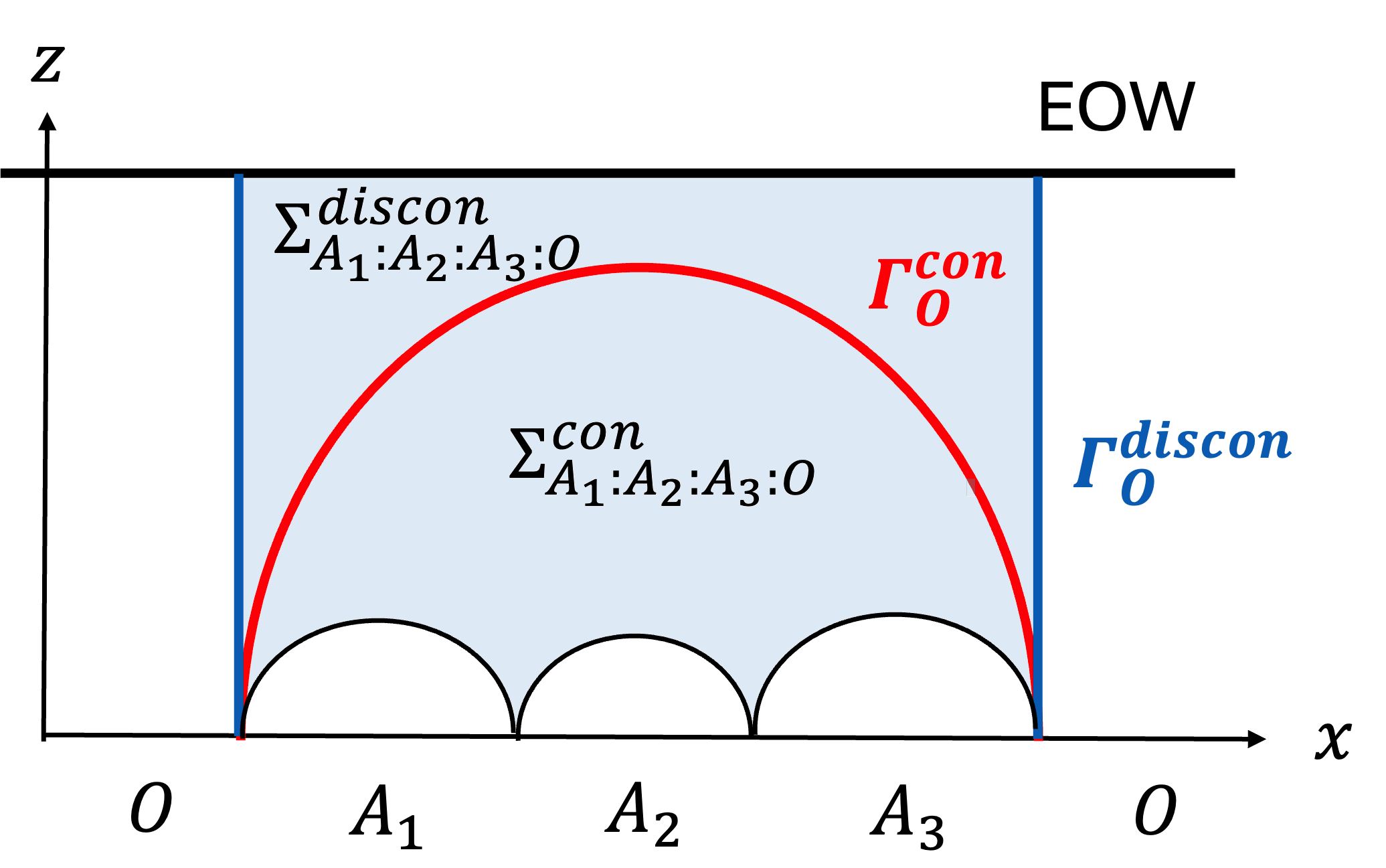}
    \caption{The EWP for the excited state $\ket{\psi}_R$ on non-compact space, $\rho_{A_1A_2A_3O}=\ket{\psi}_R \!\bra{\psi}_R$, with a connected $O$
    (on the left) and a disconnected $O$ (on the right). 
    The thin curves represent $\Gamma_{A_i}$ and the bold line the EOW brane at $r=r_h$.}
    \label{fig:BTZ-planar-pure}
\end{figure}

As a second example, we consider $q$ adjacent partitions $A_i$ with finite size $l_i$, so that the purifier $O$ is the union of two semi-infinite segments, as in the right panel of Fig.~\ref{fig:BTZ-planar-pure}. 
If $\sum_{i=1}^q l_i < \beta \log(\sqrt{2}+1)/\pi$, the geodesic $\Gamma_O$ connects the left end of $A_1$ with the right end of $A_q$. In this case, we have $V_{\{A_i\}:O} = (q-1) \pi L^2$ as in the vacuum case \eqref{eq:pure-AdS3-adjacent}.
If $\sum_{i=1}^q l_i > \beta \log(\sqrt{2}+1)/\pi$, the geodesic $\Gamma_O$ is given by two disconnected $x-$constant geodesics ending at the EOW brane.
The volume then reads $V_{\{A_i\}:O}=q \pi L^2$. In this phase, the presence of the EOW brane increases the volume of the EWP compared to vacuum.
\\

\begin{figure}[t]
  \centering
  \includegraphics[scale=0.5]{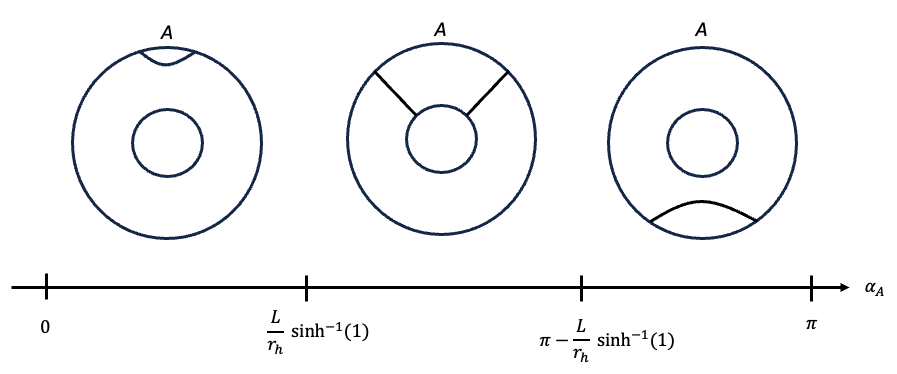}
  \caption{Configurations of the shortest geodesic $\Gamma_{A}$ in the one-sided global BTZ. The parameter $\alpha_A$ denotes half the size of subregion $A$. The central circumference represents the EOW brane at $t=0$.}
  \label{fig:BTZ-global-pure}
\end{figure}

{\bf Compact boundary $S^1$.}
Given a subregion $A$ with size $2 \a_A$ of the excited state $\ket{\Psi}_R$ of a CFT$_2$ on compact space $S^1$,
we have three possible configurations for the minimal geodesic $\Gamma_{A}$, that we show in Fig.~\ref{fig:BTZ-global-pure}. 
The shortest configuration is $\Gamma_{A} = \Gamma_{A}(\alpha_A)$ for $0<\alpha_A < L \, \arcsinh(1)/r_h$ and $\Gamma_{A} = \Gamma_{A}(\pi-\alpha_A)$ for $\pi -L \, \arcsinh(1)/r_h <\alpha_A < \pi$.
For $L \, \arcsinh(1)/r_h <\alpha_A < \pi -L \, \arcsinh(1)/r_h$, the shortest configuration is the union of two disconnected radial geodesics, each of which has one end at $\partial A$ and the other end on the EOW brane. 
Note that $L \, \arcsinh(1)/r_h < \pi/2$ constraints $r_h > 2L \, \arcsinh(1)/\pi$, that is always true in the BTZ phase $r_h>L$.

\begin{figure}[t]
  \centering
  \includegraphics[scale=0.8]{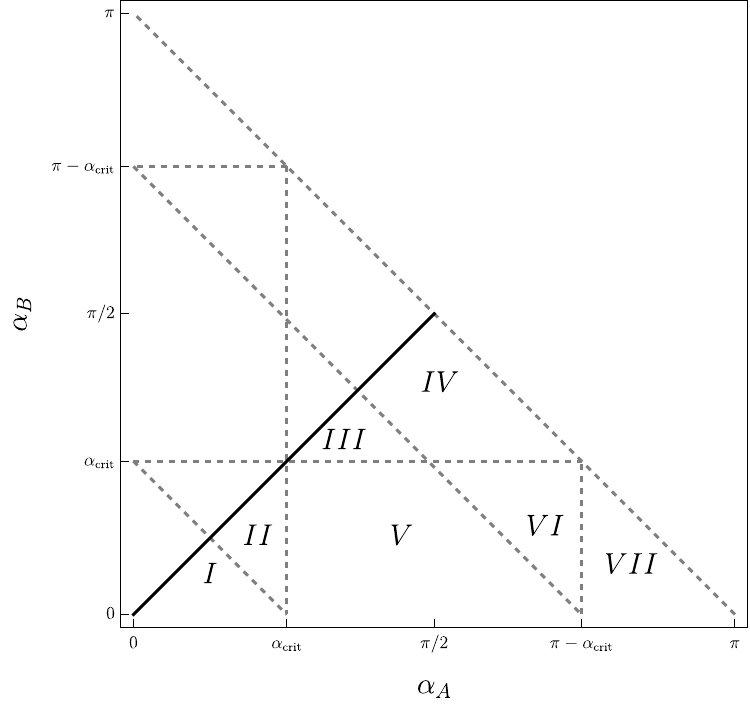}
  \caption{Regions of the parameter space $(\alpha_A,\alpha_B)$ with given value of $V_{A:B:C}$. We denote by $\alpha_{\rm crit} = L \, \arcsinh(1)/r_h$ the critical angle. The regions in $\alpha_B > \alpha_A$ are obtained by exchanging $A$ and $B$. We have $\alpha_C=\pi-\alpha_A-\alpha_B$, which becomes negative in the upper right triangle in the plot.}
  \label{fig:BTZ-global-pure-configurations1}
\end{figure}

\begin{figure}[t]
  \centering
  \includegraphics[scale=0.1]{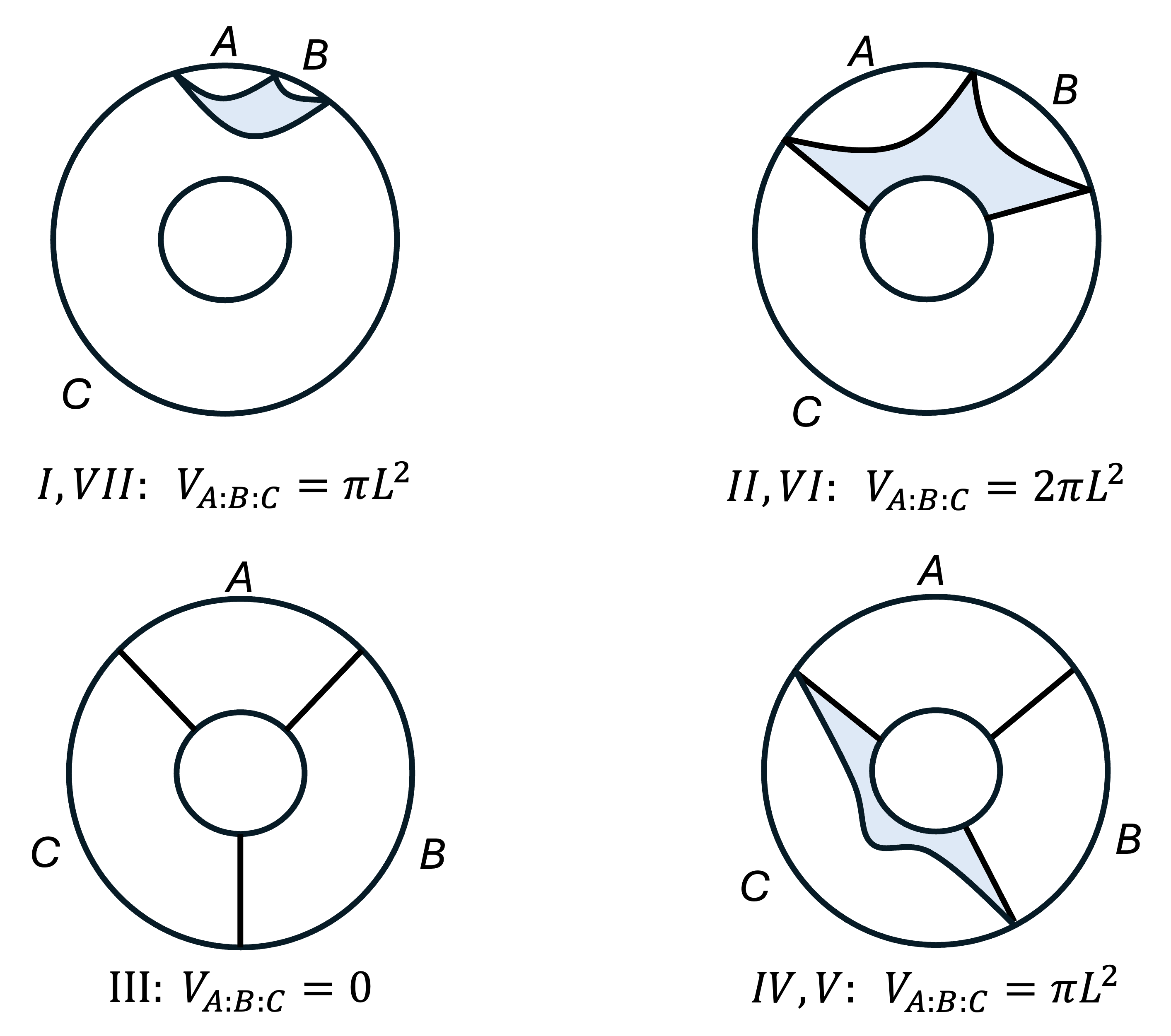}
  \caption{Allowed configurations of the EWP for the thermal state on compact space, $\rho_{ABC}=\rho_{\rm th}$, and corresponding volume $V_{A:B:C}$. The pairs of configurations $(I,VII), (II,VI), (IV,V)$ are obtained by exchanging $A$ and $C$.}
  \label{fig:BTZ-global-pure-configurations2}
\end{figure}

For simplicity, we consider a three-partition $A,B,C$ of the excited state of a CFT$_2$ on a compact space $S^1$, $\rho_{ABC} = \ket{\Psi}_R\! \bra{\Psi}_R$. 
The union $A \cup B \cup C$ is the full boundary, $\alpha_A + \alpha_B+\alpha_C = \pi$.
We study $V_{A:B:C}$ as a function of the sizes of the three subregions. 
In Figs.~\ref{fig:BTZ-global-pure-configurations1} and \ref{fig:BTZ-global-pure-configurations2} we show all the allowed configurations of the EWP and the related value of the volume, which is promptly computed by the Gauss-Bonnet theorem.
To this purpose, we stress that radial geodesics are orthogonal to the EOW brane\footnote{The tangent to a radial geodesic is entirely along the $r$-direction, while the tangent to the EOW at $t=0$ is entirely along the $\varphi$-direction.} and form an angle of zero degrees with the geodesics anchored at the boundary. This information, together with the comments in the previous sections, are enough to apply the Gauss-Bonnet theorem \eqref{eq:GB-theorem-general}.
In the high temperature limit $L/r_h <<1$ we only have region III, $V_{A:B:C}=0$, except for small windows around $\alpha_A =0,\pi$. This is consistent with the intuition that high temperature destroys the (multi-partite) correlations.
Also, we have $V_{A:B:C}=0$ for subregions with equal size at any temperature.

It is also interesting to note that the configuration in which the EWP has the topology of an annulus, like in case (a) of Fig.~\ref{fig:BTZ-global-phases}, is never favored. Indeed, for this configuration to take place, the three subregions must have sizes $\alpha_{A,B,C} \leq L \sinh^{-1}(1)/r_h$. However, $3L\sinh^{-1}(1)/r_h < \pi$ in the BTZ phase $r_h>L$, which excludes this possibility. 
Consequently, $V_{A:B:C} \leq 2 \pi L^2$ and it is quantized in units of $\pi L^2$.

Similar properties can also be seen in the multi-entropy, which is another candidate as a multi-partite entanglement measure. When the EWP ends on the black hole horizon, as shown in the upper-right and lower-right panels of Fig.~\ref{fig:BTZ-global-pure-configurations2}, the genuine tri-entropy becomes larger than its vacuum value. This can be interpreted as a production of multi-partite entanglement due to the high energetic excitation. On the other hand, when two of the subregions are sufficiently small, as shown in the upper-left panel, the genuine tri-entropy coincides with the vacuum value. When the black hole is sufficiently large so that both the bi-partite and the tri-partite geodesics attach to the black hole, as shown in the lower-left panel, the genuine tri-entropy vanishes. Appendix~B of \cite{Fujiki:2026qdt} is helpful to understand this behavior.

\subsection{Pure state in AdS$_3$/BCFT$_2$}
\label{subsec:pure_AdSBCFT}
An additional example is the gravity dual of a CFT defined on a spacetime with a boundary, i.e. a boundary conformal field theory (BCFT). According to the AdS/BCFT construction 
\cite{Takayanagi:2011zk,Fujita:2011fp}, the bulk dual is Poincar\'e AdS$_3$
\begin{equation}
    ds^2 = \frac{L^2}{z^2} \left( - dt^2 + dz^2 + dx^2 \right) \, ,
\end{equation}
with the insertion of an end-of-the-world (EOW) brane whose world-volume is given by $z=x \tan\alpha$.
This setup is dual to a two-dimensional BCFT on a half plane $x>0$.
When we take the subregion $A$ to be a segment $0\leq x\leq l_A$, the length of the shortest bulk geodesic $\Gamma_A$ anchored at $A$ is
\begin{equation}
    \frac{\mathcal{A}(\Gamma_A)}{L}= \log\frac{2l_A}{\varepsilon} + \frac{1}{2} \log\left|\frac{1-\cos\alpha}{1+\cos\alpha}\right| \, ,
    \label{eq:AdSBCFTlength}
\end{equation}
where $\varepsilon$ is the UV cutoff. 
The volume of the homology surface $r_A$ is
\begin{equation}
    \frac{\mathcal{V}_A}{L^2}=\frac{l_A}{\varepsilon}-\alpha-\frac{1}{\tan\alpha}\left(1+\log\frac{l_A\sin\alpha}{\varepsilon}\right) \, .
    \label{eq:AdSBCFTarea}
\end{equation}
For a subregion $B$ given by the interval $l_A\leq x\leq l_A+l_B$, which is adjacent to $A$, there are two possible geometric phases, which are drawn in Fig.~\ref{fig:AdSBCFTadphases}:
\begin{figure}[t]
  \centering
  \begin{tikzpicture}
    \begin{scope}[scale=1.1]
      \def\Rmax{3}
      \def\alphaAngle{55} 
      \draw[->] (-0.2,0) -- (4,0) node[right] {$x$};
      \draw[->] (0,-0.2) -- (0,3) node[above] {$z$};
      \fill[cyan!15]
        (1,0) arc (0:55:1.0)
        -- ({cos(55)},{sin(55)})
        -- ({2*cos(55)},{2*sin(55)})
        arc (55:0:2)
        -- (2,0)
        arc (0:180:0.5)
        -- cycle;
      \draw[very thick]
        (0,0) --
        ({\Rmax*cos(\alphaAngle)}, {\Rmax*sin(\alphaAngle)})
        node[pos=1.0, above right] {EOW brane};
      \draw[thick] (0,0) -- (1,0);
      \node[below] at (0.5,0) {$A$};
      \draw (1,0) arc (0:55:1.0);
      \draw[thick] (1,0) -- (2,0);
      \node[below] at (1.5,0) {$B$};
      \node[below] at (2.5,0) {$O$};
      \draw (2,0) arc (0:55:2.0);
      \draw (1,0) arc (180:0:0.5);
      \draw (0.3,0) arc[start angle=0, end angle=\alphaAngle, radius=0.3];
      \node at ({0.5*cos(\alphaAngle/2)}, {0.5*sin(\alphaAngle/2)}) {$\alpha$};
      \node[below] at (1.29,1.2) {$\Sigma_{A:B:O}$};
    \end{scope}
    \node at (6.0,1.6) {$\Longleftrightarrow$};
    \node[above] at (6.0,1.8) {\small phase transition};
    \begin{scope}[xshift=8cm,scale=1.1]
      \def\Rmax{3}
      \def\alphaAngle{55}
      \draw[->] (-0.2,0) -- (4,0) node[right] {$x$};
      \draw[->] (0,-0.2) -- (0,3) node[above] {$z$};
      \draw[very thick]
        (0,0) --
        ({\Rmax*cos(\alphaAngle)}, {\Rmax*sin(\alphaAngle)})
        node[pos=1.0, above right] {EOW brane};
      \draw[thick] (0,0) -- (1,0);
      \node[below] at (0.5,0) {$A$};
      \draw[thick] (1,0) arc (0:55:1.0);
      \draw[thick] (1,0) -- (2,0);
      \node[below] at (1.5,0) {$B$};
      \node[below] at (2.5,0) {$O$};
      \draw[thick] (2,0) arc (0:55:2.0);
      \draw (0.3,0) arc[start angle=0, end angle=\alphaAngle, radius=0.3];
      \node at ({0.5*cos(\alphaAngle/2)}, {0.5*sin(\alphaAngle/2)}) {$\alpha$};
    \end{scope}
  \end{tikzpicture}
  \caption{Connected (on the left) and disconnected (on the right) phase for adjacent subregions in AdS/BCFT. In the disconnected phase on the right panel, the EWP is the empty set because $\Gamma_A\cup\Gamma_O=\Gamma_B$.}
  \label{fig:AdSBCFTadphases}
\end{figure}
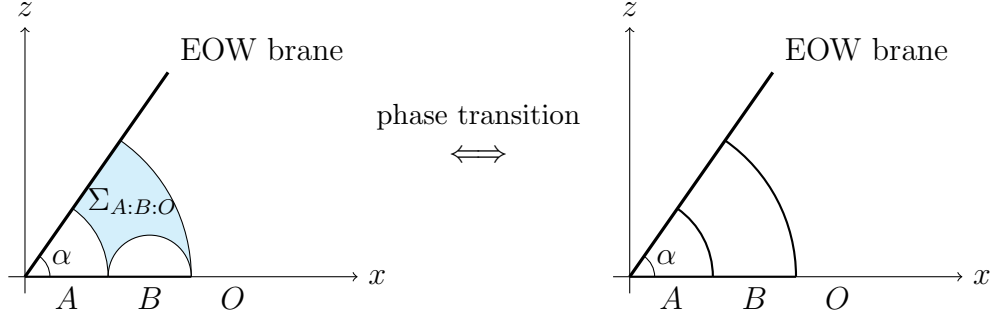
\begin{itemize}

  \item \textbf{Connected phase.}
    The connected phase is realized when
    \begin{align}
        \log\left|\frac{1-\cos\alpha}{1+\cos\alpha}\right|
        + \log\frac{4l_A(l_A+l_B)}{\varepsilon^2}
      \ge 2 \log\frac{l_B}{\varepsilon}.
    \end{align}
    In this phase, we obtain
    \begin{equation}
    \begin{aligned}
        V_{A:B:O}= L^2 \left( \pi - \frac{1}{\tan\alpha}\log\frac{l_A+l_B}{l_A} \right) \, .
    \end{aligned}
    \end{equation}
    The result can be obtained by subtraction of proper volumes in eq.~\eqref{eq:AdSBCFTarea} or by the Gauss-Bonnet theorem, as discussed in Appendix~\ref{subsec:GB-BCFT}.
    The second term can be interpreted as the brane contribution analogous to the g-function. We expect that it captures the dynamics of the boundary state. Notice that in this AdS$_3/$BCFT$_2$ example, the volume of the EWP is not quantized in the unit of $\pi L^2$, as opposed to the AdS$_3/$CFT$_2$ examples. This difference clearly arises due to the presence of the EOW brane.

    \item \textbf{Disconnected phase.}
        The disconnected phase is realized when
        \begin{align}
            \log\left|\frac{1-\cos\alpha}{1+\cos\alpha}\right|
            + \log\frac{4l_A(l_A+l_B)}{\varepsilon^2}
          \le 2 \log\frac{l_B}{\varepsilon}.
        \end{align}
        In this phase, we find
        \begin{equation}
          V_{A:B:O}=0 \, ,
        \end{equation}
        because the minimal bulk geodesic for the subregion $B$ coincides with the union of the minimal geodesics for the subregions $A$ and $O$.
\end{itemize}

We can generalize to the case of $q$ adjacent subregions $A_1,\dots ,A_q$, whose lengths are $l_{A_1},\dots,l_{A_q}$ respectively. In this case, the maximal value of the volume is
\begin{equation}
\label{eq:BCFT_pure_ad}
    V_{A_1:\ddd :A_q:O}= L^2 \left( (q-1)\pi - \frac{1}{\tan\alpha}\log\frac{\sum_{i=1}^{q}l_{A_i}}{l_{A_1}} \right) \, .
\end{equation}

In the non-adjacent case, we set the same subregion length $l$ for the $q$ subregions and the same separation $s$ between each subregion. Then, we obtain the following volume of the EWP for a fully connected $\Gamma_O$:
\begin{equation}
\label{eq:BCFT_pure_nonad}
    V_{A_1:\ddd :A_q:O}= L^2 \left( (2q-2)\pi -\frac{1}{\tan\alpha}\log\left(\frac{ql+(q-1)s}{l}\right) \right) \, .
\end{equation}
These results can be obtained easily from the Gauss-Bonnet theorem.

\section{Entanglement Wedge Polygon in Higher Dimensions}
\label{sec:Higher-d}
In this Section we study the EWP for multi-partitions of states in $d-$dimensional holographic theories, with $d\geq 3$. 
In particular, we start from CFT$_d$ on non-compact space, specializing to the vacuum state and the thermofield double state dual to a two-sided AdS black brane. We then consider a BCFT$_d$ state and a large $N$ confining gauge theory dual to the AdS$_{d+1}$ soliton.

\subsection{Vacuum state in Poincar\'e AdS$_{d+1}$}
\label{subsec:Poincare-d}
\begin{figure}[t]
    \centering
    \includegraphics[scale=1]{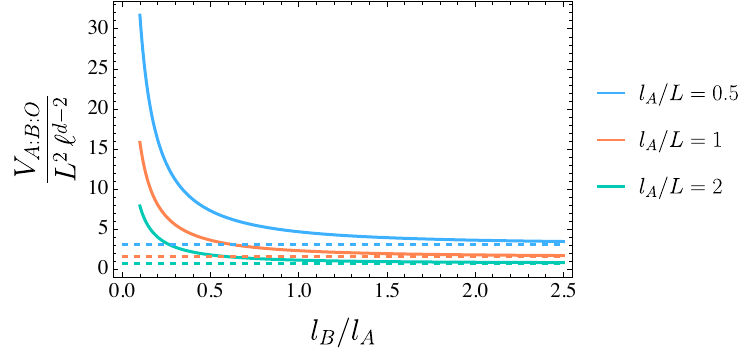}
    \caption{Volume of the EWP for a tri-partition of the vacuum state $\rho_{ABO}$ in $d=3$, with adjacent $A$ and $B$.}
    \label{fig:AdSd-pure-ad}
\end{figure}

We start from the vacuum state of a CFT$_d$ on non-compact space.
The bulk dual is Poincar\'e AdS$_{d+1}$, whose metric in cartesian coordinates is
\begin{equation}
    ds^2 = \frac{L^2}{z^2} \left( -dt^2 + dz^2 + dx^2 + \sum_{i=1}^{d-2} dy_i^2 \right) \, ,
\end{equation}
with $d \geq 3$.
Given a boundary strip $A$ with finite width $l_A$ in the $x$-direction and infinite length $\ell$ in the $y_i$-directions, 
the codimension-two minimal surface $\Gamma_A$ anchored at $A$ is described by
\begin{align}
x_{\pm}(z) &= x_* \pm \frac{l_A}{2} \mp \frac{z^d}{d\,z_*^{d-1}} \,_2F_1\left(\frac{1}{2},\frac{d}{2 (d-1)};\frac{3d-2}{2(d-1)};\left(\frac{z}{z_*}\right)^{2(d-1)}\right) \\
 &\equiv x_* + \chi_{\pm,z_*}(z) \, ,
 \label{eq:GammaA-Poincare-d}
\end{align}
where $x_*$ is the center of the boundary strip, $z_*$ is the turning point of $\Gamma_A$, and $\pm$ denote the two branches of the solution. The width $l_A$ of the boundary strip can be expressed as a function of the turning point as
\begin{equation}
 \frac{l_A}{2} = \frac{\sqrt{\pi} \, \Gamma \left( \frac{3d-2}{2(d-1)} \right)}{d \, \Gamma \left( \frac{2d-1}{2(d-1)} \right)} z_* \, .
\end{equation}
The area of $\Gamma_A$ is \cite{Ryu:2006ef}
\begin{equation}
\label{eq:area-Poincare-d}
    \frac{\mathcal{A}(\Gamma_A)}{L^{d-1}} = \frac{2}{d-2} \, \frac{\ell^{d-2}}{\varepsilon^{d-2}} - \frac{2^{d-1} \, \pi^{\frac{d-1}{2}}}{d-2} \left( \frac{\Gamma\left( \frac{d}{2(d-1)} \right)}{\Gamma\left( \frac{1}{2(d-1)} \right)} \right)^{d-1} \, \frac{\ell^{d-2}}{l_A^{d-2}} \,,
\end{equation}
where $\varepsilon$ is a UV cutoff.
The volume of the codimension-one surface $r_A$ enclosed by the strip $A$ and $\Gamma_A$ is given by \cite{Ben-Ami:2016qex}
\begin{equation}
\label{eq:volume-Poincare-d}
    \frac{\mathcal{V}_A}{L^d} = \frac{\ell^{d-2}}{d-1} \, \frac{l_A}{\varepsilon^{d-1}} - C_d \frac{\ell^{d-2}}{l_A^{d-2}} \, ,
    \qquad
    C_d = \frac{2^{d-2} \, \pi^{\frac{d-1}{2}}}{(d-1)^2} \left( \frac{\Gamma\left( \frac{d}{2(d-1)} \right)}{\Gamma\left( \frac{1}{2(d-1)} \right)} \right)^{d-3} \,  .
\end{equation}
We consider a multi-partition of the vacuum state, $\rho_{A_1 \ddd A_q O}=\rho_{\rm vac}$,
by taking $q$ connected and adjacent strips $A_1, \dots, A_q$ with finite width in the $x-$direction.
The purifier $O$ is the union of two strips with infinite width.
A convenient way to calculate the volume $V_{\{A_i\}:O}$ is by subtraction:
\begin{align}
\label{eq:V-subtraction}
    V_{\{A_i\}:O}
    &= \mathcal{V}_{A_1 \ddd A_q}-\sum_{i=1}^q \mathcal{V}_{A_i} \\
    &= L^d \, \ell^{d-2} \, C_d \left( -\frac{1}{l_{\rm tot}^{d-2}} + \sum_{i=1}^q \frac{1}{l_{A_i}^{d-2}} \right) \, ,
    \label{eq:V-adjacent-Poincare-d}
\end{align}
where $l_{\rm tot}=\sum_{i=1}^q l_{A_i}$ is the sum of the widths $l_{A_i}$ of the single subregions. 
Note that we have $V_{\{A_i\}:O} > 0$, since the codimension-one slice $r_{A_1 \ddd A_q}$, with volume $\mathcal{V}_{A_i \ddd A_q}$, contains all the codimension-one slices $r_{A_i}$, with volume $\mathcal{V}_{A_i}$.
In Fig.~\ref{fig:AdSd-pure-ad} we display the volume for $q=2$ as a function of the ratio $l_B/l_A$ for fixed value of the width $l_A$.
The volume $V_{A:B:O}$ diverges as $C_d/l_B^{d-2}$ for $l_B<<l_A$, a behavior that can be interpreted as due to the short-range entanglement among the three subregions $A,B,O$, which are getting closer to each others when $B$ is small. Note that the divergence is independent of the size of $A$.
Instead, for $l_B>>l_A$, the normalized volume approaches the constant value $C_d/l_A^{d-2}$.
Consistently with our interpretation, the smaller $A$ the closer the three subregions $A,B,O$ get, implying an increasing of the multi-partite correlations.

\begin{figure}[t]
    \centering
    \includegraphics[scale=0.9]{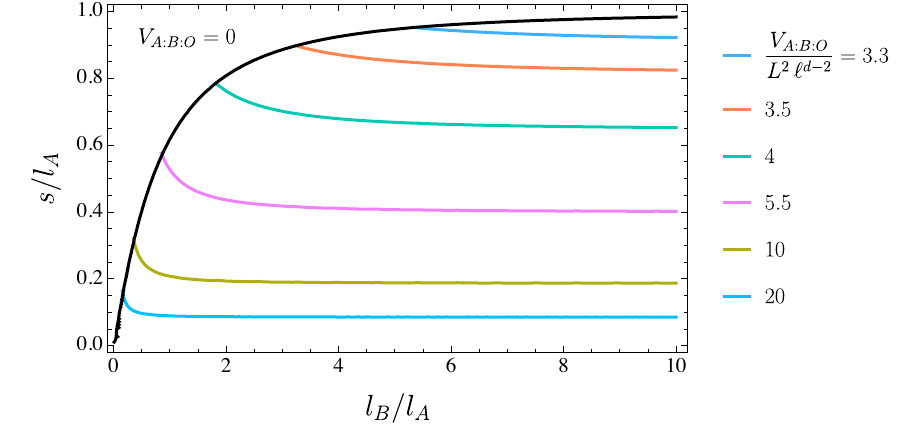}
    \caption{Contour lines of constant volume of the EWP for a tri-partition of the vacuum state $\rho_{ABO}$ in $d=3$, with non-adjacent $A$ and $B$. The black curve represents the critical separation $s_{\rm crit}^{(2)}(l_A,l_B)$ beyond which $\Gamma_O$ gets disconnected and $V_{A:B:O}$ vanishes. We have fixed $l_A/L=1$.}
    \label{fig:AdSd-pure-nonad}
\end{figure}
We now take $q$ strips with the same width $l$ and separation $s$ in the $x-$direction.
The minimal surface $\Gamma_O$ is fully disconnected for $s>s_{\rm crit}^{(q)}(l)$ and fully connected for $s<s_{\rm crit}^{(q)}(l)$ \cite{Ben-Ami:2014gsa}, as in the left panel of Fig.~\ref{fig:AdS3-Phases}.
From eq.~\eqref{eq:area-Poincare-d}, the critical distance satisfies
\begin{equation}
    \frac{1}{(q l +(q-1) s_{\rm crit}^{(q)})^{d-2}} + \frac{q-1}{\left( s_{\rm crit}^{(q)} \right)^{d-2}} = \frac{q}{l^{d-2}} \, .
\end{equation}
The volume in the two phases is
\begin{equation}
    \frac{V_{\{A_i\}:O}}{L^d} =
    \begin{cases}
        \omega_d^{(q)}(l,s) \, & \text{for $s<s_{\rm crit}^{(q)}(l)$} \\
        0 & \text{for $s>s_{\rm crit}^{(q)}(l)$}
    \end{cases} \, ,
\end{equation}
where we have defined
\begin{equation}
\label{eq:V-nonadjacent-Poincare-d}
    \omega_d^{(q)}(l,s) = \ell^{d-2} C_d \left( - \frac{1}{(q l + (q-1) s)^{d-2}} + \frac{q}{l^{d-2}} + \frac{q-1}{s^{d-2}} \right) 
\end{equation}
the dimensionless volume for $q$ strips of equal width $l$ and separation $s$.
At the transition between the two phases the volume is discontinuous, with a jump
\begin{equation}
    \omega_d^{(q)}(l,s_{\rm crit}) = \ell^{d-2} C_d \frac{2(q-1)}{s_{\rm crit}^{d-2}} \, .
\end{equation}
Starting from the fully connected configuration and increasing the separation $s_i$ between the strips $A_i$ and $A_{i+1}$, the EWP becomes the union of two fully connected surfaces. 
It is easy to see that the resulting volume $\omega_d^{(q-i)}(l,s) + \omega_d^{(i)}(l,s)$ is smaller than the initial one $\omega_d^{(q)}(l,s)$.\footnote{For $i=1,q-1$, the EWP is still connected, but its volume decreases anyway.} 
One way to see it is that, up to a translation in the $x-$direction, the fully connected surface for $q$ subregions contains the fully connected surfaces for subsets of consecutive subregions. Consequently, similarly to $d=2$,
\begin{equation}
    0 \leq \frac{V_{\{A_i\}:O}}{L^d} \leq \omega_d^{(q)}(l,s) \, ,
\end{equation}
where the equalities hold for the fully disconnected and the fully connected configuration, respectively. 
We stress that, unlike the $d=2$ case, the volume is not quantized, but it depends on the width of the $q$ strips.

In Fig.~\ref{fig:AdSd-pure-nonad} we display contour lines of constant volume $V_{A:B:O}$ for two strips $A,B$ of width $l_A,l_B$ and separation $s$.
As it is clear from the picture, for fixed $l_A$ the volume grows when decreasing the separation $s$ and the width $l_B$. The first transformation brings $A$ and $B$ closer, while the second one brings two disconnected components of $O$ closer to each others. Therefore, in both cases we expect the multi-partite entanglement among the three subregions $A,B,O$ to increase.
Since the volume scales as $(L/l_A)^{d-2}$, a similar effect takes place when decreasing the width of $A$.

\subsection{Pure state in the two-sided AdS$_{d+1}$ black brane}
\label{subsec:brane-d}
The next case is a multi-partition of the thermofield double state $\rho_{A_1 \ddd A_q O} = \ket{TFD}\!\bra{TFD}$ in the product of two non-interacting CFT$_d$ on non-compact space.
The bulk dual is a two-sided AdS$_{d+1}$ black brane, with metric
\begin{equation}
\label{eq:metric-planarBH-d}
    ds^2 = \frac{L^2}{z^2} \left( -f(z) dt^2 +\frac{dz^2}{f(z)} + dx^2 + \sum_{i=1}^{d-2} dy^2_i \right) \, ,
    \qquad
    f(z) = 1 - \frac{z^d}{z_h^d} \, ,
\end{equation}
where the black brane horizon is located at $z=z_h$ and $d\geq 3$.
The temperature of the black brane is $T= d/(4 \pi z_h)$. 
We choose as partitions strips with infinite length $\ell$ in the $y_i-$directions and given width in the $x-$direction.
For a strip $A$ with finite width $l_A$ in the $x-$direction, the minimal codimension-two surface $\Gamma_A$ can be expressed as
\begin{align}
    x_\pm(u) &= x_* \pm \frac{l_A}{2} \nonumber \\
    &\mp z_* \sum_{n=0}^{+\infty} \frac{\Gamma(n+\frac{1}{2})}{\sqrt{\pi} \, \Gamma(n+1)} \frac{u^{-d(n+1)}}{d(n+1)} \, {}_2F_1\left( \frac{1}{2}, \frac{d(n+1)}{2(d-1)};\frac{d(n+3)-2}{2(d-1)};u^{2(1-d)} \right) \left( \frac{z_*}{z_h} \right)^{n d} \\
    &\equiv x_* \pm \frac{l_A}{2} \mp z_* \sum_{n=0}^{+\infty} \frac{\Gamma(n+\frac{1}{2})}{\sqrt{\pi} \, \Gamma(n+1)} m_{d,n}(u) \left( \frac{z_*}{z_h} \right)^{n d} \, ,
\end{align}
in which $(z_*,x_*)$ is the location of the turning point of $\Gamma_A$.
We have defined $u=z_*/z$ and 
\begin{equation}
    m_{d,n}(u) = \int_u^{+\infty} \frac{dt}{t^{n d+2} \sqrt{t^{2(d-1)}-1}} \, ,
    \qquad
    \bar{m}_{d,n} \equiv m_{d,n}(1) = \frac{\sqrt{\pi} \, \Gamma(\frac{(n+1)d}{2(d-1)})}{(n d+1) \, \Gamma(\frac{nd+1}{2(d-1)})} \, .
\end{equation}
Note that $m_{d,n}(+\infty)=0$, in agreement with the boundary conditions $x_{\pm}(+\infty) = x_* \pm l_A/2$.
The width $l_A$ is related to the location $z_*$ of the turning point by
\begin{align}
    l_A &= 2 z_* \int_1^{+\infty} \frac{dt}{t^2 \sqrt{t^{2(d-1)}-1}} \left( 1- \left( \frac{z_*}{z_h} \right)^d t^{-d} \right)^{-1/2} \\
    &= 2 z_* \sum_{n=0}^{+\infty} \frac{\Gamma(n+\frac{1}{2}) \, \bar{m}_{d,n}}{\sqrt{\pi} \, \Gamma(n+1)}  \left( \frac{z_*}{z_h} \right)^{n d} \, .
    \label{eq:l-brane}
\end{align}
The series converges for $z_* < z_h$. 
Consistently, an analysis of $\Gamma_A$ shows that as the width $l_A$ increases, the turning point gets closer to the event horizon but never crosses it \cite{Hubeny:2012ry}.\
We can write down the volume functional for the codimension-one surface $r_A$ enclosed by $\Gamma_A$ and the boundary strip $A$ as
\begin{equation}
    \frac{\mathcal{V}_A}{L^d \ell^{d-2}} = 2 z_*^{1-d} \int_1^{z_*/\varepsilon} \frac{du}{u^{2-d}} \left( 1- \left( \frac{z_*}{z_h} \right)^d u^{-d} \right)^{-1/2} (x_+(u) -x_*) \, ,
\end{equation}
where $\varepsilon$ is a UV cutoff. 
The divergence in $\mathcal{V}_A$ comes from the leading term in $z_*/z_h$ and is the same as in the zero-temperature case in eq.~\eqref{eq:volume-Poincare-d}. In particular, the divergence is linear in $l_A$, leading to a finite volume of the EWP. We thus disregard this divergent term, which eventually simplifies by subtraction. We then focus on
\begin{align}
    \frac{\mathcal{V}_A^{\rm finite}}{L^d \ell^{d-2}} &=
    -l_A \frac{z_*^{1-d}}{d-1} +l_A \, z_*^{1-d} \int_1^{+\infty} \frac{du}{u^{2-d}} \sum_{n=1}^{+\infty} \frac{\Gamma(n+\frac{1}{2})}{\sqrt{\pi} \, \Gamma(n+1)} \left( \frac{z_*}{z_h} \right)^{n d} u^{-n d}  \nonumber \\
    &-2 z_*^{2-d} \int_1^{+\infty} \frac{du}{u^{2-d}} \sum_{n=0}^{+\infty} \frac{\Gamma(n+\frac{1}{2})}{\sqrt{\pi} \, \Gamma(n+1)} \left( \frac{z_*}{z_h} \right)^{n d} u^{-n d} \sum_{k=0}^{+\infty} \frac{\Gamma(k+\frac{1}{2})}{\sqrt{\pi} \, \Gamma(k+1)} m_{d,k}(u) \left( \frac{z_*}{z_h} \right)^{k d} \, .
    \label{eq:VA-brane}
\end{align}
We can obtain analytical results in the following limits:
\begin{itemize}
    \item In the low temperature regime $l_A<<z_h$, we can neglect the next-to-leading orders in $z_*/z_h$. Under this assumption, we can solve eq.~\eqref{eq:l-brane} as \cite{Fischler:2012ca}
    \begin{equation}
        z_* = \frac{l_A}{2 \bar{m}_{d,0}} \left( 1 - \frac{\bar{m}_{d,1}}{(2 \bar{m}_{d,0})^{d+1}} \left( \frac{l_A}{z_h} \right)^d + \mathcal{O}\left( \frac{l_A}{z_h} \right)^{2d} \right) \, .
    \end{equation}
    The resulting volume for the homology surface $r_A$ is
    \begin{align}
        \frac{\mathcal{V}_A^{\rm finite}}{L^d \ell^{d-2}} = &-\frac{l_A^{2-d}}{(d-1)(2 \bar{m}_{d,0})^{1-d}} - \frac{2 l_A^{2-d}}{(2 \bar{m}{d,0})^{2-d}} \int_1^{+\infty} \frac{du}{u^{2-d}} m_{d,0}(u) \nonumber \\
        &- \frac{l_A^2}{z_h^d} \frac{1}{(2 \bar{m}_{d,0})^3} \left[ 2 \bar{m}_{d,0} \bar{m}_{d,1} + \int_1^{+\infty} \frac{du}{u^{2-d}} \times \right.  \nonumber \\
        & \left.  \left( \frac{m_{d,0}(u)-\bar{m}_{d,0}}{u^d} (2 \bar{m}_{d,0}) +2(d-2) \bar{m}_{d,1} m_{d,0}(u) + 2 \bar{m}_{d,0} m_{d,1}(u) \right) \right] \nonumber \\
        &+ \mathcal{O}\left( l_A^{2-d} \left( \frac{l_A}{z_h} \right)^{2d} \right) \, . 
    \end{align}
    Note that the first line is just the finite part of the zero-temperature result \eqref{eq:volume-Poincare-d}, while the low temperature corrections scale as $l_A^2/z_h^d$.
    It is convenient to express the result as
    \begin{equation}
       \frac{\mathcal{V}_A^{\rm finite}}{L^d \ell^{d-2}} = l_A^{2-d}\left( -C_d - D_d \, \frac{l_A^d}{z_h^d} + \mathcal{O} \left( \frac{l_A^{2d}}{z_h^{2d}} \right) \right) \, ,
    \end{equation}
    with $D_d$ a $d$-dependent constant.
    \item In the high temperature regime $l_A >> z_h$, the turning point approaches the event horizon exponentially fast, $z_* \approx z_h \left( 1-a_d \exp\left(-\frac{l_A}{z_h} \sqrt{\frac{d(d-1)}{2}} \right) \right)$, with $a_d$ a $d$-dependent constant \cite{Fischler:2012ca}.
    From eq.~\eqref{eq:VA-brane}, at the leading order we get
    \begin{align}
        \frac{\mathcal{V}_A^{\rm finite}}{L^d \ell^{d-2}} &= l_A \, z_h^{1-d} \frac{\sqrt{\pi} \, \Gamma(\frac{1}{d}-1)}{d \, \Gamma(\frac{1}{d}-\frac{1}{2})} -2 z_h^{2-d} \int_1^{+\infty} \frac{du}{u^{2-d} \sqrt{1-u^{-d}}}  \times \\
        &\times \sum_{k=0}^{+\infty} \frac{\Gamma(k+\frac{1}{2})}{\sqrt{\pi} \, \Gamma(k+1)} m_{d,k}(u)
        + \mathcal{O}\left(z_h^{1-d} \exp \left( -\frac{l_A}{z_h} \right) \right) \\
        &= l_A z_h^{1-d} \left( \frac{\sqrt{\pi} \, \Gamma(\frac{1}{d}-1)}{d \, \Gamma(\frac{1}{d}-\frac{1}{2})} - E_d \frac{z_h}{l_A} + \mathcal{O}\left( \frac{1}{l_A} \exp \left( -\frac{l_A}{z_h} \right) \right) \right)  \, .
    \end{align}
    The linearity in $l_A$ can be expected from the fact that $\Gamma_A$ is well approximated by a surface wrapping the horizon $z=z_h$ except for the vicinity of $x=x_*\pm l_A/2$.
\end{itemize}
With this results in hand, we can evaluate the volume of the EWP by subtraction. 
We focus on $(q+1)-$partitions of the thermofield double state of two copies of CFT$_d$ on non-compact space, $\rho_{A_1 \ddd A_q O} = \ket{TFD}\!\bra{TFD}$. As for the $d=2$ case studied in subsection~\ref{subsubsec:two-sided_BTZ}, we assume that the purifier $O$ contains the full left boundary and, in general, part of the right one, $O = L \cup O_R$.
For $q$ adjacent strips $A_1, \dots, A_q$ with finite widths $l_{A_i}$ in the $x-$direction, $O_R$ is the union of two strips with infinite width.
The volume of the EWP is given as in eq.~\eqref{eq:V-subtraction}. We denote by $l_{\rm tot}$ the sum of the widths and consider the following limits:
\begin{itemize}
    \item In the low temperature regime $l_{\rm tot}<<z_h$, we have
    \begin{equation}
        \frac{V_{\{A_i\}:O}}{L^d \ell^{d-2}} =
        \frac{V^{(0)}_{\{A_i\}:O}}{L^d \ell^{d-2}} - \frac{D_d}{z_h^d} \left( l_{\rm tot}^2 - \sum_{i=1}^q l_{A_i}^2 \right) + \mathcal{O}\left( l_{A_i}^{2-d} \left( \frac{l_{A_i}}{z_h} \right)^{2d} \right) \, ,
    \end{equation}
    where the first term is the zero-temperature result in eq.~\eqref{eq:V-adjacent-Poincare-d}.
    \item In the intermediate regime $l_{A_i}<<z_h$ for $i=1,\dots,p$ and $l_{B_i} >>z_h$ for $i=1,\dots,q-p$, we have
    \begin{equation}
    \label{eq:V-brane-intermediateT}
         \frac{V_{\{A_i\}:O}}{L^d \ell^{d-2}} = \sum_{i=1}^p \left( \frac{C_d}{l_{A_i}^{d-2}} \right) + \frac{1}{z_h^{d-2}} \left( (q-p-1) E_d + \mathcal{O}\left( \frac{l_{A_i}}{z_h} \right) \right) \, .
    \end{equation}
    \item In the high temperature regime $l_{A_i} >>z_h$, we have
    \begin{equation}
    \label{eq:V-brane-highT}
        \frac{V_{\{A_i\}:O}}{L^d \ell^{d-2}} = (q-1) \frac{E_d}{z_h^{d-2}} + \mathcal{O} \left( \frac{1}{z_h^{d-1}} \exp \left( -\frac{l_{A_i}}{z_h} \right) \right) \, .
    \end{equation}
\end{itemize}

\begin{figure}[ht]
\center
\includegraphics[scale=1]{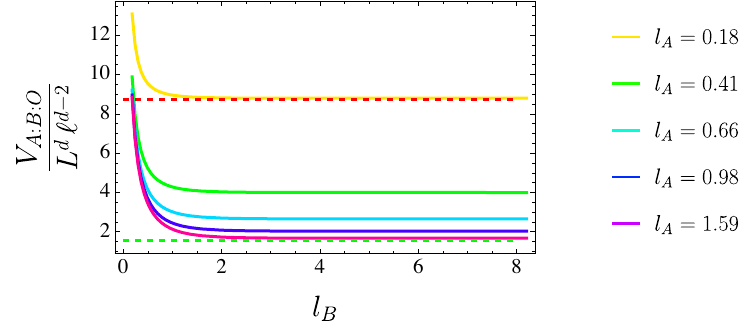}
\caption{Volume $V_{A:B:O}$ as a function of $l_B$ for fixed $l_A$, with $z_h=1$ and $d=3$.  
The dotted red line represents the intermediate temperature result~\eqref{eq:V-brane-intermediateT} for $l_A<<z_h, l_B>>z_h$ with $p=1,q=2$ and $l_A=0.18$.
The dotted green line represents the high temperature result~\eqref{eq:V-brane-highT} $E_d/z_h^{d-2}$ for $l_A,l_B >> z_h$.}
\label{fig:VABO_brane}
\end{figure}

In Fig.~\ref{fig:VABO_brane} we plot $V_{A:B:O}$ for adjacent subregions and fixed $l_A$, as a function of $l_B$.
The general behavior is in line with the zero-temperature case studied in subsection~\ref{subsec:Poincare-d}.
In the intermediate temperature regime $l_A<<z_h, \, l_B>>z_h$, the leading term scales as $1/l_A^{d-2}$, as in eq.~\eqref{eq:V-brane-intermediateT} with $p=1, \, q=2$.
In the high temperature regime, the volume $V_{A:B:O}$ is independent of the size of the subregions, in agreement with eq.~\eqref{eq:V-brane-highT}.

Let us now consider $q$ strips with the same width $l$ and separation $s$ in the $x-$direction.
For $s<s^{(q)}_{\rm crit}(l,z_h)$ the minimal surface $\Gamma_O$ is fully connected, whereas for $s>s^{(q)}_{\rm crit}(l,z_h)$ it is fully disconnected and $V_{\{A_i\}:O}=0$. The value of $s^{(q)}_{\rm crit}(l,z_h)$ has been numerically investigated in \cite{Fischler:2012uv} for $q=2$. 
We focus on the case $s<s^{(q)}_{\rm crit}(l,z_h)$, where the volume $V_{\{A_i\}:O}$ is non-trivial. We can address the problem analytically in some limits:
\begin{itemize}
    \item In the low temperature regime $l,s<<z_h$, we get
    \begin{align}
        \frac{V_{\{A_i\}:O}}{L^d \ell^{d-2}} &\approx \frac{\omega_d^{(q)}(l,s)}{\ell^{d-2}} 
        - \frac{D_d}{z_h^d} \left( (q l+(q-1)s)^2-(q-1)s^2 \right)
        \\
        &=\frac{\omega_d^{(q)}(l,s)}{\ell^{d-2}} - \frac{D_d}{z_h^d} \left( q^2 l^2 +2 q(q-1) l s +(q-1)(q-2) s^2 \right) \, ,
    \label{eq:lowT-BHd_nonad}
    \end{align}
    where the first term is the zero-temperature result introduced in eq.~\eqref{eq:V-nonadjacent-Poincare-d}.
    \item In the intermediate regime $s<<z_h<<l$, at the leading orders we have
    \begin{equation}
        \frac{V_{\{A_i\}:O}}{L^d \ell^{d-2}} \approx C_d \frac{(q-1)}{s^{d-2}} + \frac{1}{z_h^{d-2}} \left( (q-1) E_d + \mathcal{O} \left( \frac{s}{z_h} \right) \right) \, ,
    \label{eq:midT-BHd_nonad}
    \end{equation}
    where the leading term is the finite part of the zero-temperature volume $\mathcal{V}$ for $(q-1)$ strips of width $s$.
    \item In the high temperature regime $s>>z_h$, the correlations among the subregions $A_1, \dots, A_q$ are destroyed and the minimal surface $\Gamma_O$ is fully disconnected \cite{Fischler:2012uv,Ben-Ami:2016qex}. Consequently, $V_{\{A_i\}:O}=0$. 
\end{itemize}

\begin{figure}[ht]
\center
\includegraphics[scale=1]{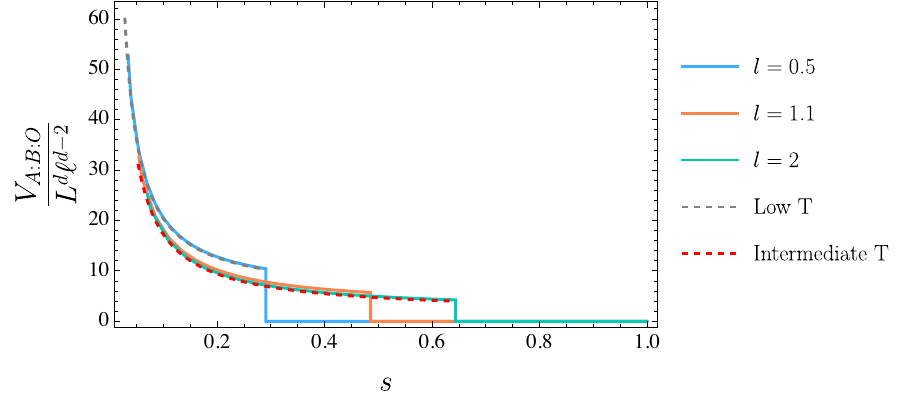}
\caption{Volume of the EWP for finite temperature and non-adjacent strips $A$ and $B$ with the same width $l$ and separation $s$. We have fixed $z_h=1$ and $d=3$.}
\label{fig:VABO_BHd_nonad}
\end{figure}

In Fig.~\ref{fig:VABO_BHd_nonad}, we show numerical results for the volume $V_{A:B:O}$ at fixed width $l$ as a function of the separation $s$ between $A$ and $B$. 
We find a perfect match with the analytical predictions for the low temperature regime in eq.~\eqref{eq:lowT-BHd_nonad} and the intermediate temperature regime in eq.~\eqref{eq:midT-BHd_nonad}, for $q=2$ and $d=3$.

\subsection{Pure state in AdS$_{d+1}$/BCFT$_d$}
\label{subsec:AdSBCFT-d}
We now turn to a BCFT$_d$ on non-compact space.
The bulk dual is Poincar\'e AdS$_{d+1}$ with an end-of-the-world (EOW) brane whose configuration is $z= x \tan\alpha$, assuming the translational symmetry along the $y_i-$directions.
We take the strip subregion $A$ with finite width $l_A$ in the $x-$direction and infinite length $\ell$ in the $y_i-$directions. We set the intersection point between the EOW brane and the minimal surface $\Gamma_A$ to be $(x,z)=(a, a \tan\alpha)$.
The minimal surface $\Gamma_A$ is given as in eq.~\eqref{eq:GammaA-Poincare-d} with the replacement $l_A \to 2 l_A$, 
\begin{align}
    x_+(z)=&\,l_A-\frac{z^d}{d\,z_*^{d-1}} \,_2F_1\left(\frac{1}{2},\frac{d}{2 (d-1)};\frac{3d-2}{2(d-1)};\left(\frac{z}{z_*}\right)^{2(d-1)}\right) \, .
\end{align}
Hence, $z_*$ is related to $a$ as follows:
\begin{equation}
\begin{aligned}
    l_A-a=&\frac{z_*(a)}{d}\left(\frac{a\tan\alpha}{z_*(a)}\right)^{d}\,_2F_1\left(\frac{1}{2},\frac{d}{2 (d-1)};\frac{3d-2}{2(d-1)};\left(\frac{a\tan\alpha}{z_*(a)}\right)^{2(d-1)}\right) \, .
\end{aligned}
\end{equation}
In the limit $a \to 0 \, , \alpha \to \pi/2$, the intersection point coincides with the turning point of $\Gamma_A$, namely $a \tan \alpha \to z_*$, and then $z_*$ coincides with the usual AdS$_{d+1}$/CFT$_d$ case.
We obtain the following area of the surface $\Gamma_A$:
\begin{align}
    \frac{\mathcal{A}^{(a)}(\Gamma_A)}{L^{d-1}} &= \frac{1}{d-2}\left(\frac{\ell}{\varepsilon}\right)^{d-2}-\frac{1}{d-2}\left(\frac{\ell}{a\tan\alpha}\right)^{d-2} \times \nonumber \\
    &\times {}_2F_1\left[\frac{1}{2},\frac{2-d}{2(d-1)};\frac{d}{2
   (d-1)};\left(\frac{a\tan\alpha}{z_*(a)}\right)^{2(d-1)}\right] \, ,
    \label{eq:higherAdSBCFTarea}
\end{align}
where $\varepsilon$ is the UV cutoff. Consistently, this area becomes half of the pure AdS$_{d+1}$ value $\mathcal{A}(\Gamma_A)$ in eq.~\eqref{eq:area-Poincare-d} when we take $\alpha\to\pi/2$.
In order to find the intersection point, we need to minimize the area with respect to $a$ by solving the equation
\begin{equation}
    \frac{\partial\mathcal{A}^{(a)}(\Gamma_A)}{\partial a}=0 \, .
\end{equation}
The volume of the codimension-one homology surface $r_A$ reads
\begin{align}
    \frac{\mathcal{V}^{(a)}_A}{L^d} &= \frac{\ell^{d-2}}{(d-1)}\left(\frac{l_A}{\varepsilon^{d-1}}-\frac{l_A}{(a\tan\alpha)^{d-1}}\right)-\frac{\ell^{d-2}}{(d-2)\tan\alpha}\left(\frac{1}{\varepsilon^{d-2}}-\frac{1}{(a\tan\alpha)^{d-2}}\right) \nonumber \\
    &-\frac{\ell^{d-2}\,a\tan\alpha}{d\,z_*(a)^{d-1}}\,_3F_2\left[\frac{1}{2},\frac{1}{2 (d-1)},\frac{d}{2 (d-1)};\frac{2d-1}{2
   (d-1)},\frac{3d-2}{2 (d-1)};\left(\frac{a\tan\alpha}{z_*(a)}\right)^{2(d-1)}\right] \, .
    \label{eq:higherAdSBCFTvolume}
\end{align}
The divergent term of $\mathcal{O}\left(\varepsilon^{2-d}\right)$ comes from the corner contribution, because it does not depend on the strip width. This volume also becomes half of the pure AdS$_{d+1}$ value $\mathcal{V}_A$ in eq.~\eqref{eq:volume-Poincare-d} when we take $\alpha\to \pi/2$. 

We now take a tri-partition $A,B,O$ of the boundary state, with $A$ and $B$ adjacent strips of widths $l_A$ and $l_B$, respectively.
We distinguish between a connected phase and a disconnected phase, drawn in Fig.~\ref{fig:higherAdSBCFTadphases}. We denote by $(b,b\tan\alpha)$ the intersection of the EOW brane and the minimal surface $\Gamma_{O}$.
\begin{figure}[t]
  \centering
  \begin{tikzpicture}
    \begin{scope}[scale=1.1]
      \def\R{1}
      \def\Rmax{3}
      \def\alphaAngle{55} 
      \def\a{\R*cos{\alphaAngle}}
      \def\b{2*\a}
      \draw[->] (-0.2,0) -- (4,0) node[right] {$x$};
      \draw[->] (0,-0.2) -- (0,3) node[above] {$z$};
      \fill[cyan!15]
        (1,0) arc (0:55:1.0)
        -- ({cos(55)},{sin(55)})
        -- ({2*cos(55)},{2*sin(55)})
        arc (55:0:2)
        -- (2,0)
        arc (0:180:0.5)
        -- cycle;
      \draw[very thick]
        (0,0) --
        ({\Rmax*cos(\alphaAngle)}, {\Rmax*sin(\alphaAngle)})
        node[pos=1.0, above right] {EOW brane};
      \draw[thick] (0,0) -- (1,0);
      \node[below] at (0.5,-0.4) {$A$};
        \draw[dashed]
          ({\a}, {\a*tan(\alphaAngle)}) -- (\a,0);
        \draw[dashed]
          (0,{\a*tan(\alphaAngle)}) -- ({\a}, {\a*tan(\alphaAngle)});
        \draw[thick]
          (-0.08,{\a*tan(\alphaAngle)}) -- (0.08,{\a*tan(\alphaAngle)});
        \node[left] at (0,{\a*tan(\alphaAngle)}) {$a\tan\alpha$};
        \draw[thick]
          (\a,-0.08) -- (\a,0.08);
        \node[below] at (\a,0) {$a$};
        (\R,-0.08) -- (\R,0.08);
      \draw (1,0) arc (0:55:1.0);
      \draw[thick] (1,0) -- (2,0);
      \node[below] at (1.5,-0.4) {$B$};
      \node[below] at (2.5,-0.4) {$O$};
        \draw[dashed]
          ({\b}, {\b*tan(\alphaAngle)}) -- (\b,0);
        \draw[dashed]
          (0,{\b*tan(\alphaAngle)}) -- ({\b}, {\b*tan(\alphaAngle)});
        \draw[thick]
          (-0.08,{\b*tan(\alphaAngle)}) -- (0.08,{\b*tan(\alphaAngle)});
        \node[left] at (0,{\b*tan(\alphaAngle)}) {$b\tan\alpha$};
        \draw[thick]
          (\b,-0.08) -- (\b,0.08);
        \node[below] at (\b,0) {$b$};
        (\R,-0.08) -- (\R,0.08);
      \draw (2,0) arc (0:55:2.0);
      \draw (1,0) arc (180:0:0.5);
      \node[below] at (1.3,1.2) {$\Sigma_{A:B:O}$};
    \end{scope}
    \node at (6.0,1.6) {$\Longleftrightarrow$};
    \node[above] at (6.0,1.8) {\small phase transition};
    \begin{scope}[xshift=8cm,scale=1.1]
      \def\Rmax{3}
      \def\alphaAngle{55}
      \draw[->] (-0.2,0) -- (4,0) node[right] {$x$};
      \draw[->] (0,-0.2) -- (0,3) node[above] {$z$};
      \draw[very thick]
        (0,0) --
        ({\Rmax*cos(\alphaAngle)}, {\Rmax*sin(\alphaAngle)})
        node[pos=1.0, above right] {EOW brane};
      \draw[thick] (0,0) -- (1,0);
      \node[below] at (0.5,-0.4) {$A$};
      \node[below] at (1,0) {$l_A$};
      \draw[thick] (1,0) arc (0:55:1.0);
      \draw[thick] (1,0) -- (2,0);
      \node[below] at (1.5,-0.4) {$B$};
      \node[below] at (2.5,-0.4) {$O$};
      \node[below] at (2,0) {$l_A+l_B$};
      \draw[thick] (2,0) arc (0:55:2.0);
      \draw (0.3,0) arc[start angle=0, end angle=\alphaAngle, radius=0.3];
      \node at ({0.5*cos(\alphaAngle/2)}, {0.5*sin(\alphaAngle/2)}) {$\alpha$};
    \end{scope}
  \end{tikzpicture}
  \caption{The EWP for adjacent strips in higher dimensional AdS/BCFT: connected phase (on the left) and disconnected phase (on the right).}
  \label{fig:higherAdSBCFTadphases}
\end{figure}
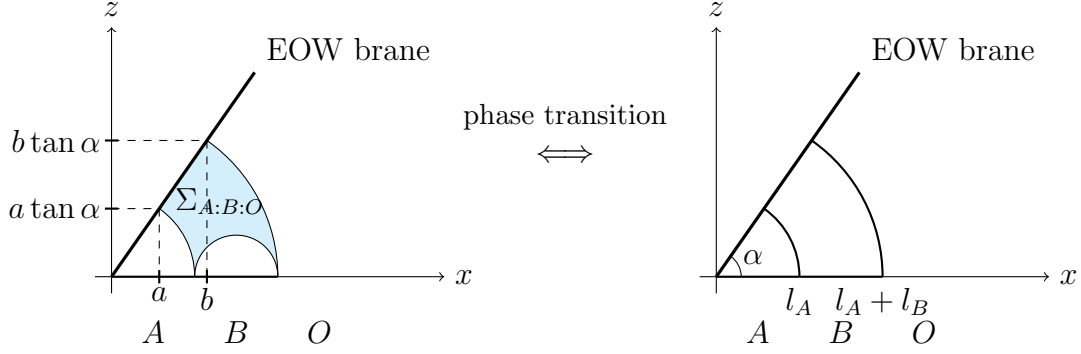
\begin{itemize}
  \item \textbf{Connected phase.}
    The minimal surface $\Gamma_B$ does not end of the EOW, namely $\mathcal{A}(\Gamma_B)\leq\mathcal{A}^{(a)}(\Gamma_A) + \mathcal{A}^{(b)}(\Gamma_O)$.
    In this phase, we obtain 
    \begin{equation}
        \begin{aligned}
            &\frac{V_{A:B:O}}{L^d} \\
            &=\frac{\ell^{d-2}}{(d-1)}\left(\frac{l_A}{(a\tan\alpha)^{d-1}}-\frac{l_A+l_B}{(b\tan\alpha)^{d-1}}\right) \\
            &-\frac{\ell^{d-2}}{(d-2)\tan\alpha}\left(\frac{1}{(a\tan\alpha)^{d-2}}-\frac{1}{(b\tan\alpha)^{d-2}}\right) \\
            &+\frac{\ell^{d-2}\,a\tan\alpha}{d\,z_*(a)^{d-1}}\,_3F_2\left[\frac{1}{2},\frac{1}{2 (d-1)},\frac{d}{2 (d-1)};\frac{2d-1}{2
           (d-1)},\frac{3d-2}{2 (d-1)};\left(\frac{a\tan\alpha}{z_*(a)}\right)^{2(d-1)}\right] \\ 
           &-\frac{\ell^{d-2}\,b\tan\alpha}{d\,z_*(b)^{d-1}}\,_3F_2\left[\frac{1}{2},\frac{1}{2 (d-1)},\frac{d}{2 (d-1)};\frac{2d-1}{2
           (d-1)},\frac{3d-2}{2 (d-1)};\left(\frac{b\tan\alpha}{z_*(b)}\right)^{2(d-1)}\right] \\
           &+\frac{2^{d-2}\pi^{\frac{d-1}{2}}}{(d-1)^2}\left(\frac{\Gamma\left( \frac{d}{2(d-1)} \right)}{\Gamma\left( \frac{1}{2(d-1)}\right)}\right)^{d-3}\left(\frac{\ell}{l_B}\right)^{d-2} \, .
        \end{aligned}
    \end{equation}    
    We remark that this result is still finite even though eq.~\eqref{eq:higherAdSBCFTvolume} has several divergent terms. Some numerical results are shown in Fig.~\ref{fig:higherAdSBCFT_plot} and Fig.~\ref{fig:higherAdSBCFT_fixed}.
    \begin{figure}
        \centering
        \includegraphics[width=0.8\linewidth]{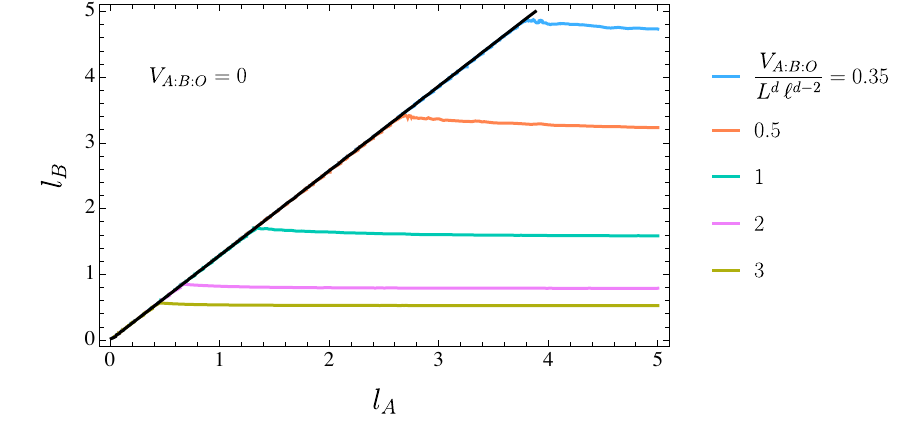}
        \caption{Contour lines of constant volume of the EWP for a tri-partition of the boundary state $\rho_{ABO}$ in the AdS/BCFT for $d=3$, with adjacent $A$ and $B$. The black curve represents the transition between the connected and the disconnected phase, $\mathcal{A}(\Gamma_B)=\mathcal{A}^{(a)}(\Gamma_A) + \mathcal{A}^{(b)}(\Gamma_O)$. We set $\alpha=\frac{\pi}{3}$.}
        \label{fig:higherAdSBCFT_plot}
    \end{figure}
    \begin{figure}
        \centering
        \includegraphics[width=0.46\linewidth]{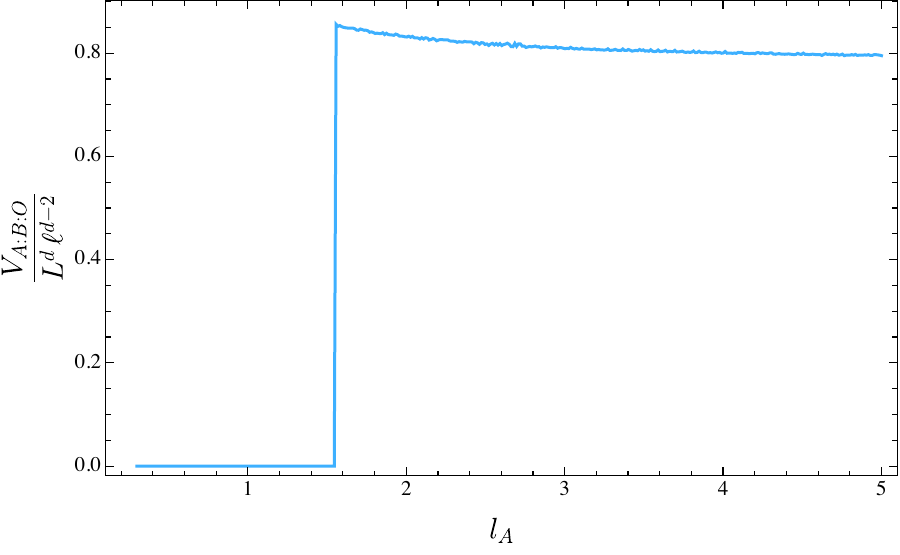}
        \qquad
        \includegraphics[width=0.46\linewidth]{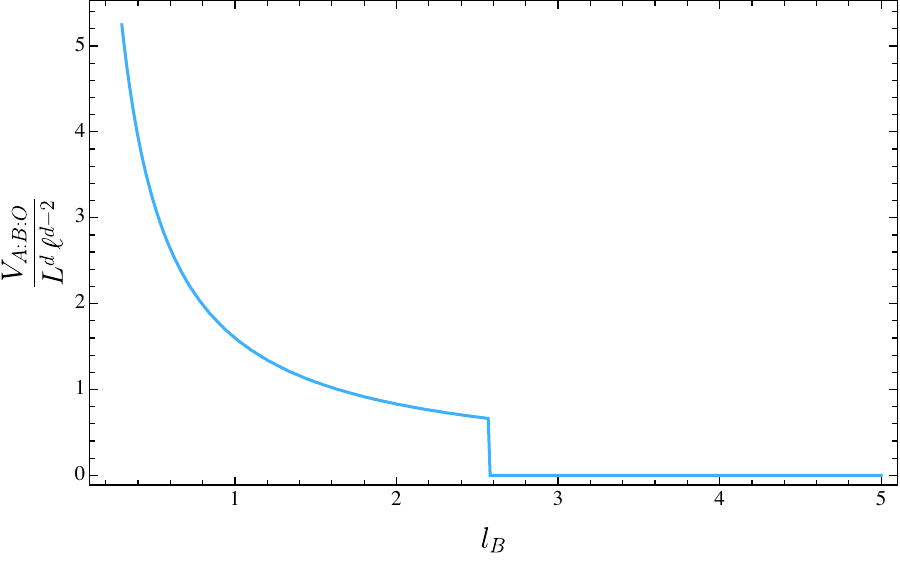}
        \caption{Volume of the EWP for the boundary state $\rho_{ABO}$ in the AdS/BCFT for $d=3$, with adjacent $A$ and $B$. We set $\alpha=\frac{\pi}{3}$ and fix $l_B=2$ (left) and $l_A=2$ (right).}
        \label{fig:higherAdSBCFT_fixed}
    \end{figure}
    \item \textbf{Disconnected phase.}
        The minimal surface $\Gamma_B$ ends on the EOW brane, namely $\mathcal{A}(\Gamma_B)\ge\mathcal{A}^{(a)}(\Gamma_A)+\mathcal{A}^{(b)}(\Gamma_O)$.
        In this phase, since $\Gamma_B = \Gamma_A \cup \Gamma_O$, we trivially have
        \begin{equation}
          V_{A:B:O}=0 \, .
        \end{equation}
\end{itemize}
In a similar way, we can easily calculate $V_{A_1:\ddd : A_q}$ for adjacent subregions $A_1,\dots ,A_q$.

\subsection{Pure state in the AdS Soliton}
\label{subsec:AdSsoliton}
In this subsection, we consider the AdS$_{d+1}$ soliton \cite{Witten:1998zw} dual to large $N$ confining gauge theories, whose metric is 
\begin{equation}
    ds^2=\frac{L^2}{r^2f(r)}dr^2+\frac{r^2}{L^2}\left(f(r)d\chi^2-dt^2+dx^2+\sum_{i=1}^{d-3}dy_i^2\right)
,\quad f(r)=1-\left(\frac{r_0}{r}\right)^d \, ,
\end{equation}
with $d \geq 3$.
The parameter $r_0$ denotes the IR cutoff in the AdS spacetime. Here we compactify the coordinate $\chi$ with a periodicity $R=\frac{4\pi L^2}{d \, r_0}$. 

\begin{figure}[httt]
    \centering
    \includegraphics[width=0.8\linewidth]{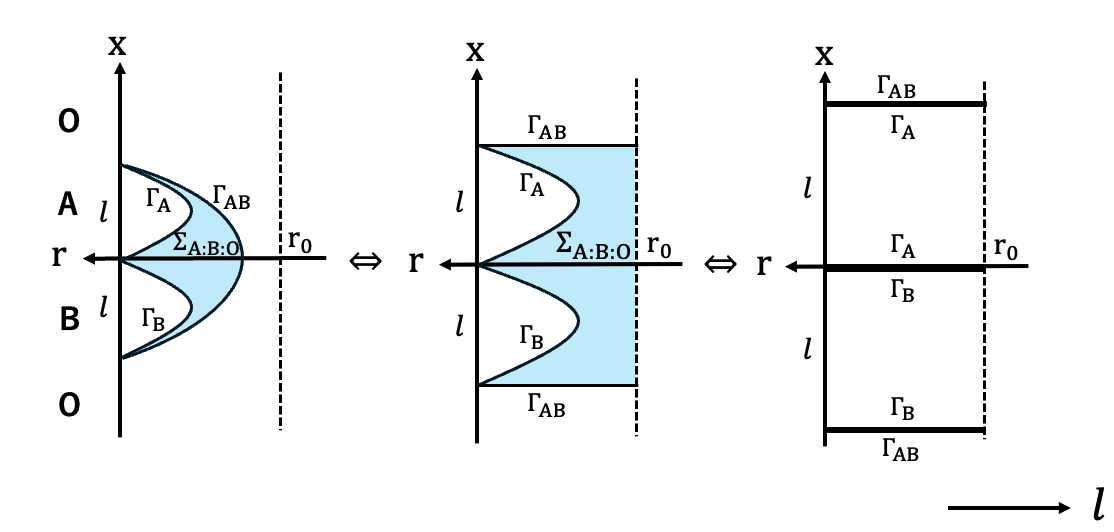}
    \caption{The three possible configurations of the EWP for the AdS soliton geometry. The phase changes as the width $l$ becomes larger. In the small $l$ phase (left), the EWP is the same as in the pure AdS case. In the intermediate $l$ phase (center), $\Gamma_{AB}$ gets disconnected, though $\Gamma_A$ and $\Gamma_B$ are connected. In the large $l$ phase (right), the EWP is the empty set.}
    \label{fig:phasetrsf_AdSsoliton}
\end{figure}

Let us take a boundary strip $A$ with a finite width $l_A$ in the $x-$direction, an infinite length $\ell$ in the $y_i-$directions, and a finite circumference $R=\frac{4\pi L^2}{d \, r_0}$ in the $\chi-$direction.
\begin{figure}[t]
  \centering
  \begin{tikzpicture}
\begin{scope}[scale=1]
  \def\rinf{0}
  \def\rstar{2.6}
  \def\rzero{3.7}
  \def\b{2.0}
  \draw[thick,->] (4.0,0) -- (-0.5,0)  node[left] {$r$};
  \draw[thick,->] (0,-2.8) -- (0,2.8) node[above] {$x$};
  \node[left] at (0,\b) {$l_A/2$};
  \node[left] at (0,-\b) {$-l_A/2$};
  \node[left] at (0,0.5) {$A$};
  \draw[very thick]
    plot[domain=-90:90,samples=120]
    ({\rstar*cos(\x)}, {\b*sin(\x)});
  \fill (\rstar,0) circle (1.5pt);
  \node[below] at (\rstar-0.3,0) {$r_\ast$};
  \node[below] at (-0.3,0) {$r_\infty$};
  \node at (2.9,1.3) {$\Gamma_A^{(\mathrm{con})}$};
  \draw[thick,dashed] (\rzero,-2.7) -- (\rzero,2.7);
  \node[below] at (\rzero-0.3,0) {$r_0$};
\end{scope}
\node at (6.2,0.3) {$\Longleftrightarrow$};
\node[above] at (6.2,0.6) {\small phase transition};
\begin{scope}[xshift=9cm,scale=1.1]
  \def\rinf{0}
  \def\rzero{3.7}
  \def\yup{1.6}
  \def\ydown{-1.6}
  \def\b{2.0}
  \draw[thick,->] (4.0,0) -- (-0.5,0)  node[left] {$r$};
  \draw[thick,->] (0,-2.8) -- (0,2.8) node[above] {$x$};
  \draw[very thick] (0,\yup) -- (3.7,\yup);
  \draw[very thick] (0,\ydown) -- (3.7,\ydown);
  \node[left] at (0,\b) {$l_A/2$};
  \node[left] at (0,-\b) {$-l_A/2$};
  \node[left] at (0,0.5) {$A$};
  \node at (1.7,2) {$\Gamma_A^{(\mathrm{dis})}$};
  \node[below] at (-0.3,0) {$r_\infty$};
  \draw[thick,dashed] (\rzero,-2.7) -- (\rzero,2.7);
  \node[below] at (\rzero-0.3,0) {$r_0$};
\end{scope}
  \end{tikzpicture}
  \caption{AdS soliton setup: connected and disconnected phases.}
  \label{fig:AdSsolitonsetup}
\end{figure}
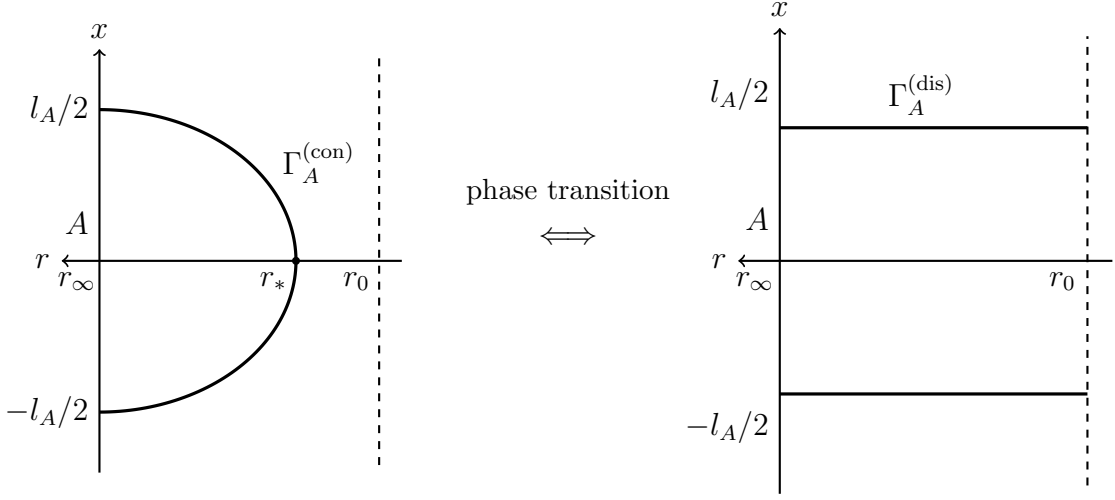
The holographic entanglement entropy shows interesting phase transition phenomena which reflect the confining/deconfining phase transition \cite{Nishioka:2006gr,Klebanov:2007ws}.
This is due to the two possible phases of minimal surfaces shown in Fig.~\ref{fig:AdSsolitonsetup}:
\begin{itemize}
  \item \textbf{Connected phase.}
    Denoting the turning point of the minimal surface by $r=r_*$ and assuming that $A$ is centered at $x=0$, the minimal surface $\Gamma_A^{\rm (con)}$ is given by
    \begin{equation}
    \begin{aligned}
        x_{\pm}(r)= \pm \int_{r_*}^{r}\frac{L^2\,d\tilde{r}}{\tilde{r}^2 \sqrt{f(\tilde{r})
   \left(\frac{\tilde{r}^{2(d-1)}f(\tilde{r})}{r_*^{2(d-1)}f(r_*)}-1\right)}} \, .
   \label{eq:x(r)-soliton}
    \end{aligned}
    \end{equation}
    Then, $r_*$ is related to the width $l_A$ by
    \begin{equation}
    \begin{aligned}
        \frac{l_A}{2}=\int_{r_*}^{r_\infty}\frac{L^2\,d\tilde{r}}{\tilde{r}^2 \sqrt{f(\tilde{r})
   \left(\frac{\tilde{r}^{2(d-1)}f(\tilde{r})}{r_*^{2(d-1)}f(r_*)}-1\right)}} \, ,
    \end{aligned}
    \end{equation}
    where $r_\infty$ is a UV cutoff which relates to $\varepsilon$ via $r_\infty=L^2/\varepsilon$.
    The area of the codimension-two surface $\Gamma_A^{\rm (con)}$ is
    \begin{equation}
    \mathcal{A}(\Gamma_A^{\rm (con)})=2R\left(\frac{\ell}{L}\right)^{d-3}\int_{r_*}^{r_\infty}\frac{r^{2d-4}\sqrt{f(r)}}{\sqrt{r^{2(d-1)}f(r)-r_*^{2(d-1)}f(r_*)}}\,dr \, .
    \end{equation}
    The volume of the codimension-one surface $r_A$ enclosed by $\Gamma_A^{\rm (con)}$ and the strip $A$ can be expressed as
    \begin{equation}
        \mathcal{V}_A^{\rm (con)} =\frac{2R\,\ell^{d-3}}{L^{d-2}}\int_{r_*}^{r_\infty}r^{d-2}\,x_+(r)\,dr \, ,
    \end{equation}
    where $x_+(r)$ is given in eq.~\eqref{eq:x(r)-soliton}.
    \item \textbf{Disconnected phase.}
    We obtain the following area of the codimension-two surface $\Gamma_A^{\rm (dis)}$:
    \begin{equation}
    \mathcal{A}(\Gamma_A^{\rm (dis)})=\frac{2R}{d-2}\left(\frac{\ell}{L}\right)^{d-3}(r_\infty^{d-2}-r_0^{d-2}) \, .
    \end{equation}
    The volume of the codimension-one surface $r_A$ is given by
    \begin{equation}
        \mathcal{V}_A^{\rm (dis)} =\frac{R\,\ell^{d-3}\,l_A}{(d-1)L^{d-2}}(r_\infty^{d-1}-r_0^{d-1}) \, .
    \end{equation}
\end{itemize}
When we take $q$ adjacent strips with the same width $l$ and consider the fully connected case, we can evaluate the volume of the EWP by subtraction of the volume of the entanglement wedges as follows:
\begin{equation}
    V_{\{A_i\}:O}=\mathcal{V}_{A_1\cdots A_q}-\sum_{i=1}^q\mathcal{V}_{A_i}.
\end{equation}
In general, we expect that three different phases can emerge, depending on the values of the strip width $l$, as depicted in Fig.~\ref{fig:phasetrsf_AdSsoliton}. 

\begin{figure}[httt]
  \centering
    \includegraphics[scale=0.8]{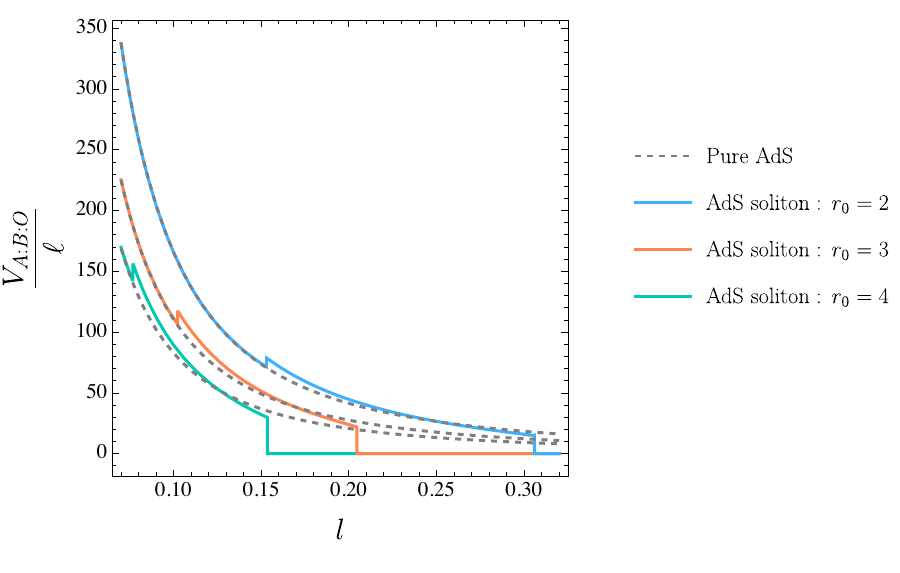}
  \caption{The volume of the EWP for the boundary state $\rho_{ABO}$ in the AdS$_5$ soliton and pure AdS$_5$. We take as $A$ and $B$ adjacent strips with the same width $l$, and we set $L=1$. As we compactify the $\chi-$direction in the AdS soliton, we do the same in pure AdS with a periodicity $R=\frac{\pi L^2}{r_0}$.}
  \label{fig:AdS5soliton}
\end{figure}

Below we focus on the EWP for the AdS$_5$ soliton, dual to three-dimensional pure Yang-Mills, and compare its volume with that for the (compactified) pure AdS$_5$ geometry. We consider two adjacent strips $A$ and $B$ with the same width $l$. The numerical results for comparison of the volumes of the EWP are shown in Fig.~\ref{fig:AdS5soliton}.
From this figure, we find that there are three phases, corresponding to the minimal surface phase transition shown in Fig.~\ref{fig:phasetrsf_AdSsoliton}. In the small $l$ region, we get a result very similar to that in the pure AdS case. For the intermediate values of $l$, the volume of the EWP for the AdS soliton turns out to exceed that for pure AdS. From the gravity point of view, this is obvious because the EWP in the first phase is contained into the EWP in the second phase. This might suggest that multi-partite entanglement enhances slightly before the confining/deconfining phase transition. Finally, in the large $l$ phase, the volume of the EWP vanishes because disconnected branches of $\Gamma_{A},\Gamma_B,$ and $\Gamma_{AB}$ coincide two by two. The result suggests that multi-partite entanglement vanishes in the confinement phase, as we naturally expect.

\section{Entanglement Wedge Polygon in Time-dependent Backgrounds}
\label{sec:quenches}
\subsection{One-sided BTZ}
\label{subsec:EOW-quench}
We consider the excited state $\ket{\Psi}_R$ in a CFT$_2$ introduced in subsection~\ref{subsubsec:one-sided_BTZ} and we evolve it in Lorentzian time, getting $\ket{\Psi(t_b)}_R = e^{-i H t_b} \ket{\Psi}_R$.
This corresponds to the quench process described in \cite{Calabrese:2005in}, that in the holographic context can be seen as the formation of a black hole from a pure state.
The bulk dual of this state is a one-sided BTZ black hole with an end-of-the-world (EOW) brane cutting the Penrose diagram of a two-sided BTZ black hole into half \cite{Hartman:2013qma}.
The spacetime metric is thus the same as in eq.~\eqref{eq:BTZ-metric}. 
We distinguish between a planar boundary, for which $\varphi=x/L$ is non-compact, and an $S^1$ boundary, for which the space coordinate $\varphi$ is compact.
\\

{\bf Planar boundary $\mathbb{R}$.}
We consider the time-evolved excited state in a CFT$_2$ on non-compact space.
For a boundary segment $A$ of length $l_A >> \beta$ and initial times, the extremal codimension-two surface $\Gamma_A$ anchored at $A$ enters the black hole horizon and ends on the EOW brane. When this happens, $\Gamma_A$ is the union of two disconnected geodesics, whose length depends on time as \cite{Hartman:2013qma}
\begin{equation}
    A(t_b) = 2 L \log \left( \frac{\beta}{\pi \varepsilon} \right) + 2 L \log \left( \cosh{\left( \frac{2\pi t_b}{\beta} \right)} \right) \, ,
\end{equation}
with $\varepsilon$ a UV cutoff.
This has to be compared with the length of the minimal geodesic outside the black hole
\begin{equation}
    A_{\rm out} = 2L \log \left( \frac{\beta}{\pi \varepsilon} \right) + 2L \log \left( \sinh{\left( \frac{\pi l_A}{\beta} \right)} \right) \, .
\end{equation}
The length $A(t_b)$ grows linearly at late times and becomes greater than $A_{\rm out}$ at $t_b \sim l_A/2$, when we have a phase transition.
At large times, we can thus approximate the regularized length of $\Gamma_A$ as
\begin{equation}
    A^{\rm fin}(\Gamma_A) = \frac{2\pi L}{\beta} \times
    \begin{cases}
       2t_b \, & \text{for $t_b<l_A/2$} \\
        l_A & \text{for $t_b>l_A/2$}
    \end{cases}
    \, .
\end{equation}

The volume of the homology surface $r_A$ bounded by the HRT surface and the boundary subregion $A$, after subtracting the value in the thermal state, can be approximated by the volume of the spatial slice behind the event horizon. 
We thus estimate
\begin{equation}
    \mathcal{V}_A - \mathcal{V}_A^{\rm (thermal)} = \frac{2\pi L^2}{\beta^2} \times
    \begin{cases}
       l_A \, t_b \, & \text{for $t_b<l_A/2$} \\
        0 & \text{for $t_b>l_A/2$}
    \end{cases} \, .
\end{equation}

Let us now take two segments $A$ and $B$ with lengths $l_B>l_A>>\beta$ and $O$ the union of two disconnected semi-infinite segments, so that $A,B,O$ is a tri-partition of the full boundary, $\rho_{ABO}(t_b) = \ket{\Psi(t_b)}_R \! \bra{\Psi(t_b)}_R$.
Following the above estimate, the volume of the EWP behaves as
\begin{equation}
\label{eq:VABO-quench-noncompact}
    V_{A:B:O} = \frac{2 \pi L^2}{\beta^2} \times
    \begin{cases}
        0 & \text{for $t_b<l_A/2$} \\
        l_A \, t_b \, & \text{for $l_A/2<t_b<l_B/2$} \\
        (l_A+l_B) t_b \, & \text{for $l_B/2<t_b<(l_A+l_B)/2$} \\
        \frac{\beta^2}{2} \, & \text{for $t_b>(l_A+l_B)/2$}
    \end{cases} \, .
\end{equation}
Note that the volume finally saturates at the constant value $V_{A:B:O}=\pi L^2$, fixed by the Gauss-Bonnet theorem. The result simply generalizes to the case of $q$ adjacent segments with $l_{A_q}>\dots > l_{A_1} >> \beta$. At $t_b = l_{A_i}/2$ the volume undergoes a phase transition and the growth rate increases. At $t_b = \sum_{i=1}^q l_{A_i}/2$ the volume saturates at the value $V_{\{A_i\}:O}=(q-1)\pi L^2$.

Another simple example is given by choosing $B$ with a finite length $l_B>>\beta$ and $A,O$ as semi-infinite lines, such that $A,B,O$ is a tri-partition of the whole boundary.
Similarly to the previous case, the volume of the EWP is given by
\begin{equation}
    V_{A:B:O} = \frac{2\pi L^2}{\beta^2} \times
    \begin{cases}
        0 & \text{for $t_b<l_B/2$} \\
        l_B \, t_b \, & \text{for $t_B>l_B/2$}
    \end{cases}
    \, ,
\end{equation}
where the phase transition is due to the change of dominance between the geodesic $\Gamma_B$ given by two disconnected components reaching the EOW brane and the connected geodesic $\Gamma_B$ external to the event horizon.
\\

{\bf Compact boundary $S^1$.}
We now turn to the time-evolved excited state in a CFT$_2$ on compact space $S^1$, and we assume that $\beta<<R$, with $R$ the size of the spatial circle. 
We consider a boundary subregion $A$ at time $t_b$ with size $l_A=R \, \alpha_A/(2\pi)$, where $\alpha_A$ is the opening angle of $A$. For a proper $l_A$, as discussed in subsection~\ref{subsubsec:one-sided_BTZ}, and initial times, the extremal codimension-two surface $\Gamma_A$ anchored at $A$ enters the black hole horizon and ends on the EOW brane. 
With the growing of time $t_b$, the length behind the black hole horizon grows, until at $t_b \sim l_A/2$ the shortest geodesic becomes the connected one outside the black hole, whose length is the thermal one.

The volume of the homology surface $r_A$ bounded by the HRT surface and the boundary subregion $A$ has been estimated in \cite{Haah:2025hyf,Fan:2025moc}.
In \cite{Haah:2025hyf}, after subtracting the value $\mathcal{V}_A$ in the thermal state, the regularized volume has been approximated by the volume of the spatial slice behind the event horizon. 
As a result, for a subregion of size $l_A<R/2$, it has been argued that
\begin{equation}
    \mathcal{V}_A - \mathcal{V}_A^{\rm (thermal)} = \frac{2\pi L^2}{\beta^2} \times
    \begin{cases}
       l_A \, t_b \, & \text{for $t_b<l_A/2$} \\
        0 & \text{for $t_b>l_A/2$}
    \end{cases} \, ,
\end{equation}
and
\begin{equation}
    \mathcal{V}_{\bar{A}} - \mathcal{V}_{\bar{A}}^{\rm (thermal)} = \frac{2\pi L^2}{\beta^2} \times
    \begin{cases}
       (R-l_A) \, t_b \, & \text{for $t_b<l_A/2$} \\
        R \, t_b & \text{for $t_b>l_A/2$}
    \end{cases} \, ,
\end{equation}
for the complementary subregion $\bar{A}$ of length $R-l_A>R/2$.

Following the same approximation, we estimate the time evolution of the EWP for three subregions $A,B,O$ whose union is the full boundary $S^1$. We denote by $l_O > l_B > l_A$ their sizes, with $l_O=R-l_A-l_B$. 
Assuming that at initial time $\Gamma_A, \Gamma_B,$ and $\Gamma_O$ are all in the disconnected phase, the resulting volume of the EWP is
\begin{equation}
\label{eq:VABO-quench-compact}
    V_{A:B:O} = \frac{2\pi L^2}{\beta^2} \times
    \begin{cases}
        0 & \text{for $t_b<l_A/2$} \\
        l_A \, t_b \, & \text{for $l_A/2<t_b<l_B/2$} \\
        (l_A+l_B) t_b \, & \text{for $l_B/2<t_b<(l_A+l_B)/2$} \\
        R \, t_b \, & \text{for $t_b>(l_A+l_B)/2$}
    \end{cases}
    \, .
\end{equation}
Contrary to eq.~\eqref{eq:VABO-quench-noncompact}, at later times the result is just the volume of the full spatial slice behind the black hole horizon. 
The growth rate of $V_{A:B:O}$ at late times is indeed half the growth rate of volume complexity of the thermofield double state, 
$dV/dt_b = 8 \pi G_N L M$, with $M$ the black hole mass \cite{Susskind:2014rva,Stanford:2014jda,Carmi:2017jqz}.

The result trivially generalizes to the case of $q$ adjacent segments with $l_{A_q}>\dots > l_{A_1}$. At $t_b = l_{A_i}/2$ the growth rate increases due to the phase transition. At $t_b = (R-l_O)/2$ the growth rate reaches the complexity value $4 \pi G_N L M$.

\subsection{Thermofield double state}
\label{subsec:TFD-time}
A thermofield double state (TFD) of the product of two identical two-dimensional theories $CFT_L \times CFT_R$,
is dual to an eternal two-sided AdS$_3$ black hole. 
In order to introduce time-dependence, we assume that the boundary time $t_b$ grows in the same direction on both the left and right boundaries, as in \cite{Hartman:2013qma}.
The boundary state is then
$\ket{TFD(t_b)} = \sum_n e^{-\beta E_n/2 -2i E_n t_b} \ket{E_n}_L\ket{E_n}_R$ up to a normalization.
We comment on the time evolution of the volume of the EWP for three partitions $A,B,O$ whose union is the whole right and left boundaries at given time $t_b$.

If $O$ is the union of the full right boundary with a, possibly empty, subregion of the left, the EWP lies outside the black hole in the left region of spacetime, and is thus time-independent.
In the three-dimensional case, the Gauss-Bonnet theorem applies and the volume is topological, as described in subsection~\ref{subsec:pure_BTZ}.

A simple choice for which the volume $V_{A:B:O}$ is time-dependent is given by taking disconnected $A,B,O$ with one component on the right boundary and an identical component on the left one.
In particular, let us denote by $l_A,l_B,l_O$ the sizes of the connected components on each boundary, with $l_O>l_B>l_A>>\beta$.
In this case, $V_{A:B:O}$ is simply given by twice the volume obtained in the previous subsection, due to the reflection symmetry of the geometry in the right and left sides of spacetime.

\subsection{Global quench in Vaidya AdS$_3$}
\label{subsec:Vaidya}
Another holographic model describing the formation of a black hole from a pure state is provided by the Vaidya spacetime.
The Vaidya-AdS$_3$ metric is
\begin{equation}
    ds^2 = \frac{1}{z^2} \left[ -f(v,z) dv^2 -2 dz \, dv + dx^2 \right] \, ,
\end{equation}
where $v$ is the ingoing null coordinate, coinciding with $t$ at the boundary $z=0$, and $x$ is a non-compact coordinate. In this subsection we fix $L=1$ for simplicity.
We consider the blackening factor \cite{Hubeny:2007xt}
\begin{equation}
    f(v,z) = 1 - m(v) z^2 \, , \qquad
    m(v) = \frac{M}{2} \left( 1 + \tanh \frac{v}{v_0} \right) \, .
\end{equation}
The metric describes the formation of a BTZ black hole with mass $M$ triggered by the gravitational collapse of null matter falling from the boundary of vacuum AdS$_3$. The collapsing matter follows a null trajectory represented by a shell of thickness parameterized by $v_0$. At $v \to -\infty$ we recover AdS$_3$, whereas at $v \to +\infty$ we have a BTZ spacetime. This time-dependent geometry provides a holographic model for a thermal global quench, namely a unitary evolution from the vacuum state triggered by a global excitation on a time scale $v_0$.

In the zero-thickness limit $v_0 \to 0$, corresponding to a sudden injection of energy, the bulk geometry is a gluing of the AdS$_3$ and the BTZ solution along the null trajectory $v=0$:
\begin{equation}
    m(v) = M \, \theta(v) \, ,
\end{equation}
with $\theta(v)$ the Heaviside step function.
This allows us to have some analytic control on the problem.

\begin{figure}[ht]
\center
\includegraphics[scale=0.7]{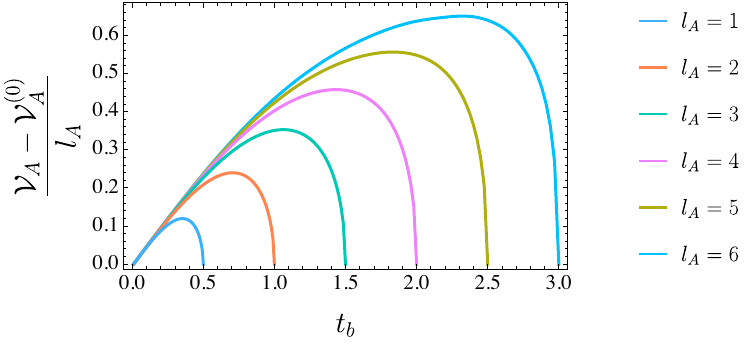}
\qquad
\includegraphics[scale=0.8]{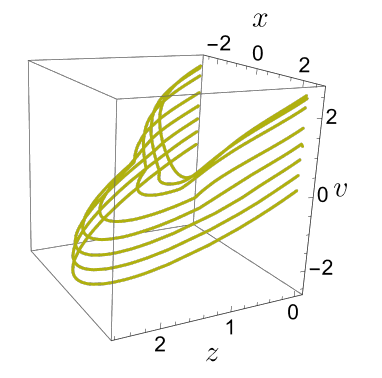}
\caption{Left: Volume $\mathcal{V}_A$ given by the $x$-independent ansatz. We have subtracted the volume at $t_b=0$ (volume $\mathcal{V}_A^{(0)}$ for Poincar\'e AdS$_3$) and normalized by the length $l_A$ of the subregion. 
Right: Geodesics $\Gamma_A$ at which the maximal surface is anchored for $l_A=5$ and varying boundary time.
A refraction happens at $v=0$, where the gluing between the AdS$_3$ and the BTZ spacetime takes place.
We have set $M=1$ and $L=1$.
}
\label{fig:VolA_Vaidya}
\end{figure}

We take as boundary subregion $A$ a strip of length $l_A$ in the $x$-direction at constant boundary time $t_b$. The codimension-two minimal surfaces $\Gamma_A$ have been studied in \cite{Hubeny:2007xt,Balasubramanian:2011ur} and the volume of the enclosed codimension-one surfaces $r_A$ in \cite{Chen:2018mcc,Auzzi:2019mah}. 
It is convenient to parametrize the codimension-one homology surface $r_A$ anchored at $\Gamma_A \cup A$ as $z=z(v,x)$. The volume functional reads
\begin{equation}
    \mathcal{V}_A = \int dv \, dx \, \mathit{V} \, , \qquad
    \mathit{V} = \frac{\sqrt{-(2 z_v + f(v,z)) - z_x^2}}{z^2} \, ,
\end{equation}
in which $z_v=\partial_v z$ and $z_x=\partial_x z$.
The surface $r_A$ with maximal volume satisfies the equation of motion
\begin{equation}
\label{eq:eom-volume-Vaidya}
    \partial_v \frac{\partial \mathit{V}}{\partial z_v} + \partial_x \frac{\partial \mathit{V}}{\partial z_x} - \frac{\partial \mathit{V}}{\partial z} =0 \, ,
\end{equation}
with boundary conditions specified by the boundary subregion $A$ and the corresponding codimension-two minimal surface $\Gamma_A$.
As it was shown in \cite{Auzzi:2019mah}, the $x$-independent ansatz $z = z(v)$ minimizing the area functional is a solution to the volume equations of motion \eqref{eq:eom-volume-Vaidya} only in the time-independent regime (before the quench $t_b<0$ and after thermalization $t_b>l_A/2$). In the intermediate time regime $0< t_b < l_A/2$, the $x$-independent ansatz violates eq.~\eqref{eq:eom-volume-Vaidya}, but still provides a reasonable (lower bounding) estimate of the maximal volume $\mathcal{V}_A$, especially at early times $t_b \gtrsim 0$ and shortly before thermalization $t_b \lesssim l_A/2$.
In Fig.~\ref{fig:VolA_Vaidya} we show the time evolution of $\mathcal{V}_A$ obtained by the $x$-independent ansatz.

When $l_A \sqrt{M}>1$, we can distinguish among three main regimes \cite{Auzzi:2019mah}:
\begin{itemize}
    \item Early times, $0 <t_b \leq \mathcal{O}(\log(l_A \sqrt{M}))$: 
    \begin{equation}
    \label{eq:VA-Vaidya-early}
        \mathcal{V}_A =  \frac{l_A}{\varepsilon} + \frac{M}{2} \,l_A \, t_b + \mathcal{O}(l_A^0) \, .
    \end{equation}
    \item Intermediate times, $\mathcal{O}(\log(l_A \sqrt{M})) < t_b < \frac{l_A}{2} - \frac{0.53}{\sqrt{M}}$:
    The $x$-independent ansatz fails, but it is still useful as a lower bound to the maximal volume.
    \item Late times, $\frac{l_A}{2} - \frac{0.53}{\sqrt{M}} < t_b < \frac{l_A}{2}$: 
    \begin{equation}
    \label{eq:VA-Vaidya-late}
        \mathcal{V}_A = \frac{l_A}{\varepsilon} + l_A^2 \left( \frac{\sqrt{M}}{2} \right)^{9/4} \left( \frac{l_A}{2} - t_b \right)^{1/4} + \mathcal{O}(l_A^0) \, .
    \end{equation}
\end{itemize}

\begin{figure}[t]
\center
\includegraphics[scale=1]{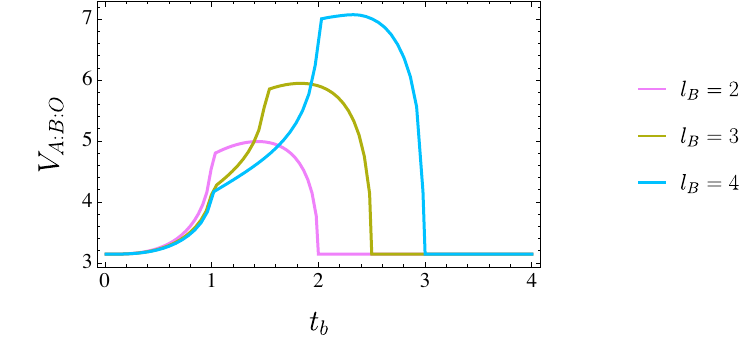}
\caption{Estimate of the volume of the EWP for a tri-partition of the state dual to a Vaidya spacetime. The plot has been obtained by subtraction of volumes of the maximal homology surfaces for the $x-$independent ansatz. We have set $l_A=2$, $M=1$, and $L=1$.
}
\label{fig:Vaidya-trg}
\end{figure} 
Let us take three adjacent strips $A,B,O$ whose union is the full boundary.
We assume that $A$ and $B$ has finite lengths $l_A\leq l_B$, thus $O$ is the union of two semi-infinite segments.  
Contrary to the static cases, where eq.~\eqref{eq:V-subtraction} can be applied, in the time-dependent case it is in general not possible to obtain $V_{A:B:O}$ by subtractions of the volumes $\mathcal{V}$ for single subregions. This happens because the codimension-one surface for a single strip does not lie on a constant-time slice in the bulk. Consequently, at $0 < t_b < (l_A +l_B)/2$, the homolgy surface with maximal volume $r_{AB}$ does not necessarily contain the homology surfaces $r_{A,B}$ with maximal volume, so that a subtraction of volumes is not correct. 
Nonetheless, in some time regimes a subtraction of volumes provides a good estimate of the actual volume $V_{A:B:O}$. In particular, this happens at early and late times, when the $x-$independent ansatz also applies.
Starting by the above mentioned results for a single strip, we thus conclude that:
\begin{itemize}
    \item Early times, $0 < t_b \leq \mathcal{O}\left(\log\left(l_A\sqrt{M}\right) \right)$:
    the three homology surfaces with maximal volume $r_A,r_B,r_{AB}$ mainly lie in the AdS$_3$ part of spacetime, thus at constant time in the bulk, with deviations close to the boundary. Therefore, from eq.~\eqref{eq:VA-Vaidya-early} we deduce that the volume is time-independent
    \begin{equation}
        \frac{V_{A:B:O}}{L^2} = \mathcal{O}(1) \, .
    \end{equation}
    \item Late times, $\frac{l_A+l_B}{2} - \frac{0.53}{\sqrt{M}} < t_b < \frac{l_A + l_B}{2}$:
    from eq.~\eqref{eq:VA-Vaidya-late}, the volume of the EWP decreases as
    \begin{equation}
        \frac{V_{A:B:O}}{L^2} = (l_A+l_B)^2 \left( \frac{\sqrt{M}}{2} \right)^{9/4} \left( \frac{l_A+l_B}{2} - t_b \right)^{1/4} + \mathcal{O}(1) \, .
    \end{equation}
    Due to the Gauss-Bonnet theorem, the volume drops down to the starting (vacuum) value $V_{A:B:O}=\pi L^2$ at $t_b=(l_A+l_B)/2$. We expect a maximum followed by a sharp decrease.
\end{itemize}

In Fig.~\ref{fig:Vaidya-trg} we show a rough estimate of the time evolution of $V_{A:B:O}$ in Vaidya spacetime for given value of $l_A$ and different values of $l_B$. The numerical plot has been obtained by subtraction of the volumes $\mathcal{V}_{AB}-\mathcal{V}_A - \mathcal{V}_B$ for the $x-$independent ansatz. We stress that the result is tentative for two reasons: the $x-$independent ansatz is not precisely valid throughout the full time evolution and the subtraction of volumes is not in general valid in the covariant case. However, we expect that this estimate captures the qualitative behavior of the actual volume of the EWP in time.
We leave a more precise numerical analysis for future work.

\section{Entanglement Wedge Polygon for mixed states}
\label{sec:def-mixed}
\begin{figure}[htbp]
		\centering
		\includegraphics[width=13cm]{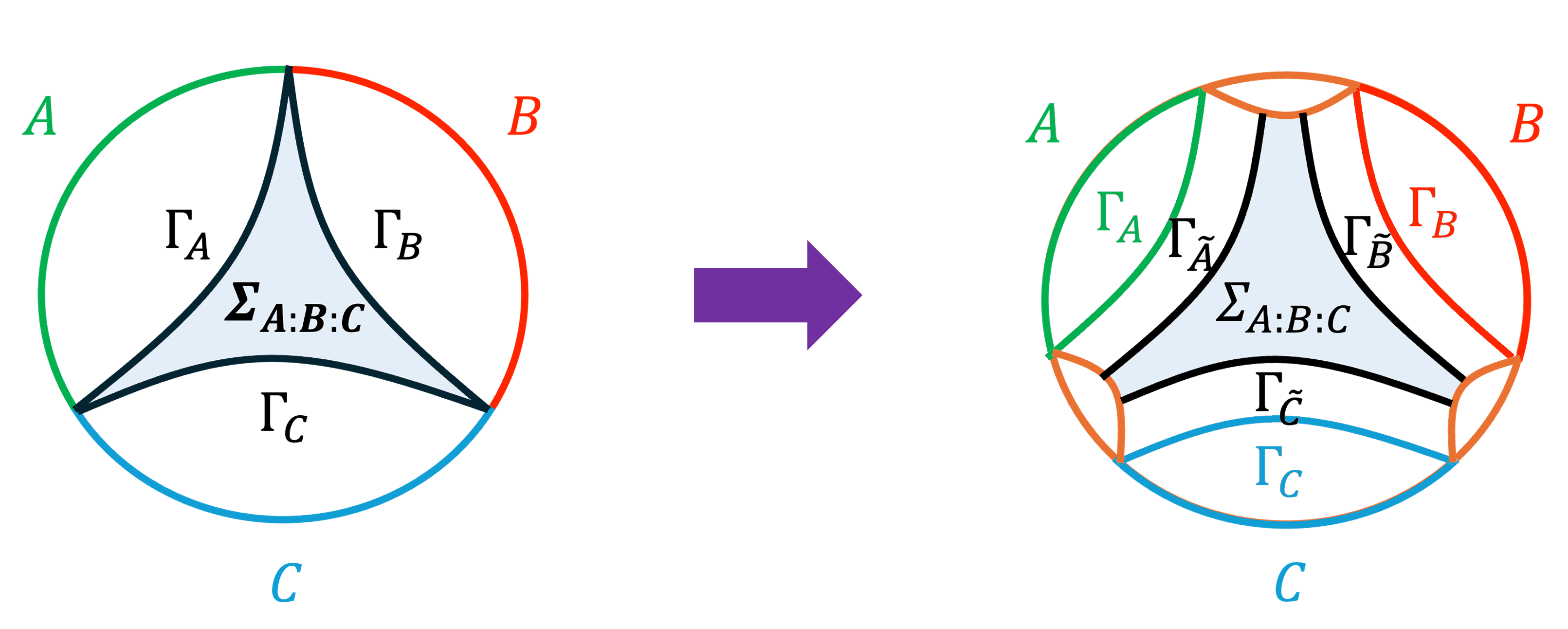}
        		\caption{The mixed state generalization of the entanglement wedge polygon.} 
		\label{fig:MixedstateEWP}
\end{figure}

Our previous definition of $V(\Sigma_{A_1:\ddd :A_q})$ can be applied to the case where $\rho_{A_1\ddd A_q}$ is a mixed state. 
However, the definition \eqref{eq:def-Sigma} applied to a mixed state is at odds with our interpretation of the volume of the EWP as a measure of the $q-$partite entanglement.
In particular, if we define $A_{q+1} = \ov{A_1\ddd A_q}$, we have $V(\Sigma_{A_1:\ddd :A_q})=V(\Sigma_{A_1:\ddd :A_{q+1}})$ . 
This property appears strange.
For example, we have $V(\Sigma_{A_1:A_2})\neq 0$ if $A_3$ is not empty, which contrasts with the fact that $V$ excludes the bi-partite entanglement. In other words, we want to find a definition that always gives $V(\Sigma_{A_1:A_2})=0$, even if $\rho_{A_1A_2}$ is mixed. 

Motivated by this, we introduce a new definition of
$V_{\{A_i\}}$ for mixed states. 
Rather than taking the extremal $\Gamma_{A_i}$ as boundary for $\Sigma_{\{A_i\}}$, we consider the \textit{entanglement wedge cross section} for $A_i$, that we denote by $\Gamma_{\ti{A_i}}$, see Fig.~\ref{fig:MixedstateEWP}.
Namely, $\Gamma_{\ti{A}_i}$ is the extremal codimension-two surface ending on $\Gamma_{\ov{A_{1}\ddd A_{q}}}$ and homologous to $A_i$ (assuming $\Gamma_{\ov{A_{1}\ddd A_{q}}}$ is homologously trivial). 
In other words, $\Gamma_{\ti{A_i}}$ is contained into the EW for $A_1 \ddd A_q$ and divides $A_i$ from the other subregions.
The area of $\Gamma_{\ti{A}_i}$, obtained by extremizing with respect to the location of its ends,  was introduced to compute the \textit{entanglement of purification} $E_P(\rho_{A_i:A_1\ddd A_{i-1}A_{i+1}\ddd A_q})$ \cite{Takayanagi:2017knl,Nguyen:2017yqw}, a measure of the correlation between $A_i$ and the other subregions. 

A reasoning for replacing $\Gamma_{A_i}$ with $\Gamma_{\ti{A_i}}$ in the mixed state case is the following.
The state $\rho_{A_1\ddd A_q}$ is dual to the EW$(A_1\ddd A_{q})$, which is the region surrounded by $\Gamma_{\ov{A_{1}\ddd A_{q}}}$. To compute the volume of the EWP as we did for the pure state case, we need to subtract the regions dual to $A_i$ ($i=1,\ddd,q$). For a pure state, such regions correspond to the EW$(A_i)$. For a mixed state, it is natural to identify the "entanglement wedge" of $A_i$ inside EW$(A_1\ddd A_{q})$ with the region between the entanglement wedge cross section $\Gamma_{\ti{A}_i}$ and the subregion $A_i$.
Given this, we define the volume of the EWP for a \emph{mixed} state as follows.
\\

{\bf Definition $M$.}
Let us consider $q$ \emph{disjoint} subregions that can share boundaries.
Given the purifier $O=\ov{A_{1}\ddd A_{q}}$, the entanglement wedge cross section for subregion $A_i$ is the space-like codimension-two extremal surface homologous to $\ti{A_i} =A_i \cup \Gamma_{O}^{(A_i)}$, where $\Gamma_O = \Gamma_{O}^{(A_i)} \cup \Gamma_{O}^{(A_1 \ddd A_{i-1} A_{i+1} \ddd A_q)}$.
Note that $\Gamma_O^{(A_i)}$ is in general disconnected.
The EWP is then defined as the codimension-one maximal surface $\Sigma_{\{A_i\}}$ satisfying
\begin{equation}
\label{eq:def-Sigma-mixed}
    \partial \Sigma_{\{A_i\}} = \left( \bigcup_{i=1}^q \Gamma_{\tilde{A}_i} \right) \cup \Gamma_O \setminus \bigcup_{i=1}^q \Gamma_{O}^{(A_i)} \, .
\end{equation}

Note that definition~\eqref{eq:def-Sigma-mixed} reduces to definition~\eqref{eq:def-Sigma} when the state is pure. 
Indeed, if $\rho_{A_1 \ddd A_q}$ is pure, $\Gamma_{\tilde{A}_i}=\Gamma_{A_i}$ and $\Gamma_O,\Gamma_O^{(A_i)} = \emptyset$.

In the static case, we can define the codimension-one homology surface anchored at $\Gamma_{\ti{A_i}} \cup \ti{A_i} = \Gamma_{\ti{A_i}} \cup \Gamma_O^{(A_i)} \cup A_i$, that we denote by $r_{\ti{A_i}}$.
This generalizes the EW$(A_i)$ to the case of mixed states.
Then, definition~\eqref{eq:def-Sigma-mixed} is equivalent to removing the homology surfaces $r_{\ti{A_i}}$ from the EW$(A_1 \ddd A_q)$, namely
\begin{align}
\label{eq:def-Sigma-mixed-static}
    \Sigma_{A_1: \ddd :A_q} &= r_{(A_1 \ddd A_q)} \setminus \bigcup_{i=1}^q r_{\ti{A_i}} \, .
\end{align}
As for pure states, definition \eqref{eq:def-Sigma-mixed} applies more generally to the covariant case.
Similarly, the properties $P1$ and $P2$ that hold for pure states promptly generalize to mixed states.
\\

{\bf Property $M1$.}
The volume of the EWP vanishes for bi-partitions of a mixed state:
\begin{equation}
    V_{A:B} = 0 \, .
\end{equation}
\begin{proof}
For a bi-partite mixed state $\rho_{AB}$, we have
$\Gamma_{\tilde{A}}=\Gamma_{\tilde{B}}$ and $\Gamma_O \setminus \left( \Gamma_{O}^{(A)} \cup \Gamma_{O}^{(B)} \right) = \emptyset$, so $\Sigma_{A:B}=\emptyset$.
\end{proof}

{\bf Property $M2$.}
The volume of the EWP increases for finer grained partitions of mixed states:
\begin{equation}
\label{eq:inequality-mixed}
    V_{(AB):C:D} \leq V_{A:B:C:D} \, ,
\end{equation}
where the equality holds if and only if 
$\Gamma_{\tilde{(AB)}} = \Gamma_{\tilde{A}} \cup \Gamma_{\tilde{B}}$, i.e. $E_P(\rho_{AB:CD})=E_P(\rho_{A:BCD}) + E_P(\rho_{B:ACD})$.\footnote{While $\Gamma_{\tilde{(AB)}} = \Gamma_{\tilde{A}} \cup \Gamma_{\tilde{B}}$ implies $E_P(\rho_{AB:CD})=E_P(\rho_{A:BCD}) + E_P(\rho_{B:ACD})$, similar loopholes as for the entanglement entropy apply to the reverse.}
\begin{proof}
If $\Gamma_{\ti{(AB)}} = \Gamma_{\ti{A}} \cup \Gamma_{\ti{B}}$, meaning that $E_P(\rho_{AB:CD})=E_P(\rho_{A:BCD}) + E_P(\rho_{B:ACD})$, we also have  $\Gamma_O^{(AB)} = \Gamma_O^{(A)} \cup \Gamma_O^{(B)}$. Therefore, from the definition \eqref{eq:def-Sigma-mixed} we clearly have $\Sigma_{(AB):C:D}=\Sigma_{A:B:C:D}$.

We now prove that if $\Gamma_{\ti{(AB)}} \neq \Gamma_{\ti{A}} \cup \Gamma_{\ti{B}}$, we have $\Sigma_{(AB):C:D} \subset \Sigma_{A:B:C:D}$.

Let us start from the static case, and denote by $r_{\tilde{A}}$ the maximal codimension-one surface anchored at $\Gamma_{\tilde{A}} \cup \tilde{A} = \Gamma_{\ti{A}} \cup \Gamma_O^{(A)} \cup A$. 
First, we observe that $\Gamma_{\tilde{(AB)}}$ is external to both $\Gamma_{\tilde{A}}$ and $\Gamma_{\tilde{B}}$.
Indeed, if this were not case then $\Gamma_{\tilde{(AB)}}$ would necessarily intersect at least one of $\Gamma_{\tilde{A}}$ and $\Gamma_{\tilde{B}}$.
By the same arguments as in theorem 17 of \cite{Wall:2012uf}, such intersections would contradict the minimality of the entanglement wedge cross sections.
Consequently, we have $r_{\tilde{A}} \cup r_{\tilde{B}} \subset r_{\tilde{(AB)}}$. 
Then, it follows from eq.~\eqref{eq:def-Sigma-mixed-static} that
$\Sigma_{(AB):C:D} = r_{(ABCD)} \setminus \left( r_{\tilde{(AB)}} \cup r_{\tilde{C}} \cup r_{\tilde{D}} \right) \subset r_{(ABCD)} \setminus \left( r_{\tilde{A}} \cup r_{\tilde{B}} \cup r_{\tilde{C}} \cup r_{\tilde{D}} \right) =\Sigma_{A:B:C:D}$, which completes the proof in the static case.

In the covariant case, we assume that the entanglement wedge cross-section $\Gamma_{\ti{A}}$ can be built by the maximin prescription \cite{Wall:2012uf}.
Then, by adapting theorem 17 and corollary 17 (h) of \cite{Wall:2012uf}, since $\ti{A},\ti{B} \subset \ti{AB} \subset (ABCD) \cup \Gamma_O$ and $\ti{A},\ti{B},\ti{C},\ti{D}$ are disjoint, we conclude that $\Gamma_{\ti{A}},\Gamma_{\ti{B}},\Gamma_{\ti{C}},\Gamma_{\ti{D}},\Gamma_{\ti{(AB)}},$ and $\Gamma_O$ are all minimal on the same achronal codimension-one slice $\Sigma$.
It is now convenient to define the sub-slice $r_{\ti{A}} \subset \Sigma$ bounded by $\Gamma_{\ti{A}} \cup \ti{A}$.
We can introduce $\hat{\Sigma}_{A:B:C:D} = r_{(ABCD)} \setminus \left( r_{\ti{A}} \cup r_{\ti{B}} \cup r_{\ti{C}} \cup r_{\ti{D}} \right)$ and $\hat{\Sigma}_{(AB):C:D} = r_{(ABCD)} \setminus \left( r_{\ti{AB}} \cup r_{\ti{C}} \cup r_{\ti{D}} \right)$, with $\hat{\Sigma}_{A:B:C:D}, \hat{\Sigma}_{(AB):C:D} \subset \Sigma$.
Similarly to the static case, we then have $\hat{\Sigma}_{(AB):C:D} \subset \hat{\Sigma}_{A:B:C:D}$.
The rest of the proof parallels the pure state case.
\end{proof}

Besides these, we have additional properties.
The first one provides a comparison between the EWP for a mixed state and for its purification.
The second one tells us that if a subregion is not correlated with the remaining ones, then it plays no role in determining the volume of the EWP.
\\

{\bf Property $M3$.}
The volume of the EWP for a mixed state $\rho_{ABC} = \Tr_O (\rho_{ABCO})$ is smaller than the volume of the EWP for the purification $\rho_{ABCO}$:
\begin{equation}
\label{eq:mixed_vs_pure}
    V_{A:B:C} \leq V_{A:B:C:O} \, .
\end{equation}
\begin{proof}
By definition of entanglement wedge cross section, $\Gamma_{\tilde{A}},\Gamma_{\tilde{B}},\Gamma_{\tilde{C}} \subset \text{EW}(ABC)$.
Let us assume that $A_i \subset \ti{A_i}$ for at least one of the subregions.
From theorem 17 of \cite{Wall:2012uf}, since $A_i \subset \ti{A_i}$, then $\Gamma_{\ti{A_i}}$ is external to $\Gamma_{A_i}$,\footnote{There are particular situations in which this is not true. An example is a partition of the thermal state on compact space $\rho_{\rm th} = \Tr_L(\ket{TFD}\!\bra{TFD})$, where the purifier is a full copy of the system. In this case, for $\alpha_i > \alpha_{\rm crit}$, we have $\ti{A_i} = A_i \cup \Gamma_O \supset A_i$, but  
$\Gamma_{\ti{A_i}}$ coincides with the branch of $\Gamma_{A_i}$ anchored at the boundary. In this case, $\Sigma_{A:B:C:O} = \Sigma_{A:B:C}$ and $V_{A:B:C} = V_{A:B:C:O}$.
We omit such kind of configurations from the  proof. We will comment more on these examples later.} and the two surfaces lie on the same achronal codimension-one slice $\Sigma$.
Moreover, since $A,B,C,O$ are disjoint, all the $\Gamma_{A_i}, \Gamma_{\ti{A_i}},$ and $\Gamma_O$ also lie on the same slice $\Sigma$. 
We can then define $\hat{\Sigma}_{A:B:C:O} = r_{(ABC)} \setminus \left( r_A \cup r_B \cup r_C \right)$ and $\hat{\Sigma}_{A:B:C} = r_{(ABC)} \setminus \left( r_{\ti{A}} \cup r_{\ti{B}} \cup r_{\ti{C}} \right)$, both sub-slices of $\Sigma$.
Since $\Gamma_{\ti{A_i}}$ is external to $\Gamma_{A_i}$, we have $r_{A_i} \subset r_{\ti{A_i}}$.
Consequently, $\hat{\Sigma}_{A:B:C} \subset \hat{\Sigma}_{A:B:C:O}$, also see Fig.~\ref{fig:MixedstateEWP}.

Now, if both $\hat{\Sigma}_{A:B:C} = \Sigma_{A:B:C}$ and $\hat{\Sigma}_{A:B:C:O} = \Sigma_{A:B:C:O}$, meaning that both sub-slices are maximal (as it happens in static case), the proof is concluded.

If instead $\hat{\Sigma}_{A:B:C}$ and $\hat{\Sigma}_{A:B:C:O}$ are not maximal, we can define $\hat{\Sigma}_{\rm comp} = \hat{\Sigma}_{A:B:C:O} \setminus \hat{\Sigma}_{A:B:C}$. By locally deforming $\hat{\Sigma}_{A:B:C}$ with its boundary $\partial\hat{\Sigma}_{A:B:C}$ held fixed, we get
\begin{equation}
    V_{A:B:C} = \max_{\partial \tilde{\Sigma}_{A:B:C} = \partial\hat{\Sigma}_{A:B:C}} V( \tilde{\Sigma}_{A:B:C}) 
    < \max_{\partial \tilde{\Sigma}_{A:B:C} = \partial\hat{\Sigma}_{A:B:C}} V( \tilde{\Sigma}_{A:B:C:O}) 
     \leq V_{A:B:C:O} \, ,
\end{equation}
where the second inequality follows by observing that $V_{A:B:C:O}$ is the maximal volume among all possible deformations that keep $\partial\Sigma_{A:B:C:O}$ fixed, but we are only considering a specific one.
\end{proof}

{\bf Property $M4$.}
If $\Gamma_O = \Gamma_{A_i} \cup \Gamma_{A_1 \dots A_{i-1} A_{i+1} \dots A_q}$, or in other words $A_i$ is not correlated with the remaining subregions, then $V_{\{A_i\}}= V_{A_1: \ddd: A_{i-1}: A_{i+1}: \ddd: A_q}$. As a direct consequence, if $q-2$ subregions or more are uncorrelated, then $V_{\{A_i\}}=0$.

\section{Entanglement Wedge Polygon in AdS$_3$ for mixed states}
\label{sec:AdS3-mixed}

\subsection{Mixed state in AdS$_3$}
\label{subsec:mixed_AdS3}
For $q-$partitions of a reduced vacuum state, $\rho_{A_1 \ddd A_q} = \Tr_O(\rho_{\rm vac})$, we have to apply the definition of the EWP for mixed state.
We here consider the vacuum state $\rho_{\rm vac}$ of a CFT$_2$ on both non-compact and compact space.
For $q$ subregions with finite size,
in the non-compact space the purifier $O$ is disconnected, whereas in the compact space it can be connected.

\begin{figure}[t]
  \centering
  \includegraphics[scale=0.09]{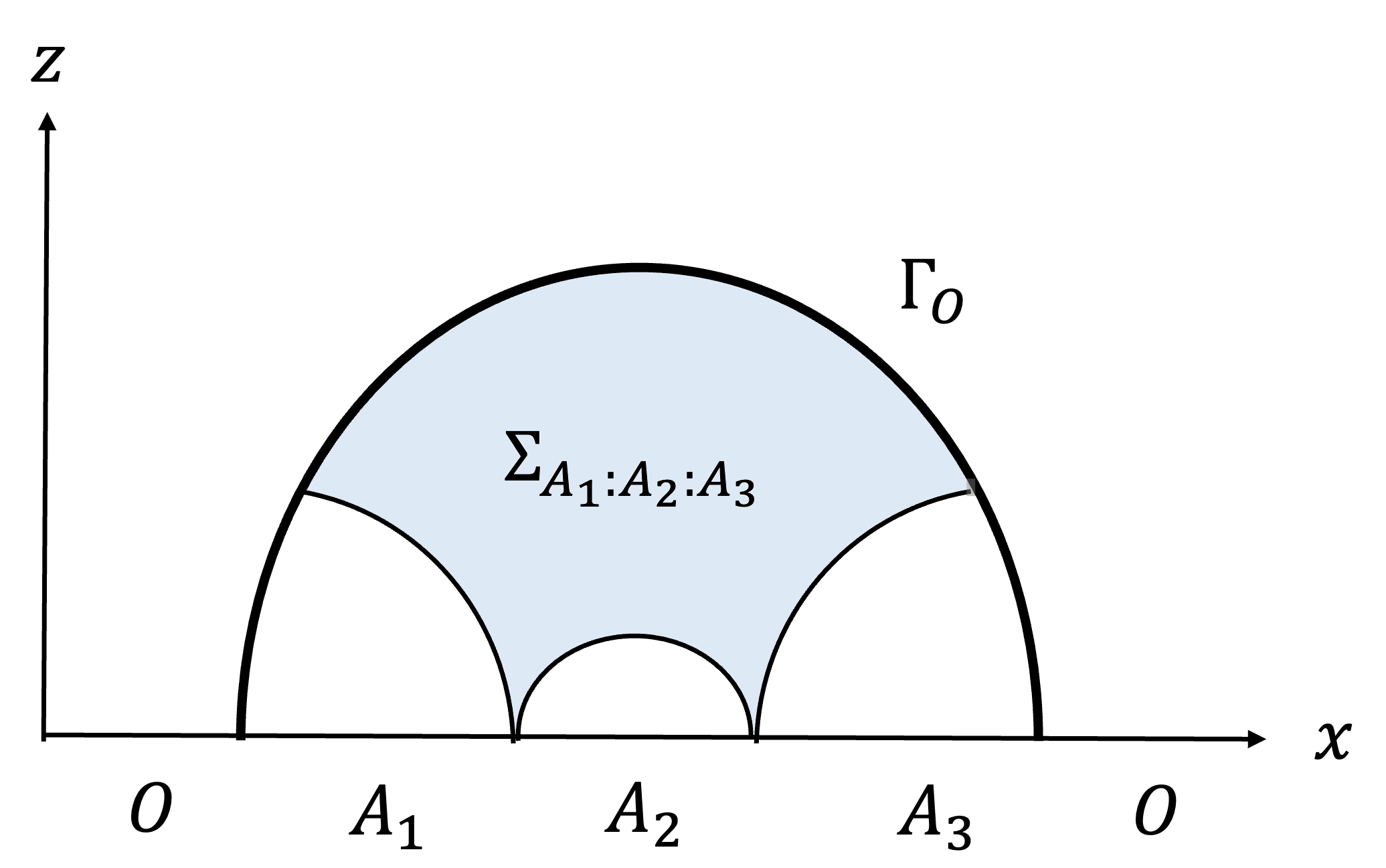}
  \qquad
  \includegraphics[scale=0.09]{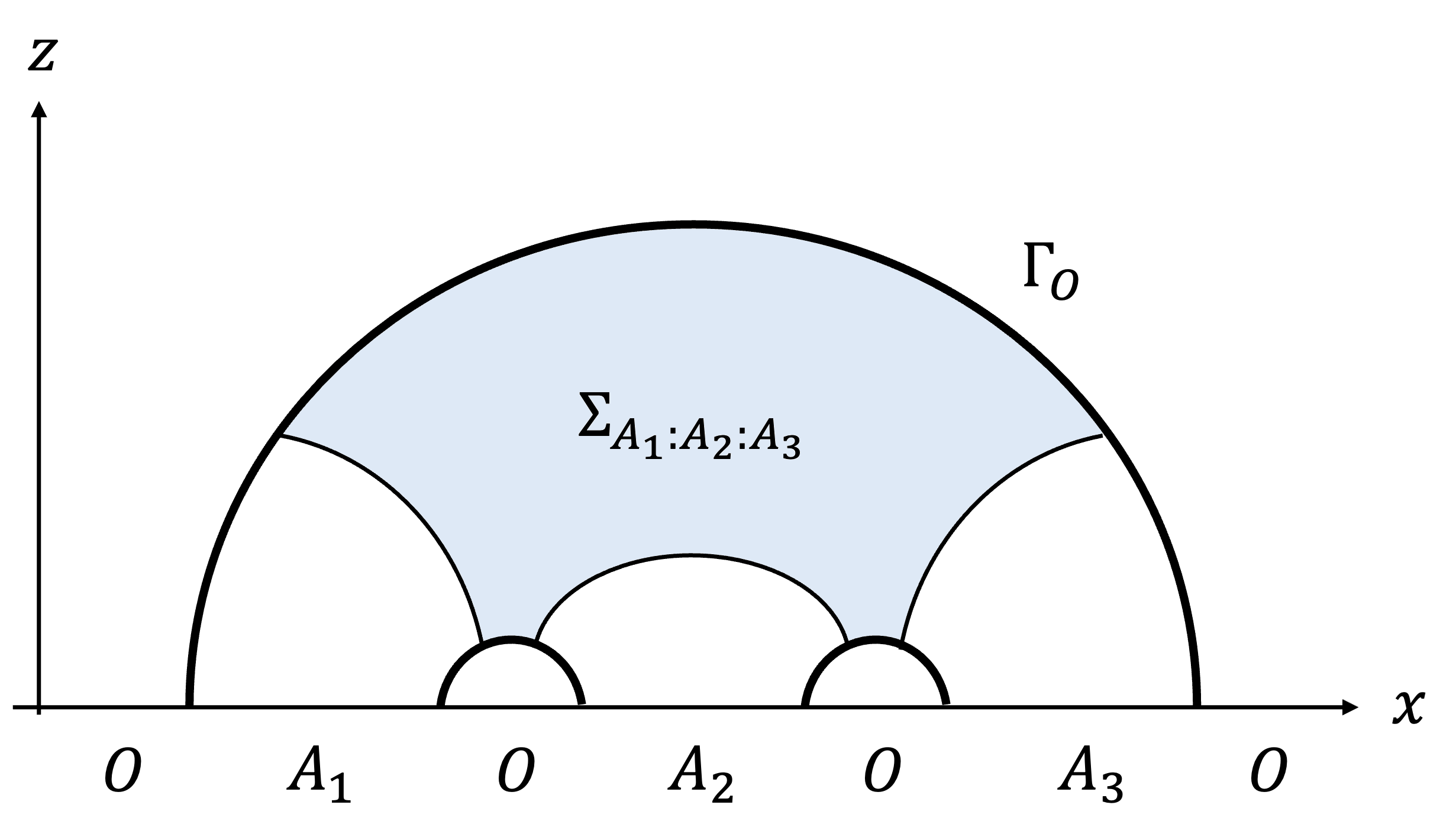}
  \caption{The EWP for a reduced vacuum state on non-compact space, $\rho_{A_1A_2A_3}=\Tr_O(\rho_{\rm vac})$, for adjacent regions (on the left) and non-adjacent regions (on the right). The bold curves represent $\Gamma_O$ and the thin ones $\Gamma_{\tilde{A_i}}$.}
  \label{fig:AdS3-mixed}
\end{figure}

Let us start from adjacent $A_1, \dots, A_q$ with finite size.
In Poincar\'e AdS$_3$ the purifier is given by the union of two disconnected subregions with infinite size, while in global AdS$_3$ it is a connected subregion with finite size.
In both cases we have $\Gamma_{\tilde{A_i}}=\Gamma_{A_i}$ for $i=2, \dots, q-1$.
Instead, $\Gamma_{\tilde{A_1}}$ and $\Gamma_{\tilde{A_q}}$ are geodesics with one end anchored at the boundary and the other end orthogonal to $\Gamma_O$, as displayed in Fig.~\ref{fig:AdS3-mixed}.
Therefore, by applying the Gauss-Bonnet theorem in eq.~\eqref{eq:GB-theorem-general} with $\chi=1$, we obtain
\begin{equation}
    V_{\{A_i\}} = \left( (q-1) \pi + 2 \times \frac{\pi}{2} - 2\pi \right) L^2 
    =(q-2) \pi L^2 \, .
\end{equation}
Note that, compared to $V_{\{A_i\}:O}= (q-1) \pi L^2$, there is a difference of $\pi L^2$, due to the lacking of the $\pi L^2/2$ volumes subtended by $\Gamma_{A_{1,2}}, \Gamma_{\tilde{A}_{1,2}},\Gamma_O^{(A_{1,2})}$.

For non-adjacent $A_1, \dots, A_q$ with finite size, 
the purifier $O$ is disconnected in both Poincar\'e and global AdS$_3$.
The highest volume of the EWP is obtained for a fully connected $\Gamma_O$, as in the right panel of Fig.~\ref{fig:AdS3-mixed}.
All the geodesics $\Gamma_{\tilde{A_i}}$ have both ends attached to $\Gamma_O$ and are orthogonal to it. As a result, from the Gauss-Bonnet theorem we get
\begin{equation}
    V_{\{A_i\}} = \left( 2q \times \frac{\pi}{2} - 2\pi \right) L^2 = (q-2)\pi L^2 \, .
\end{equation}
Compared to $V_{\{A_i\}:O}=(2q-2)\pi L^2$, there is a difference of $q \pi L^2$, arising from the volumes subtended by $\Gamma_{\tilde{A_i}},\Gamma_{A_i},\Gamma_O^{(A_i)}$, equal to $\pi L^2$ each.

If we keep increasing the separation between $A_j$ and $A_{j+1}$, $\Gamma_O$ gets partially disconnected. 
Consequently, $V_{\{A_i\}}= (q-3) \pi L^2$ or $V_{\{A_i\}}=(q-4) \pi L^2$, if only one subregion or more than one, respectively, gets disconnected from the others.
We then deduce that
\begin{equation}
    0 \leq V_{\{A_i\}} \leq (q-2) \pi L^2 \, ,
\end{equation}
and the volume is quantized in units of $\pi L^2$.
The upper bound is saturated when $\Gamma_O$ is fully connected, while the lower bound is saturated when $\Gamma_O$ connects two or less subregions.

\subsection{Mixed state in BTZ}
\label{subsec:mixed_BTZ}
We now investigate the EWP for $q-$partitions $A_i$ of a thermal state and its reduction.
In particular, we start from a thermofield double state $\ket{TFD}$, a pure state for the product of two entangled non-interacting CFT$_2$ on the left and right boundaries of a two-sided eternal BTZ black hole.
The state of a single boundary theory, say the right one, is the thermal state $\rho_{\rm th} = \Tr_L(\ket{TFD}\!\bra{TFD})$, obtained by tracing out the degrees of freedom of the other boundary theory, say the left one.
In what follows, we consider partitions of the (reduction of a) thermal state, assuming that the purifier $O$ contains one full boundary, $L \subseteq O$, and possibly it has a component $O_R$ on the other boundary, where all the subregions $A_i$ lies.

As for the pure state in one-sided BTZ, let us distinguish between a black brane with non-compact boundary and a black hole with compact boundary.
\\

\begin{figure}
    \centering
    \includegraphics[scale=0.09]{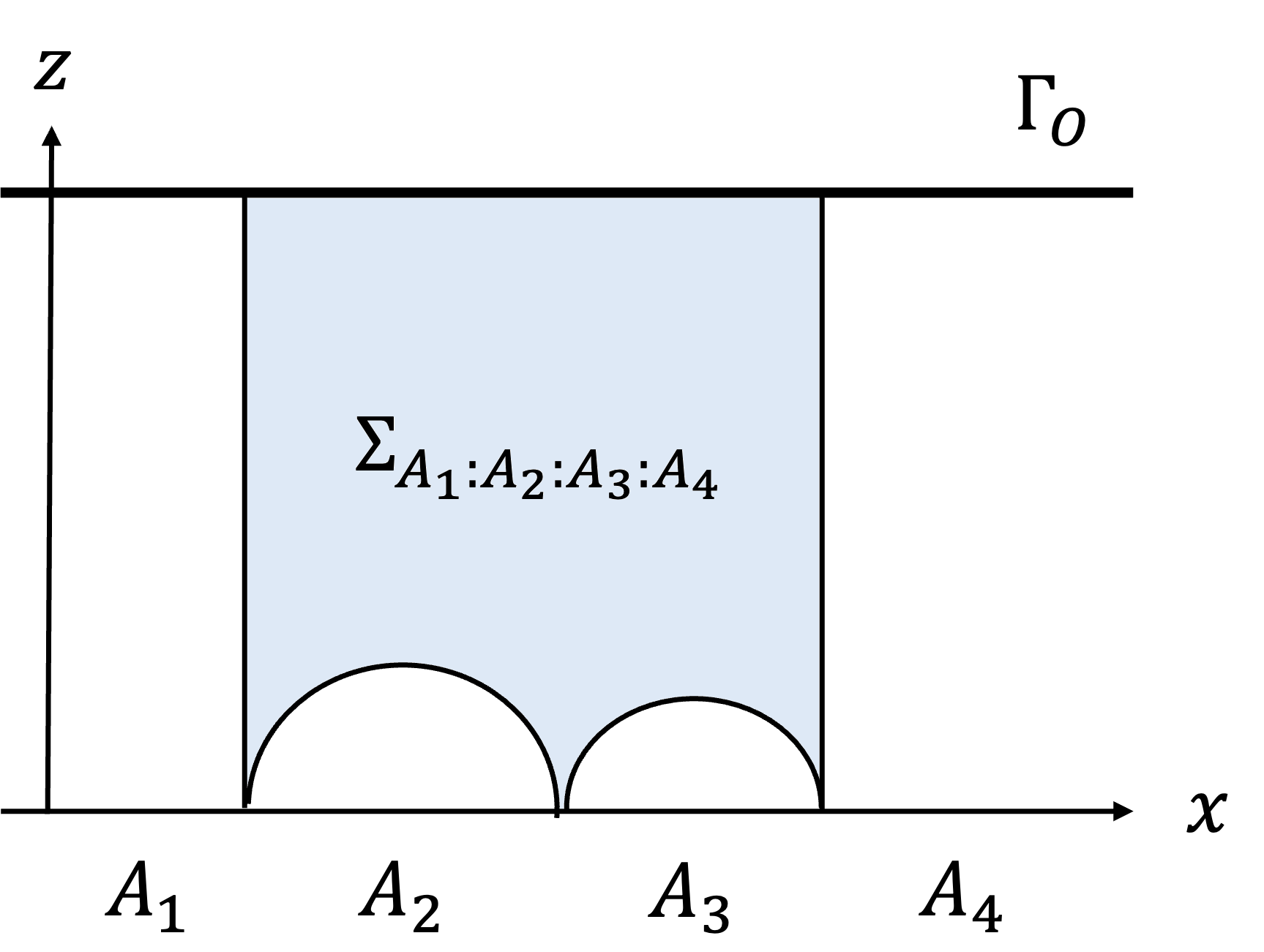}
    \qquad
    \includegraphics[scale=0.09]{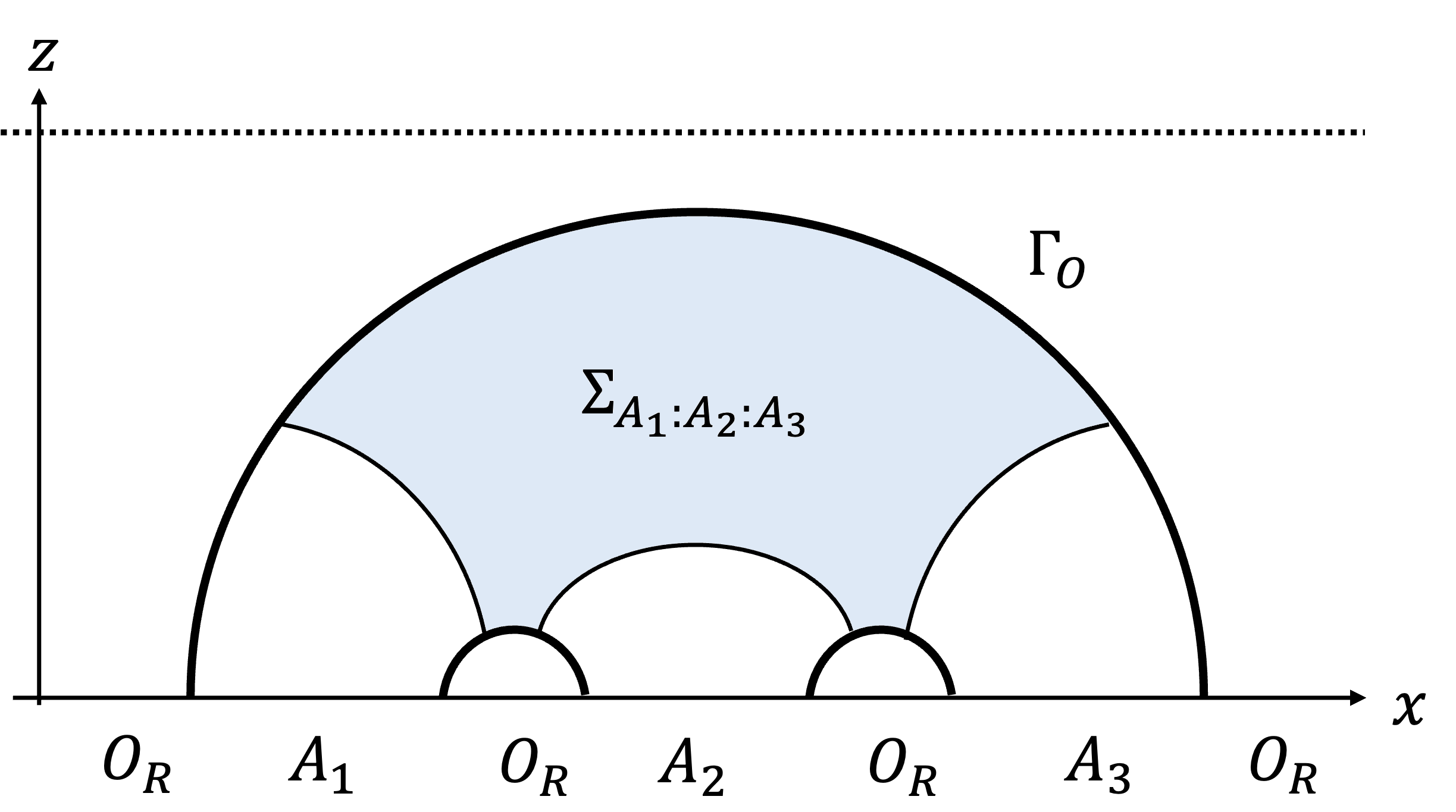}
    \caption{The EWP for the thermal state on non-compact space, $\rho_{A_1A_2A_3A_4}=\rho_{\rm th}$ (on the left) and for a reduced thermal state on non-compact space, $\rho_{A_1A_2A_3}=\Tr_{O_R}(\rho_{\rm th})$ (on the right). The bold curves represent $\Gamma_O$ and the thin ones $\Gamma_{\tilde{A_i}}$. The event horizon is the dotted line.}
    \label{fig:BTZ-planar-mixed}
\end{figure}

{\bf Planar BTZ.}
We consider a $q-$partition of the thermal state of a CFT$_2$ on non-compact space, $\rho_{A_1 \ddd A_q}=\rho_{\rm th}$.
This is realized by taking $q$ connected and adjacent subregions $A_i$ whose union is the full right boundary of a two-sided black brane.
Therefore, $A_1$ and $A_q$ are semi-infinite segments, whereas $A_i$ are segments with finite size $l_i$.
The purifier $O$ is the full left boundary,
$\rho_{\rm th} = \Tr_L(\ket{TFD}\!\bra{TFD})$,
and $\Gamma_O$ is the event horizon.
Consequently, $\Gamma_{\tilde{A_1}}$ and $\Gamma_{\tilde{A_q}}$ are $x$-constant geodesics with one end at the boundary and the other end at the event horizon, or $\Gamma_O$.
For the remaining $A_i$, if $l_i < \beta \log(\sqrt{2}+1)/\pi$, we have $\Gamma_{\tilde{A_i}} = \Gamma_{A_i}$, connecting the left and right endpoints of $A_i$.
If instead $l_i > \beta \log(\sqrt{2}+1)/\pi$, $\Gamma_{\tilde{A_i}}$ is disconnected and each branch is attached to the boundary and the event horizon $\Gamma_O$ \cite{Takayanagi:2017knl}.
Therefore, the configuration of the EWP is the same as for the one-sided planar BTZ with an EOW brane on the horizon, see subsection \ref{subsubsec:one-sided_BTZ}.
The volume is a quantized multiple of $\pi L^2$, with 
\begin{equation}
  0 \leq V_{\{A_i\}} \leq (q-2) \pi L^2 \, ,
\end{equation}
where the largest volume is obtained when $l_i < \beta \log(\sqrt{2}+1)/\pi$ for $2\leq i \leq q-1$, see the left panel of Fig.~\ref{fig:BTZ-planar-mixed}.

Next, we consider a $q-$partition of a reduced thermal state, $\rho_{A_1 \ddd A_q} = \tr_{O_R}(\rho_{\rm th})$.
The purifier $O$ is thus the union of the full left boundary and a (non-connected) subregion on the right boundary, $O=L \cup O_R$.
The largest volume is achieved when $\Gamma_O$ is fully connected, as in the right panel of Fig.~\ref{fig:BTZ-planar-mixed}.
In this case, the external branch of $\Gamma_O$ connects the left end of $A_1$ to the right end of $A_q$.
This configuration is analog to the Poincar\'e AdS$_3$ case and leads to the same result 
$V_{\{ A_i \}} = (2q \times \pi/2 - 2 \pi )L^2 = (q-2) \pi L^2$.
If we increase the separation between two subregions, eventually the EWP gets disconnected and the volume decreases of a $\pi L^2$ or $2\pi L^2$ contribution, as in the Poincar\'e AdS$_3$ case.
This leads to the same bounds and quantization discussed in subsection~\ref{subsec:mixed_AdS3}. 
\\

{\bf Global BTZ.}
Given a subregion $A$ with size $2 \a_A$ of a thermal state of a CFT$_2$ on compact space $S^1$, the three possible configurations for the entanglement wedge cross section $\Gamma_{\tilde{A}}$ are the same as shown in Fig.~\ref{fig:BTZ-global-pure}, with the EOW brane replaced by $\Gamma_O$ \cite{Nguyen:2017yqw}.

Let us consider a three-partition $A,B,C$ of the thermal state of a CFT$_2$ on compact space $S^1$, $\rho_{ABC} = \rho_{\rm th}$.
The purifier $O$ is the full left boundary of a two-sided black hole, $O=L$, so $\Gamma_O$ is the event horizon. 
The union $A \cup B \cup C$ is the full right boundary, $\alpha_A + \alpha_B+\alpha_C = \pi$.
The value of the volume $V_{A:B:C}$ and the allowed configurations of the EWP are the same as for the one-sided global BTZ, see Figs.~\ref{fig:BTZ-global-pure-configurations1} and \ref{fig:BTZ-global-pure-configurations2}. 
We recall that the configuration in which the EWP has the topology of an annulus is never favored. As a consequence, the maximum value of $V_{A:B:C}$ is smaller than the maximum value of $V_{A:B:C:L}$ for the TFD purification, in agreement with the property \eqref{eq:mixed_vs_pure}.
Nonetheless, there are configurations for which $V_{A:B:C}=V_{A:B:C:L}$.
An example is the case $\alpha_C > L \coth^{-1} \left( 2 \coth(\pi r_h/L)-1 \right)/r_h > \pi- L \sinh^{-1}(1)/r_h$ and $\alpha_{A,B}<L \sinh^{-1}(1)/r_h$, a particular instance of configuration I in Fig.~\ref{fig:BTZ-global-pure-configurations2} and case (c) in subection~\ref{subsubsec:two-sided_BTZ}. Here we have $V_{A:B:C:L}=\pi L^2 = V_{A:B:C}$.
The fact that $L \coth^{-1} \left( 2 \coth(\pi r_h/L)-1 \right)/r_h > \pi- \alpha_{\rm crit}$ also implies that when configuration II (or VI) is favored, the EWP for the TFD purification has the annulus topology and higher volume, $V_{A:B:C:L} = 3 \pi L^2 > V_{A:B:C}$.
Therefore, property \eqref{eq:mixed_vs_pure} is always consistent.

\begin{figure}[t]
  \centering
  \includegraphics[scale=0.7]{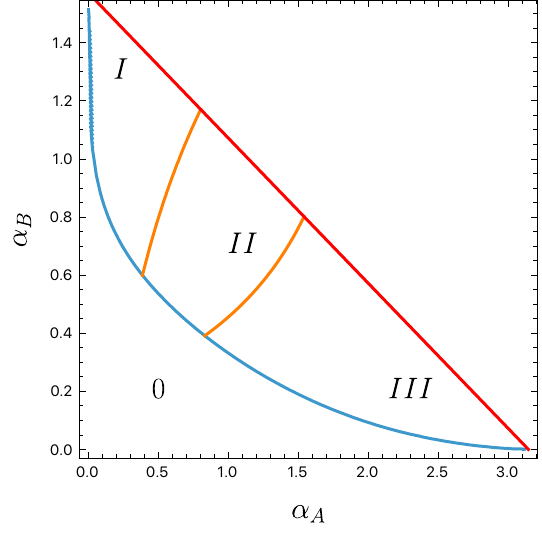}
  \qquad
  \includegraphics[scale=0.7]{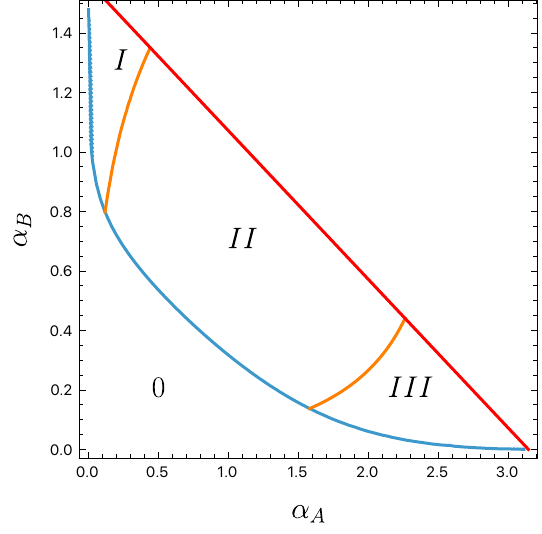} 
  \caption{Volume of the entanglement wedge triangle $V_{A:B:C}$ as a function of the sizes of $A$ and $B$ for $r_h/L=1.1$ (left) and $r_h/L=2$ (right). The blue curve represents the critical angle $\alpha_s =\alpha_{\rm crit}$. Namely, $\Gamma_O$ is fully connected in the region above the blue curve. Below the blue curve, $\Gamma_O$ is fully disconnected, as we have checked numerically by considering all the possible configurations of $\Gamma_O$ including the ones where only $B$ and $C$ are connected \cite{Ben-Ami:2014gsa}. The red curve denotes $\alpha_s=0$.}
  \label{fig:BTZ-global-mixed-V2}
\end{figure}

\begin{figure}[t]
  \centering
  \includegraphics[scale=0.1]{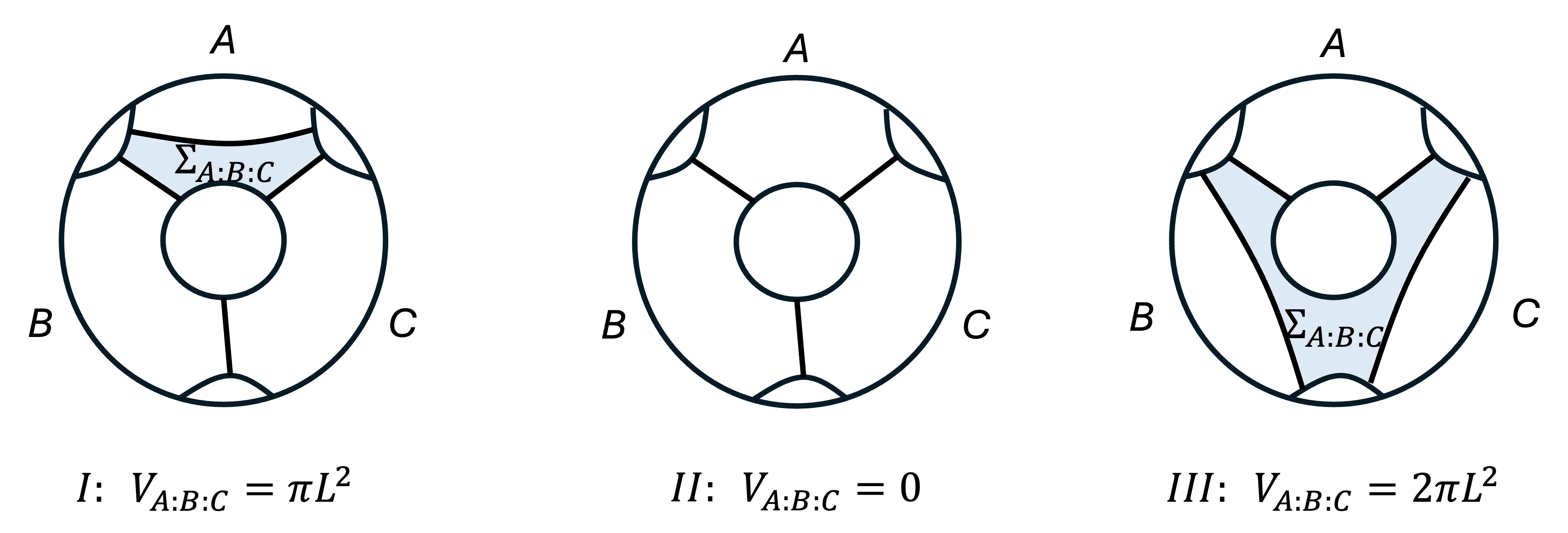} 
  \caption{Allowed configurations of the EWP for the reduced thermal state on compact space, $\rho_{ABC}=\Tr_{O_R}(\rho_{\rm th})$, and related volume.}
  \label{fig:BTZ-global-mixed-V3}
\end{figure}

We now consider a three-partition of a reduced thermal state on $S^1$, $\rho_{ABC}= \Tr_{O_R}(\rho_{\rm th})$, with $A,B,C$ not adjacent.
For simplicity, we take $\alpha_B=\alpha_C$ and we assume that the three subregions have the same separations $2\alpha_s$. 
This assumption fixes $\alpha_s = (\pi-\alpha_A-2 \alpha_B)/3$.
In Figs.~\ref{fig:BTZ-global-mixed-V2} and \ref{fig:BTZ-global-mixed-V3} we display the parameter space $(\alpha_A,\alpha_B)$ with the configurations of the EWP and the corresponding volume $V_{A:B:C}$, computed by the Gauss-Bonnet theorem.
In Appendix~\ref{app:mixed-BTZ} we show that, in the BTZ phase, if the three subregions have the same size, $V_{A:B:C}=0$.
Anoter interesting observation is that the volume of the EWP is not monogamous.
A way to show this is to fix subregion $A$ and increase the value of $\alpha_B$.
Consequently, we obtain two larger subregions $B' \supset B$ and $C' \supset C$.
The three subregions $A,B',C'$ have the same separation $\alpha_s' < \alpha_s$.
From Fig.~\ref{fig:BTZ-global-mixed-V2} it is clear that, with an opportune choice of $\alpha_A$, this process eventually leads to a transition $III \to II$ or $II \to I$. In the first case we have $V_{A:B':C'} < V_{A:B:C}$, while in the second case we get $V_{A:B':C'} > V_{A:B:C}$. 
Therefore, the volume of the EWP can either decrease or increase under increasing the size of the subregions. In the next Section, we will see that the same property is true for a reduction of the vacuum state of a CFT$_d$.

\subsection{Mixed state in AdS$_3$/BCFT$_2$}
\label{subsec:mixed_AdS/BCFT}
\begin{figure}
    \centering
    \includegraphics[width=1\linewidth]{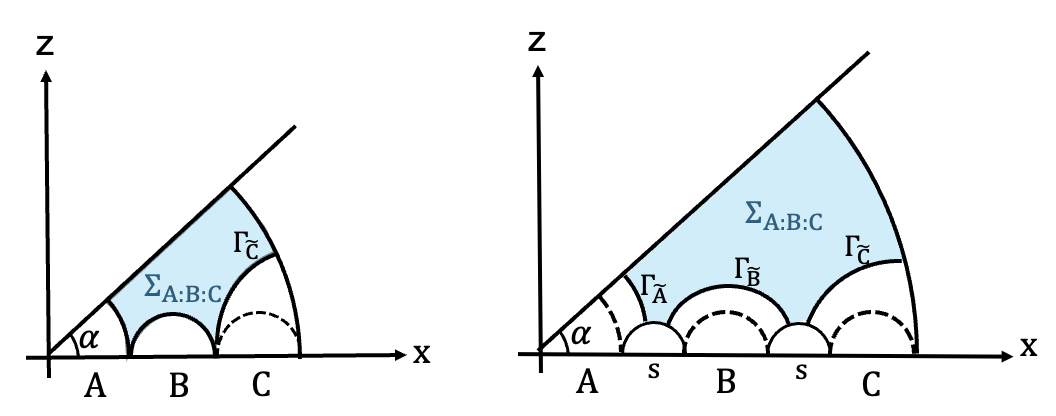}
    \caption{The EWP for a mixed state $\rho_{ABC}$ in a BCFT for adjacent (on the left) and non-adjacent (on the right) subregions, with $q=3$.}
    \label{fig:mixed_AdSBCFT}
\end{figure}

We consider a multi-partition of a mixed state in the AdS$_3$/BCFT$_2$ setup. First, we take adjacent subregions $A_1, \dots, A_q$ with lengths $l_{A_1}, \dots, l_{A_q}$, as drawn in the left panel of Fig.~\ref{fig:mixed_AdSBCFT}.
Using the Gauss-Bonnet theorem, we obtain the volume of the EWP for a fully connected $\Gamma_O$,
\begin{equation}
    V_{\{A_i\}}= L^2 \left( \left(q-\frac{3}{2}\right)\pi -\frac{1}{\tan\alpha}\log\left(\frac{\sum_{i=1}^{q}l_{A_i}}{l_{A_1}}\right) \right) \, .
\end{equation}
Starting from the state $\rho_{A_1 \ddd A_q O}$ on the half line $x>0$, we obtain the mixed state $\rho_{A_1 \ddd A_q}$ by tracing out the Hilbert space of the semi-infinite subregion $O$. Compared to the volume of the EWP for the purification $\rho_{A_1 \ddd A_q O}$ in eq.~\eqref{eq:BCFT_pure_ad}, the volume of the EWP for the $q-$partite mixed state $\rho_{A_1 \ddd A_q}$ is smaller of a contribution $\pi L^2/2$ due to the volume subtended by $\Gamma_C$, $\Gamma_{\ti{C}}$, and $\Gamma_O^{(C)}$. 

Next, we consider non-adjacent subregions with the same length $l$ and the same separation $s$, as in the right panel of Fig.~\ref{fig:mixed_AdSBCFT}. 
The shortest geodesic $\Gamma_{\ti{A}}$ is orthogonal to the branch of $\Gamma_O$ connecting the right end of $A$ with the left end of $B$, and also orthogonal to the EOW brane. By imposing the orthogonality to $\Gamma_O$, the geodesic $\Gamma_{\ti{A}}$ is given by $z^2 +x^2=\ti{l}^2$, with $\ti{l}=\sqrt{l(l+s)}$. With this result in hand, we can promptly compute the volume of the EWP by the Gauss-Bonnet theorem, similarly to Appendix~\ref{subsec:GB-BCFT}.
The volume of the EWP in the fully connected phase is 
\begin{equation}
    V_{\{A_i\}}= L^2 \left( \left(q-\frac{3}{2}\right)\pi -\frac{1}{\tan\alpha}\log\left(\frac{ql+(q-1)s}{\sqrt{l(l+s)}}\right) \right) \, .
\end{equation}
The difference with the value of the maximal volume for the purification $V_{\{A_i\}:O}$ in eq.~\eqref{eq:BCFT_pure_nonad} is $\pi/2 - \ln(\sqrt{l(l+s)}/l)/\tan(\alpha) + (q-1)\pi$, that is the total volume subtended by $\Gamma_{\ti{A_i}}$, $\Gamma_{A_i}$, and $\Gamma_O^{(A_i)}$.
This result also reflects the reduction of degrees of freedom compared to the pure state case.

\section{Entanglement Wedge Polygon in Poincar\'e AdS$_d$ for mixed states}
\label{sec:Higher-d-mixed}

Next, we consider a multi-partition of a reduced vacuum state of a CFT$_d$ on non-compact space, $\rho_{A_1 \ddd A_q}=\Tr_O(\rho_{\rm vac})$.
An interesting question is whether the volume of the EWP is monotonic under a change of size of a boundary subregion. 
After computing the volume $V_{\{A_i\}}$, we address this point.

\subsection{Check of monotonicity}
\label{subsec:mixed-monotonicity}

Let us take three strips $A,B,C$.
For simplicity, we assume that $A$ and $C$ have the same width $l_A$ in the $x-$direction.
We denote by $l_B$ the width of $B$
and by $\ell$ the infinite length of the three strips in the $y_i-$directions.
We assume that the three strips have the same separation $s$.
It is not restrictive to center $B$ at $x=0$, 
so that the configuration is symmetric under $x \to -x$. In Appendix~\ref{app:monotonicity_def1} we provide the details of the computation of $V_{A:B:C}$.

\begin{figure}[H]
\center
\includegraphics[scale=0.55]{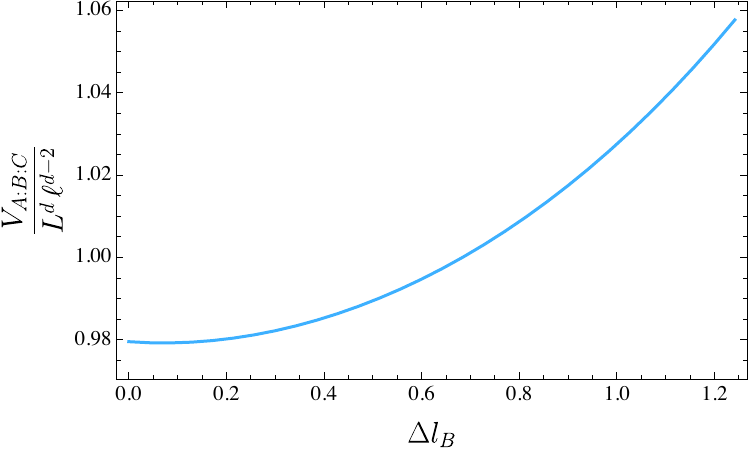}
\qquad
\includegraphics[scale=0.55]{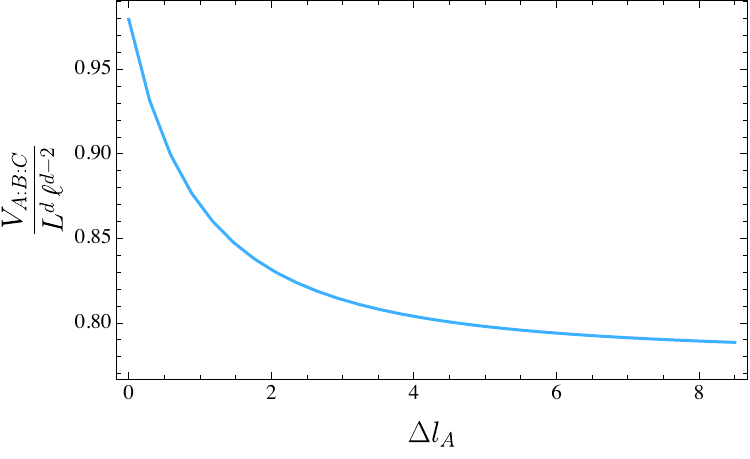}
\caption{On the left: Volume $V_{A:B:C}$ as a function of $\Delta l_B$.
On the right: Volume $V_{A:B:C}$ as a function of $\Delta l_A$ and $\Delta s=0$.
We have set $l_A=l_B=1$, $s=s^{(q=3)}_{\rm crit}(l_A,l_B)$, and $d=3$.}
\label{fig:VABC_Monotonicity_Poincare1}
\end{figure}

In order to study the monotonicity of $V_{A:B:C}$, we first send $l_B \to l_B + \Delta l_B$ and $s \to s - \Delta l_B/2$, keeping $l_A$ fixed. Then, for $\Delta l_B >0$ the resulting $B' \supset B$.
In the left panel of Fig.~\ref{fig:VABC_Monotonicity_Poincare1}, we show the behavior of the volume $V_{A:B':C}$ as a function of $\Delta l_B$ in $d=3$.
From the plot, we find $V_{A:B:C}>V_{A:B':C}$ for small $\Delta l_B$, and $V_{A:B:C}<V_{A:B':C}$ for large $\Delta l_B \lesssim 2s$.
As an additional analysis, we send $l_A \to l_A + \Delta l_A + \Delta s$ and $s \to s - \Delta s$, keeping $l_B$ fixed. For $\Delta l_A, \Delta s \geq 0$ but not both vanishing, we have $A' \supset A$ and $C' \supset C$.
In the right panel of Fig.~\ref{fig:VABC_Monotonicity_Poincare1}, we display $V_{A':B:C'}$ for $\Delta s=0$.
The lowest value of $V_{A':B:C'}$ is attained by $\Delta s \to 0,\Delta l_A \to +\infty$, in which case $\Gamma_{\tilde{A}}$ and $\Gamma_{\tilde{C}}$ are $x-$constant surfaces at $x=\pm (l_B+s)/2$, respectively.
We find $V_{A:B:C} > V_{A':B:C'}$.
We thus conclude that an increasing of the strip width may cause an increasing or a decreasing of $V_{A:B:C}$, according to the transformation performed.
Therefore, the volume of the EWP does not satisfy any monotonicity property.

\subsection{A different definition of the Entanglement Wedge Polygon for mixed states}
\label{subsec:alternative-mixed-state}

\begin{figure}[H]
\center
\includegraphics[scale=0.15]{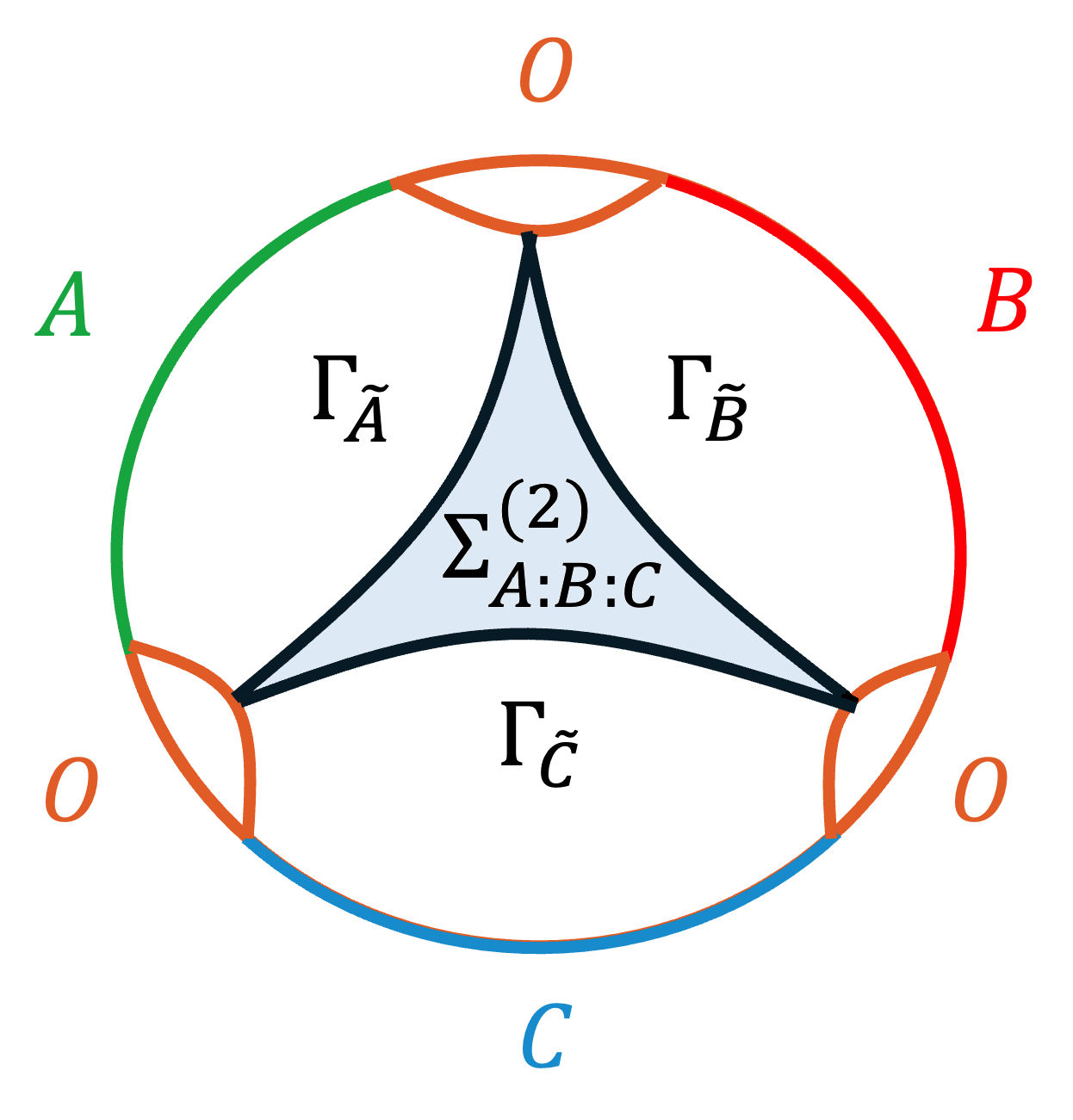}
\caption{An alternative definition of entanglement wedge polygon for mixed states.}
\label{fig:EWmixed2}
\end{figure}

\begin{figure}[H]
\center
\includegraphics[scale=0.068]{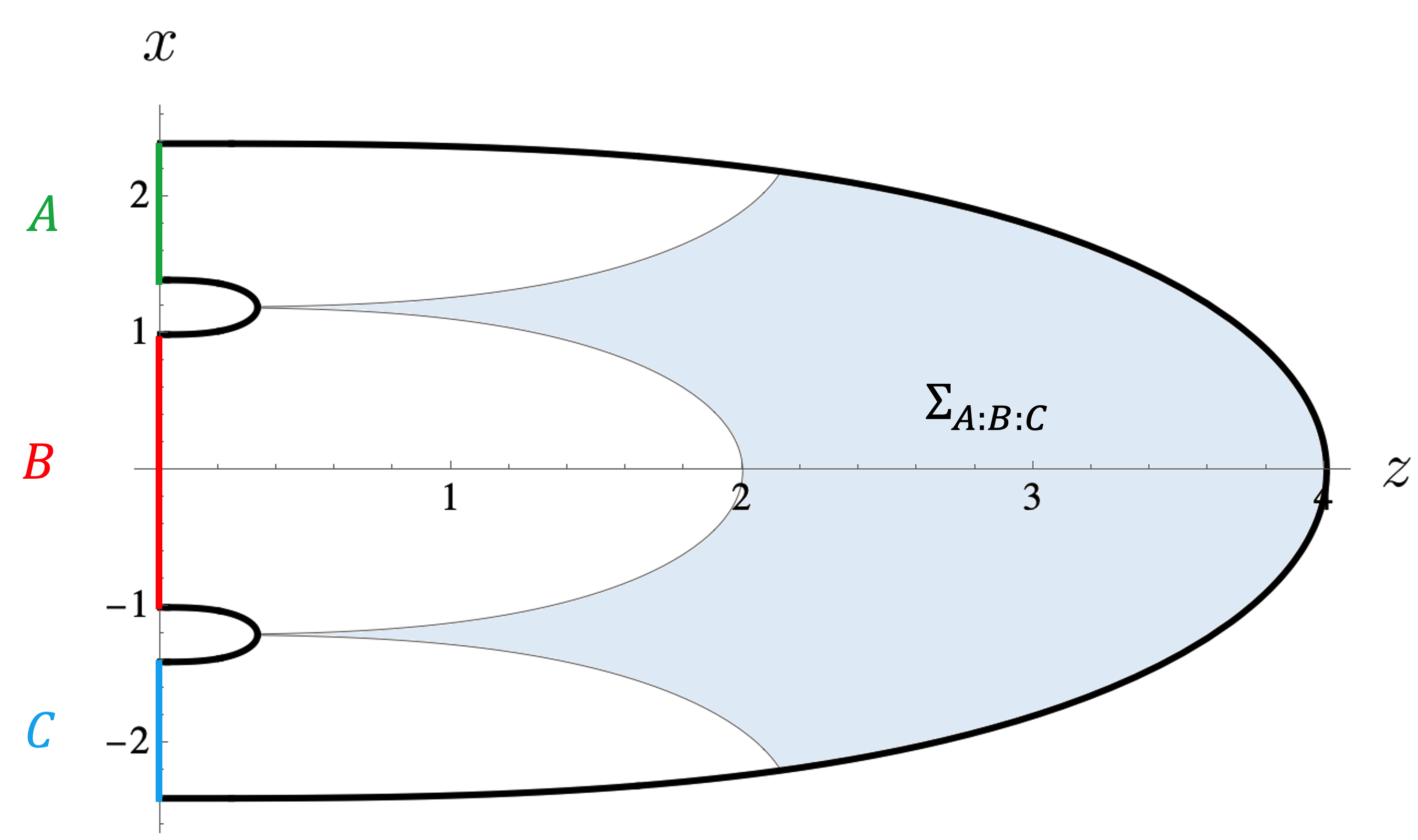}
\qquad
\includegraphics[scale=0.068]{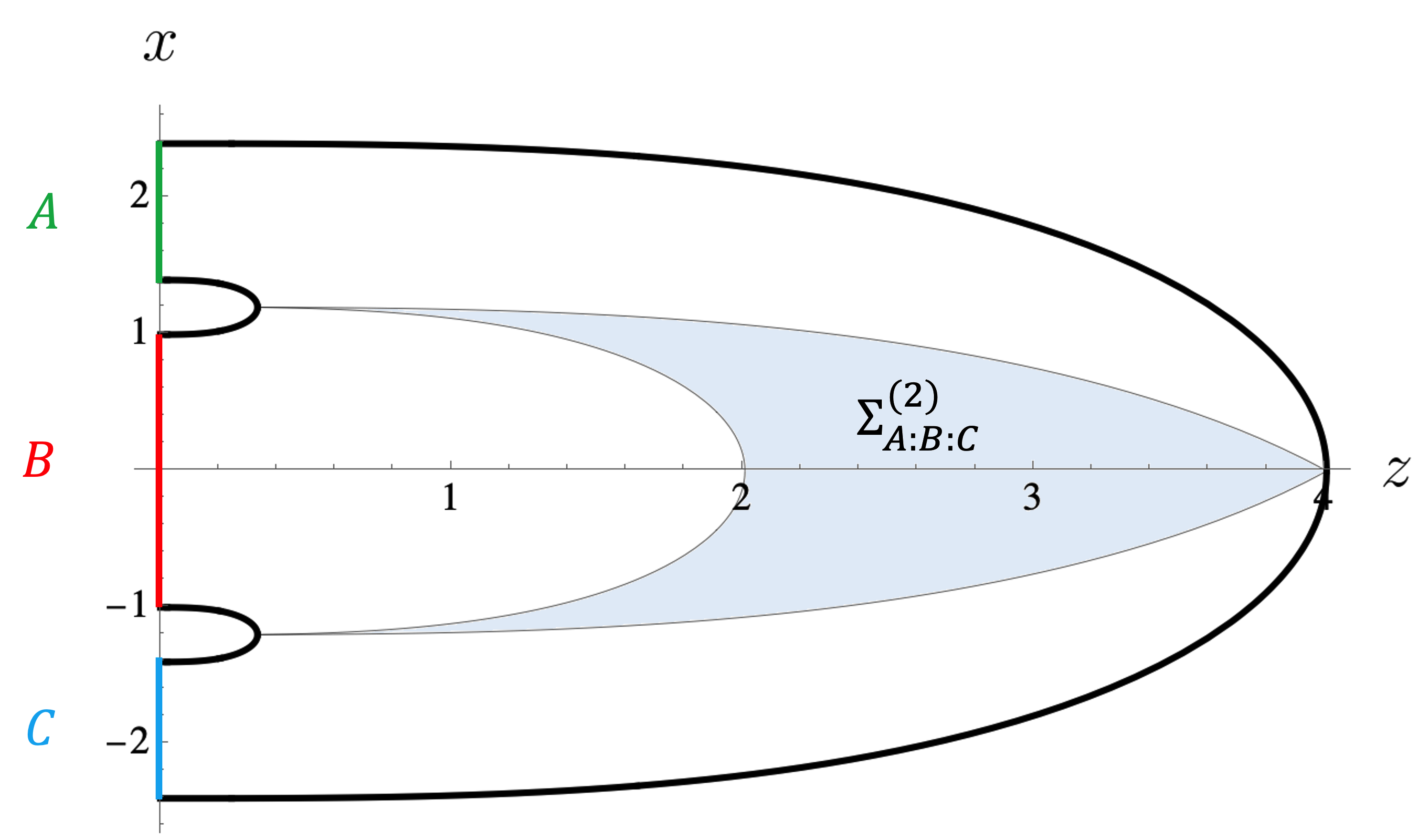}
\caption{The EWP for a mixed state $\rho_{ABC}$ in Poincar\'e AdS$_{d+1}$ for definition~\eqref{eq:def-Sigma-mixed} (on the left) and for definition~\eqref{eq:def-Sigma-mixed-2} (on the right). The bold curves represent $\Gamma_O$ and the dotted ones $\Gamma_{\tilde{A_i}}$.
We have set $l_A=1,l_B=2, s= 0.4,L=1$.}
\label{fig:EWP-Poincared-mixed}
\end{figure}

One may look for a definition for the EWP of a mixed state which satisfies monotonicity.
A possibility is to start from the generalization of the entanglement wedge cross-section for multi-partite states $\rho_{A_1 \ddd A_q}$ introduced in \cite{Bao:2018gck,Umemoto:2018jpc}. In particular, one considers codimension-two surfaces $\Gamma_{\ti{A_i}}$ such that consecutive $\Gamma_{\ti{A_i}}$ share codimension-three boundaries on $\Gamma_O$, and then minimizes the sum of their areas under this constraint. Therefore, we could define the EWP as 
\begin{equation}
\label{eq:def-Sigma-mixed-2}
    \partial \Sigma^{(2)}_{\{A_i\}} = \bigcup_{i=1}^q \Gamma_{\ti{A_i}} \, ,
\end{equation}
with $\Gamma_{\ti{A_i}}$ given by this prescription. Refer to Fig.~\ref{fig:EWmixed2} for a sketch in the $q=3$ case in global AdS$_3$. Note that this definition differs from the one we have consider so far in eq.~\eqref{eq:def-Sigma-mixed}, since our $\Gamma_{\ti{A_i}}$ in general do not meet at $\Gamma_O$.
A comparison between the two different definitions of EWP is shown in Fig.~\ref{fig:EWP-Poincared-mixed} for three strips $A,B,C$ on the boundary of Poincar\'e AdS$_{d+1}$ with the same separation $s$ and widths $l_A=l_C \neq l_B$.

\begin{figure}[H]
\center
\includegraphics[scale=0.55]{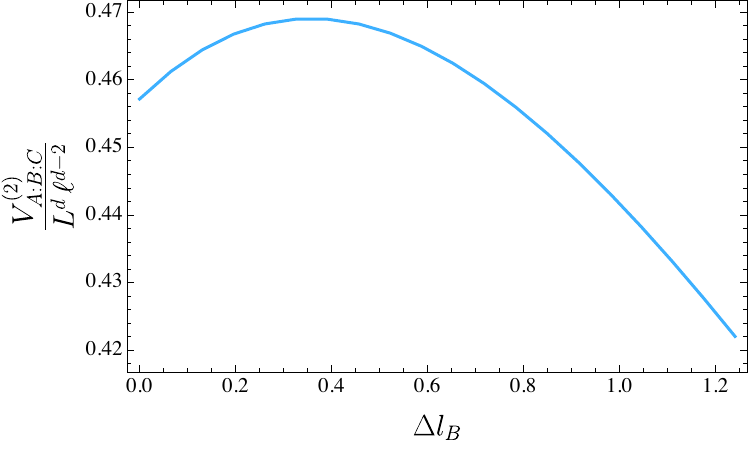}
\qquad
\includegraphics[scale=0.55]{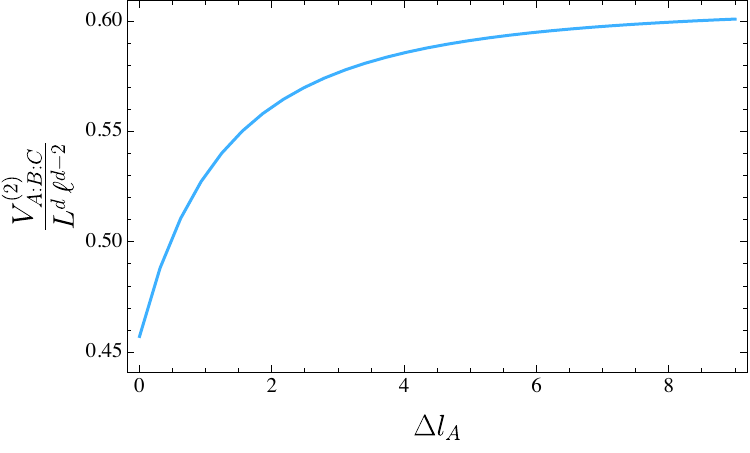}
\caption{On the left: Volume $V_{A:B:C}$ as a function of $\Delta l_B$.
On the right: Volume $V_{A:B:C}$ as a function of $\Delta l_A$ and $\Delta s=0$.
We have set $l_B=l_A=1$, $s=s^{(q=3)}_{\rm crit}(l_A,l_B)$, and $d=3$.}
\label{fig:VABC_Monotonicity_Poincare2}
\end{figure}

In Appendix~\ref{app:monotonicity_def2} we provide details on the computation of the volume $V^{(2)}_{A:B:C}$ for the configuration in the right panel of Fig.~\ref{fig:EWP-Poincared-mixed}.
In order to check the monotonicity property of $V^{(2)}_{A:B:C}$, we consider the same transformations as in the previous subsection.
First, we send $l_B \to l_B + \Delta l_B$ and $s \to s- \Delta l_B/2$, holding $l_A$ fixed. For $\Delta l_B>0$ we thus get $B' \supset B$. As it is clear from the left panel of Fig.~\ref{fig:VABC_Monotonicity_Poincare2}, $V^{(2)}_{A:B:C} < V^{(2)}_{A:B':C}$ for small $\Delta l_B$, while $V^{(2)}_{A:B:C} > V^{(2)}_{A:B':C}$ for large $\Delta l_B \lesssim 2s$.
Second, we take $l_A \to l_A + \Delta l_A + \Delta s$ and $s \to s-\Delta s$, with $l_B$ unvaried. For $\Delta l_A, \Delta s \geq 0$ and not both vanishing, we obtain $A' \supset A$ and $C' \supset C$.
In the right panel of Fig.~\ref{fig:VABC_Monotonicity_Poincare2} we display the volume $V^{(2)}_{A':B:C'}$ as a function of $\Delta l_A$ for $\Delta s=0$. We find $V^{(2)}_{A:B:C}<V^{(2)}_{A':B:C'}$.
We thus conclude that the new definition of the EWP for mixed state does not provide a monotonous volume as well.

\section{Conclusions and Discussions}
\label{sec:conclusions}

In holography an important line of work has been the relation of geometric quantities in holographic spacetimes to information measures in the boundary CFT. In this article we have continued this program by considering a particular codimension-1 volume of the bulk spacetime which we have called the entanglement wedge polygon (EWP). While initially defined for pure states we have also provided a suitable generalization for case of mixed states. In particular our focus was on its precise definition in the bulk spacetime along with its basic properties and several explicit examples in varying dimensions, leaving its precise meaning in quantum information theory as a future problem.

In three dimensions when the timeslice has constant negative curvature, as in the case of vacuum AdS$_3$, the EWP is topological as a direct result of the Gauss-Bonnet theorem. This in turn implies that the volume of the EWP becomes quantized and is completely determined just based on the topology of the time slice and the number of boundary regions. 
This property remains true for highly excited states, that compared to the vacuum state display a richer variety of configurations of the EWP. In this case, analogies between the volume of the EWP and multi-entropy, a candidate as a measure of multi-partite entanglement, have been found \cite{Fujiki:2026qdt}.  
It would be very interesting to better understand these features, especially quantization, via an independent calculation in the boundary 2d CFT. We also observed that in the presence of the end-of-the-world-brane dual to BCFTS, this topological property is no longer correct even in three dimensions. 

In higher dimensions, the volume of the EWP loses the topological and quantized nature of the three-dimensional case, and depends on the size of the partitions. We have studied this dependence for various holographic states on non-compact space, such as the vacuum state, the thermofield double state, a BCFT state, and for a large $N$ confining gauge theory dual to an AdS soliton geometry.
In general, we have found that for high temperature the volume of the EWP tends to decrease, in agreement with the expectation that the high temperature destroys the (multi-partite) correlations.
Another natural expectation confirmed by our analysis is that the volume of the EWP vanishes in the confinement phase of a large $N$ gauge theory, as the multi-partite entanglement is expected to do.

We have then initiated an explicit analysis of the volume of the EWP for time-dependent scenarios, such as the time-evolution of a thermofield double state and two kinds of global quenches, dual to a one-sided AdS black hole with the insertion of an end-of-the-world brane behind the event horizon and the AdS Vaidya spacetime. 
The time-evolution of the EWP has been estimated to manifest various transitions with discontinuous changes in its growth rate. At late times, the volume of the EWP has been found to either saturate to a constant value or grow linearly, according to the choice of boundary partitions. 
We leave a more detailed analysis for future work.

More generally for holographic spacetimes of all dimensions we were able to derive a number of key properties which the EWP satisfies. For pure states it was found that:

{\bf Property $P1$.}
The volume of the EWP vanishes for bi-partitions of pure states:
\begin{equation}
    V_{A:\ov{A}} =  0 \, .
\end{equation}

{\bf Property $P2$.}
The volume of the EWP increases for finer grained partitions of pure states:
\begin{equation}
    V_{(AB):C:O} \leq V_{A:B:C:O} \, ,
    \qquad
    O= \ov{ABC} \, .
\end{equation}

In the case of mixed states it was found that the EWP has the following properties:

{\bf Property $M1$.}
The volume of the EWP vanishes for bi-partitions of a mixed state:
\begin{equation}
    V_{A:B} = 0 \, .
\end{equation}

{\bf Property $M2$.}
The volume of the EWP increases for finer grained partitions of mixed states:
\begin{equation}
    V_{(AB):C:D} \leq V_{A:B:C:D} \, .
\end{equation}

For the conditions under which the equality in properties $P2$ and $M2$ holds, see the comments around eq.~\eqref{eq:inequality-pure} and \eqref{eq:inequality-mixed}, respectively.
The properties $P1,M1$ indicate that the EWP is vanishing in the case of bi-partitions giving some evidence that the EWP captures only the
multi-partite entanglement among the boundary subregions while neglecting the bi-partite contributions. It is also important that our mixed state generalization preserves the property that the volumes for AdS$_3$ setups become topological.

Compared to the volume of the EWP for a pure state, we have also proven that the volume of the EWP for a mixed state obtained by partial trace is lower or equal. We have confirmed the proof by some concrete examples. 

While tensor network constructions further suggest that the EWP may be associated with some quantification of multipartite entanglement there is still much work to be done to more fully establish this claim. In particular it would be interesting to provide an independent boundary CFT calculation in terms of density matrices in order to more fully elucidate a precise information measure and more concretely determine its potential connection. Similarly one would like to better understand its quantum information properties such as inequalities and relations to other measures. Unfortunately, as shown in subsection~\ref{subsec:mixed-monotonicity} the EWP is in general not monotonic, implying that it is not a correlation measure. Still it is possible that the EWP can provide another explicit example of a holographic connection between boundary measures of information and geometric bulk quantities.


\section*{Acknowledgements}
We are grateful to Pawel Caputa, Kenya Tasuki and Zixia Wei for useful discussions. This work is supported by MEXT KAKENHI Grant-in-Aid for Transformative Research Areas (A) through the ``Extreme Universe'' collaboration: Grant Number 21H05187. TT is also supported by Inamori Research Institute for Science and by JSPS Grant-in-Aid for Scientific Research (B) No.~25K01000.
The work of NZ is also supported by a JSPS Postdoctoral Fellowship for Research in Japan No.~25KF0224. KF is also supported by Grant-in-Aid for JSPS Fellows No.~26KJ1554.

\appendix
\section{Gauss-Bonnet theorem}
\label{sec:GB-details}
We collect here our conventions for the application of the Gauss-Bonnet theorem, mainly taken from \cite{spivak1999comprehensive3}. First, we consider a two-dimensional submanifold $\Sigma$ with induced metric $h_{\alpha \beta}$ and a curve $\gamma(s)$ on $\Sigma$, with $s$ a parameter.
We denote the coordinates on $\Sigma$ by $(u,v)$.
Therefore, the curve $\gamma(s)$ is described by $x^{\alpha} = (u(s),v(s))$ and its tangent vector $\dot{\gamma}(s)$ by $T^{\alpha} = (\dot{u}(s),\dot{v}(s))$.
We assume that the tangent vector is normalized, 
$h_{\alpha \beta} T^{\alpha} T^{\beta}=1$.
We then define the unit normal to the curve $\gamma$ by $h_{\alpha \beta} T^{\alpha} n^{\beta} =0$ and $h_{\alpha \beta} n^{\alpha} n^{\beta} =1$. 
In our applications, $\gamma$ belongs to the boundary $\partial \Sigma$ of the submanifold, so we can fix the orientation of the unit normal such that $n^{\alpha}$ points towards the interior of the surface $\Sigma$ whenever $s$ runs counterclockwise on $\partial \Sigma$.
With this choice, the geodesic curvature of $\gamma$ is defined as
\begin{equation}
\label{eq:def-kg}
    k_g(s) = h_{\alpha \beta} n^{\alpha} T^{\d} \nabla_{\d} T^{\beta} \, .
\end{equation}
Note that the sign of $k_g(s)$ is completely fixed by the orientation of the curve, and so by the orientation of $n^{\alpha}$.
Clearly, when $\gamma(s)$ is a geodesic, the geodesic equations $T^{\d} \nabla_{\d} T^{\beta} =0$ are satisfied and the geodesic curvature vanishes.

Next, let us assume that the boundary $\partial \Sigma$ is piecewise smooth and there is a simple closed curve $\gamma(s)$, with $s \in \left[ a,b\right]$, that is smooth in every interval $t_{i-1} \leq s \leq t_{i}$ for some partition $a=t_0 < \dots < t_{n+1}=b$. Under the assumption $\dot{\gamma}(b^-)=\dot{\gamma}(a^+)$, the boundary $\partial \Sigma$ has vertices in $s=t_1, \dots, t_n$, where $\gamma(s)$ is continuous and its left and right derivatives are well-defined.
By working with a positively oriented $\dot{\gamma}(s)$, for every vertex we define the tangent vectors $V_1 = \dot{\gamma}(t_i^+)$ and $V_2 = -\dot{\gamma}(t_i^-)$.
The internal angle $\theta_i$ at the $i$-th vertex is then the oriented angle from $V_1$ to $V_2$, with $\theta_i \in \left[ 0, 2\pi \right]$.
If $V_1$ and $V_2$ points in the same direction, we can instead define $\tilde{V}_1$ as the tangent to the geodesic from $\gamma(t_i)$ to $\gamma(t_i + \varepsilon)$ and $\tilde{V}_2$ as the tangent to the geodesic from $\gamma(t_i)$ to $\gamma(t_i - \varepsilon)$, with $0<\varepsilon<<1$. Then, $\theta_i=0$ if the angle from $\tilde{V}_1$ to $\tilde{V}_2$ is positively oriented and $\theta_i=2\pi$ otherwise.
Finally, we define the discontinuity at the $i$-th vertex as $\d_i = \pi - \theta_i$, with $\d_i \in \left[ -\pi, \pi \right]$.

The Gauss-Bonnet theorem for a surface $\Sigma$ with a piecewise smooth boundary and $n$ vertices can be expressed as
\begin{align}
    \int_{\Sigma} K dA &= 
    - \int_{\partial \Sigma} k_g \, ds - \sum_{i=1}^n \d_i + 2\pi \chi(\Sigma) \\
    &= - \int_{\partial \Sigma} k_g \, ds + \sum_{i=1}^n \theta_i + \pi (2\chi(\Sigma)-n) \, ,
    \label{eq:GB-2}
\end{align}
where $K$ is the Gaussian curvature of $\Sigma$ and $\chi(\Sigma)$ its Euler characteristic.

\subsection{Gauss-Bonnet theorem for an infinite segment in Poincar\'e AdS$_3$}
\label{subsec:GB-infinite}
We partition the boundary of Poincar\'e AdS$_3$ into three segments $A,B,O$. Let us assume that $A,O$ are semi-infinite segments and that $B$ has a finite length $l$.
Then, both $\Gamma_A$ and $\Gamma_O$ are $x-$constant geodesics.
Therefore, $\Sigma_{A:B:O}$ ends at $z=+\infty$.
Since $z-$constant curves are not geodesics, we need to evaluate the geodesic curvature $k_g$ for the curve $z=+\infty$.
We start by the induced metric on $\Sigma_{A:B:O}$, 
\begin{equation}
    ds^2 = \frac{L^2}{z^2} (dz^2 + dx^2) \, .
\end{equation}
The tangent vector and unit normal at $z=+\infty$ are
\begin{equation}
    T_{\infty}^{\a} = \left( 0,-\frac{z}{L} \right) \, , 
    \qquad
    n_{\infty}^{\a} = \left( -\frac{z}{L},0 \right) \, ,
\end{equation}
respectively.
The acceleration vector is
\begin{equation}
    T_{\infty}^{\d} \nabla_{\d} T_{\infty}^{\b} = 
    \left( - \frac{1}{L},0 \right) \, .
\end{equation}
From this, the geodesic curvature \eqref{eq:def-kg} is $k_g = z^{-1} |_{z=+\infty} =0$.
Consequently, the only boundary contribution from $z=+\infty$ comes from the angles at its intersection with $\Gamma_A$ and $\Gamma_O$.  
Here we have
\begin{equation}
    V_1^\a = T_\infty^\a \, ,
    \qquad
    V_2^\a = -\dot{\Gamma}_O = \left( - \frac{z}{L}, 0 \right) \, ,
\end{equation}
which are orthogonal to each others, giving $\theta_\infty = \pi/2$.
Therefore, from eq.~\eqref{eq:GB-2} we get
\begin{equation}
    V_{A:B:O} = - \left( 2 \times \frac{\pi}{2} + \pi (2 \times 1 -4) \right) L^2 = \pi L^2 \, ,
\end{equation}
where we have used $\chi=1$.

\subsection{Gauss-Bonnet theorem in AdS/BCFT}
\label{subsec:GB-BCFT}
The induced metric on $\Sigma_{A:B:O}$ is 
\begin{equation}
    ds^2 = \frac{L^2}{z^2} (dz^2 + dx^2) \, .
\end{equation}
The EOW brane is located at $z=x \tan(\alpha)$. 
The value of the radial coordinate $z$ at the intersection of the brane with a geodesic $x^2 + z^2 = l^2$ is $z=l \sin(\alpha)$.
The EOW brane has tangent vector and unit normal
\begin{equation}
    T^{\alpha}_{\rm EOW} = \left( -\frac{z}{L} \sin(\alpha), -\frac{z}{L} \cos(\alpha) \right) \, , 
    \qquad
    n^{\alpha}_{\rm EOW} = \left( -\frac{z}{L} \cos(\alpha), \frac{z}{L} \sin(\alpha) \right) \, .
\end{equation}
The acceleration vector reads
\begin{equation}
    T^{\d}_{\rm EOW} \nabla_{\d} T^{\beta}_{\rm EOW} = \left( \frac{z}{L^2} \cos^2(\alpha), -\frac{z}{L^2} \sin(\alpha) \cos(\alpha) \right) \, ,
\end{equation}
from which the geodesic curvature \eqref{eq:def-kg} is $k_g = - \cos(\alpha)/L$.
Therefore, the boundary contribution from the EOW brane gives
\begin{equation}
    \int_{EOW} k_g \, ds = 
    \int_{l_A \sin(\alpha)}^{(l_A+l_B) \sin(\alpha)} \frac{dz \, L \, k_g}{z} \sqrt{\frac{1}{\tan^2(\alpha)}+1} = -\frac{1}{\tan(\alpha)} \log \frac{l_A +l_B}{l_A} \, .
\end{equation}
At the intersection between $\Gamma_O$ and the EOW we have
\begin{equation}
    V_1^{\alpha} = \left. T^{\alpha}_{\rm EOW} \right|_{z=(l_A+l_B) \sin(\alpha)} \, ,
    \qquad
    V_2^{\alpha} = - \dot{\Gamma}_O = \frac{l_A+l_B}{L} \left( - \sin(\alpha) \cos(\alpha), \sin^2(\alpha) \right) \, ,
\end{equation}
which are orthogonal to each others, leading to $\theta=\pi/2$. Note that the orthogonality holds independently of the length of the anchoring subregion, implying that $\Gamma_A$ is also orthogonal to the EOW brane.

\subsection{Gauss-Bonnet theorem in BTZ with an EOW brane}
\label{subsec:GB-EOW}
Let us consider a tri-partition $A,B,O$ of the excited state $\ket{\Psi}_R$ of a CFT$_2$ dual to a one-sided BTZ with an EOW brane. We work at $t=0$ and take $A$ and $O$ as semi-infinite segments.
It is convenient to express the spacetime metric as\footnote{The metric is brought in the form \eqref{eq:BTZ-metric} by the change of coordinates $\tanh(\rho)=\sqrt{r^2-1}/r$ and $L=1$. The value of the temperature $T=1/(2\pi)$ is then obtained by fixing $z_h=1$.}
\begin{equation}
    ds^2 = - \sinh^2 (\rho) dt^2 + d\rho^2 + \cosh^2 (\rho) d\varphi^2 \, .
\end{equation}
At $t=0$, the EOW brane is located at the horizon $\rho=0$.
The induced metric on $\Sigma_{A:B:O}$ at $t=0$ is
\begin{equation}
    ds^2 = d\rho^2 + \cosh^2(\rho) d\varphi^2 \, .
\end{equation}
The tangent vector and the unit normal at the EOW brane are
\begin{equation}
    T^\alpha_{\rm EOW} = \left( 0, 1 \right) \, ,
    \qquad
    n^{\alpha}_{\rm EOW} = \left( 1,0 \right) \, .
\end{equation}
Since the acceleration vector vanishes,
\begin{equation}
    T^{\delta}_{\rm EOW} \nabla_{\delta} T^{\beta}_{\rm EOW} = \left. \left( -\cosh(\rho) \sinh(\rho), 0 \right) \right|_{\rho=0} \, ,
\end{equation}
the geodesic curvature on the EOW brane is also vanishing.
Obviously, $\Gamma_O$, a $\varphi-$constant geodesic, and the EOW brane, at constant $\rho$, are orthogonal, so $\theta=\pi/2$.
Putting all together, from eq.~\eqref{eq:GB-2} we find
\begin{equation}
    V_{A:B:O} = -\left( 2 \times \frac{\pi}{2} + \pi (2 \times 1 - 4) \right) L^2 = \pi L^2 \, .
\end{equation}

\section{Details on mixed state in BTZ}
\label{app:mixed-BTZ}
Let us consider three subregions $A,B,C$ with the same size $2 \alpha$ and equal separation $2 \alpha_s$.
In this case, $O$ is the union of the full left boundary and part of the right boundary.
The surface $\Gamma_O$ is fully disconnected for $\alpha_s >\alpha_{s, \rm{crit}}$ and fully connected for $\alpha_s <\alpha_{s, \rm{crit}}$, in which case it also includes the event horizon.
It is straightforward to compute the critical value\footnote{The length of the minimal geodesic anchored at a subregion $A$ in compact BTZ is $\mathcal{A}(\Gamma_A)=2 L \log \left( \frac{2 L^2}{r_h \varepsilon} \sinh{\left( \frac{r_h}{L} \a_A \right)} \right)$, where $\varepsilon$ is a UV cutoff.
The critical opening angle is obtained by solving
\begin{equation}
    6L \log \left( \frac{2L^2}{r_h \varepsilon} \sinh \left( \frac{r_h}{L} \alpha \right) \right) = 6L \log \left( \frac{2L^2}{r_h \varepsilon} \sinh \left( \frac{r_h}{L} \alpha_{s,\rm{crit}} \right) \right) + 2 \pi r_h \, .
\end{equation}
}
\begin{equation}
    \alpha_{s, \rm{crit}} = \frac{L}{2 r_h} \log \left( \frac{2}{1+ e^{-2 \pi r_h/(3L)}} \right)
    = \frac{\pi}{6} -\frac{L}{2 r_h} \log \left( \cosh \left( \frac{\pi r_h}{3L} \right) \right) \, .
\end{equation}
When $\alpha_s > \alpha_{s,\rm{crit}}$, the volume of the entanglement wedge triangle vanishes.
When $\alpha_s < \alpha_{s,\rm{crit}}$, there are two possible configurations for $\Gamma_{\tilde{A}}$. The first one, which we denote by $\Gamma_{\tilde{A}}^{(1)}$, is a connected geodesic anchored at both ends to $\Gamma(2\alpha_s)$. The second one, denoted by $\Gamma_{\tilde{A}}^{(2)}$, is the union of two disconnected geodesics anchored at one end to $\Gamma(2\alpha_s)$ and at the other end to the black hole horizon.
Let us compare their length for given $\alpha$ and $\alpha_s$.
The length for the first configuration is
\begin{align}
    \mathcal{A}\left( \Gamma_{\tilde{A}}^{(1)} \right) &= 
    2 \int_0^{\theta_{0}} \frac{r^2(\theta)}{r_*} d\theta
    = 2 \frac{r_h^2}{r_*} \int_0^{\theta_0} \frac{d\theta}{1-\frac{\cosh^2(r_h \theta/L)}{\cosh^2(r_h \alpha_{\tilde{A}}/L)}} \\
    &= 2 L \arctanh \left( \coth \left( \frac{r_h \alpha_{\tilde{A}}}{L} \right) \tanh \left( \frac{r_h \theta_0}{L} \right) \right) \, ,
    \label{eq:area-BTZ-mixed-1}
\end{align}
where we have defined by $\theta=\theta_0$ the intersection of the geodesic with $\Gamma(2\alpha_s)$ and by $\alpha_{\tilde{A}}$ half size of the boundary subregion $\tilde{A} \supset A$ to which $\Gamma_{\tilde{A}}^{(1)}$ is anchored. For simplicity, we have assumed that $A$ is centered at $\theta=0$. We have also used 
$r_* = r_h \coth(r_h \alpha_{\tilde{A}}/L)$.
The value of $\theta_0$ can be obtained by solving 
$r_{\tilde{A}}(\theta_0) = r_{\alpha_s}(\theta_0)$, namely
\begin{equation}
    r_h \left( 1-\frac{\cosh^2(r_h \theta_0/L)}{\cosh^2(r_h \alpha_{\tilde{A}}/L)} \right)^{-1/2} =
    r_h \left( 1-\frac{\cosh^2(r_h (\theta_0 -\alpha -\alpha_s)/L)}{\cosh^2(r_h \alpha_s/L)} \right)^{-1/2} \, ,
\end{equation}
which gives 
\begin{equation}
    \theta_0 = \frac{\pi}{6} + \arctanh \left( \coth\left( \frac{\pi r_h}{6 L} \right) \frac{\cosh(r_h \alpha_{\tilde{A}}/L) - \cosh(r_h \alpha_s/L)}{\cosh(r_h \alpha_{\tilde{A}}/L) + \cosh(r_h \alpha_s/L)} \right) \, .
\end{equation}
Plugging the solution into eq.~\eqref{eq:area-BTZ-mixed-1} and minimizing with respect to $\alpha_{\tilde{A}}$, we find
\begin{equation}
    \alpha_{\tilde{A}}^{(\rm{min})} = \frac{L}{r_h} \arccosh \left( \frac{\cosh(\pi r_h/(3L))}{\cosh(r_h \alpha_s/L)} \right) \, .
\end{equation}
As a sanity check, note that when $\alpha_s \to 0$, $\alpha_{\tilde{A}}^{(\rm{min})} \to \pi/3$.
Given this result, for the shortest geodesic $\Gamma_{\tilde{A}}^{(1)}$ we get
\begin{equation}
    \theta_0^{(\rm{min})} = \frac{\pi}{6} + \arctanh \left( \coth\left( \frac{\pi r_h}{6 L} \right) \frac{\cosh(r_h \pi/(3L)) - \cosh^2(r_h \alpha_s/L)}{\cosh(r_h \pi/(3L)) + \cosh^2(r_h \alpha_s/L)} \right) \, ,
\end{equation}
that leads to
\begin{equation}
    \mathcal{A}^{(\rm{min})}\left( \Gamma_{\tilde{A}}^{(1)} \right) =
    L \log \left( \frac{\sinh\left( \frac{\pi r_h}{3L} \right) + \sqrt{\cosh^2 \left( \frac{\pi r_h}{3L} \right) -\cosh^2 \left( \frac{\alpha_s r_h}{L} \right)}}{\sinh\left( \frac{\pi r_h}{3L} \right) -\sqrt{\cosh^2 \left( \frac{\pi r_h}{3L} \right) -\cosh^2 \left( \frac{\alpha_s r_h}{L} \right)}} \right) \, .
\end{equation}
Let us now focus on the length of $\Gamma_{\tilde{A}}^{(2)}$. For a radial geodesic anchored at the horizon and at $r=r_0$ on $\Gamma(2\alpha_s)$, we have
\begin{equation}
    \mathcal{A}\left( \Gamma_{\tilde{A}}^{(2)} \right) 
    = 2L \int_{r_h}^{r_0} \frac{dr}{\sqrt{r^2 -r_h^2}} 
    = 2L \log \left( \frac{r_0 + \sqrt{r_0^2 -r_h^2}}{r_h} \right) \, .
\end{equation}
Clearly, the length is minimized when $r_0$ is minimal, which happens at the turning point of $\Gamma(2\alpha_s)$, namely $r_0^{(\rm{min})} = r_h \coth(r_h \alpha_s/L)$.
Therefore, the length of the minimal geodesic is
\begin{equation}
    \mathcal{A}^{(\rm{min})}\left( \Gamma_{\tilde{A}}^{(2)} \right) 
    = 2L \log \left( \coth \left( \frac{r_h \alpha_s}{2L} \right) \right) \, .
\end{equation}
Let us compare the lengths of the two configurations.
In the low temperature limit $r_h/L <<1$, we have 
\begin{align}
    \mathcal{A}^{(\rm{min})}\left( \Gamma_{\tilde{A}}^{(1)} \right) &= L \log \left( \frac{\pi+\sqrt{\pi^2 -9 \alpha_s^2}}{\pi-\sqrt{\pi^2 -9 \alpha_s^2}} \right) + \mathcal{O}\left( \left(\frac{r_h}{L} \right)^2 \right) \, , \\
    \mathcal{A}^{(\rm{min})}\left( \Gamma_{\tilde{A}}^{(2)} \right) &= 2 L \log \left( \frac{2L}{r_h \alpha_s} \right) + \mathcal{O}\left( \left(\frac{r_h}{L} \right)^2 \right) \, ,
\end{align}
so $\Gamma_{\tilde{A}}^{(1)}$ is shorter for any $\alpha_s$, as expected.
In this regime, the Gauss-Bonnet theorem gives $V_{A:B:C}= (6 \times \pi/2-0) L^2 = 3 \pi L^2$.
In the high temperature limit $r_h/L >>1$, we have
\begin{align}
    \mathcal{A}^{(\rm{min})}\left( \Gamma_{\tilde{A}}^{(1)} \right) &= 2 \left( \frac{\pi}{3}-\alpha_s \right) \frac{r_h}{L} + \log(4) + \mathcal{O}\left( e^{-2 \left( \frac{\pi}{3}-\alpha_s \right)r_h/L} \right) \, , \\
    \mathcal{A}^{(\rm{min})}\left( \Gamma_{\tilde{A}}^{(2)} \right) &= 4 e^{-\alpha_s r_h/L} + \mathcal{O}\left( e^{-3\alpha_s r_h/L} \right) \, ,
\end{align}
so $\Gamma_{\tilde{A}}^{(2)}$ is shorter for any $\alpha_s$.
Consequently, consecutive branches of the entanglement wedge cross sections for $A,B,$ and $C$ overlap and $V_{A:B:C}=0$.
In the general case, the two configurations have the same length for
\begin{equation}
    \bar{\alpha}_s = - \frac{\pi}{3} + \frac{L}{r_h} \log \left( \frac{1}{2} \left( -1 + e^{2 \pi r_h/(3L)} + \sqrt{1-6 e^{2 \pi r_h/(3L)} + e^{4 \pi r_h/(3L)}} \right) \right) \, .
\end{equation}
Note that $\bar{\alpha}_s$ is defined only for $r_h \geq \frac{3L}{\pi} \log(1+\sqrt{2}) \approx 0.84 \, L$, when the term under the square root is positive. Also, the solution grows monotonically with $r_h$ from $\bar{\alpha}_s=0$ to the asymptotic value $\bar{\alpha}_s = \pi/3$ for $r_h>>L$.
Moreover, we have $\bar{\alpha}_s > \alpha_{s, \rm{crit}}$ for $r_h > a_0 \, L$ with $a_0 \approx 0.87$.
Therefore, in the BTZ phase $r_h>L$ ($\beta < 2\pi L$, high temperature), whenever $\Gamma_O$ is fully connected $\Gamma_{\tilde{A}}^{(2)}$ is always shorter and $V_{A:B:C} =0$.

\section{Details on mixed state in Poincar\'e AdS$_{d+1}$}
\label{app:monotonicity}
In this Appendix we compute the volume of the EWP for a tri-partition of a reduced vacuum state of a CFT$_d$ on non-compact space, $\rho_{ABC}= \Tr_O(\rho_{\rm vac})$.
For simplicity, we assume that $A,C$ have the same width $l_A$ and that the three strips $A,B,C$ have the same separation $s$ in the $x-$direction.
We employ both the definitions of EWP for a mixed state in eqs.~\eqref{eq:def-Sigma-mixed} and \eqref{eq:def-Sigma-mixed-2}, that we refer to as definition $1$ and definition $2$, respectively.

\subsection{Computation for definition $1$}
\label{app:monotonicity_def1}
We here refer to the configuration in the left panel of Fig.~\ref{fig:EWP-Poincared-mixed}.
By symmetry, $\Gamma_{\tilde{B}}$ is connected and centered at $x=0$.
Therefore, in order to determine it, we just have to minimize in a one-dimensional parameter space.
By the minimality condition, $\Gamma_{\tilde{B}}$ has to be orthogonal to $\Gamma_O$.
The orthogonality condition can be expressed as follows.
First, the tangent vector to a minimal surface $\Gamma$ is given by
\begin{equation}
 V^{\mu} = \left( V^z,V^x, V^{y_i} \right) =
 N \, \left( 1, \left. \frac{dx}{dz} \right|_{z_*} , \vec{0} \right) \, ,
 \qquad
 \frac{dx}{dz} = \pm \frac{z^{d-1}}{\sqrt{z_*^{2(d-1)} - z^{2(d-1)}}} \, ,
\end{equation}
where $N$ is a normalization and $z_*$ is the turning point of $\Gamma$.
The $\pm$ sign refers to the branch $\mp$ of the solution.
In order for two $\Gamma$s to be orthogonal, we have to consider branch $+ (-)$ of the first and branch $- (+)$ of the second.
The orthogonality condition then reads
\begin{equation}
 0 = V_1 \cdot V_2
 \propto 1 + \left. \frac{dx}{dz} \right|_{z_{*1}} \left. \frac{dx}{dz} \right|_{\tilde{z}_*} 
 = 1 - \frac{z_1^{2(d-1)}}{\sqrt{z_{*1}^{2(d-1)} - z_1^{2(d-1)}} \sqrt{\tilde{z}_*^{2(d-1)} - z_1^{2(d-1)}}} \, ,
\end{equation}
with $z_1$ the intersection point and $z_{*1}, \tilde{z}_*$ the turning points.
Clearly, we have the constraint $z_1 < \min({z_{*1},\tilde{z}_*})$. 
We can express the turning point of one minimal surface in terms of the turning point of the second and their intersection point as
\begin{equation}
\label{eq:orthogonality_GammaB}
 \tilde{z}_* = \frac{z_{*1} z_1}{(z_{*1}^{2(d-1)} - z_1^{2(d-1)})^{\frac{1}{2(d-1)}}} \, .
\end{equation}
To uniquely fix $\Gamma_{\tilde{B}}$, it is enough to impose that the intersection point belongs to both the minimal surfaces, namely from eq.~\eqref{eq:GammaA-Poincare-d}
\begin{equation}
 \chi_{+,\tilde{z}_*}(z_1) = \frac{l_B+s}{2} + \chi_{-,z_{*1}}(z_1) \, .
\end{equation} 
By solving for $z_1$ and plugging it into eq.~\eqref{eq:orthogonality_GammaB}, we obtain the $\Gamma_{\tilde{B}}$ with minimal area.
The corresponding minimal area reads
\begin{align}
\label{eq:area_Gamma_Btilde}
\mathcal{A}(\Gamma_{\tilde{B}}) &= 2 L^{d-1} \ell^{d-2} \int_{z_1}^{\tilde{z}_*(z_1)} \frac{\tilde{z}_*(z_1)^{d-1}}{z^{d-1} \sqrt{\tilde{z}_*(z_1)^{2(d-1)} - z^{2(d-1)}}} dz \nonumber \\
&= \left. 2 L^{d-1} \ell^{d-2} \frac{z^{2-d}}{2-d} \, {}_2 F_1 \left( \frac{1}{2}, \frac{2-d}{2(d-1)}; \frac{d}{2(d-1)}; \left( \frac{z}{\tilde{z}_*(z_1)} \right)^{2(d-1)} \right) \right|_{z_1}^{\tilde{z}_*(z_1)} \, .
\end{align}

We now turn to the computation of $\Gamma_{\tilde{A}}$.
By symmetry, $\Gamma_{\tilde{C}}$ is given by the same solution with $x \to -x$.
In this case, we have to minimize in a two-dimensional parameter space.
This is equivalent to impose orthogonality along the two intersections between $\Gamma_{\tilde{A}}$ and $\Gamma_O$.
The first condition fixes the turning point as in eq.~\eqref{eq:orthogonality_GammaB}.
In this case, the intersection $z_1$ satisfies
\begin{equation}
\frac{\sqrt{\pi} \, \Gamma \left( \frac{3d-2}{2(d-1)} \right)}{d \, \Gamma \left( \frac{2d-1}{2(d-1)} \right)} \tilde{z}_* + \frac{l_B}{2} + s_x + \chi_{-,\tilde{z}_*}(z_1) = \frac{l_B+s}{2} + \chi_{+,z_{*1}}(z_1) \, ,
\end{equation}
where $0 \leq s_x \leq s$.
The second orthogonality condition can be obtained in the same way as the first one, and fixes the second intersection point $z_2$ as
\begin{equation}
z_2 = \frac{z_{*2} \tilde{z}_*}{(\tilde{z}_{*}^{2(d-1)} + z_{*2}^{2(d-1)})^{\frac{1}{2(d-1)}}} \, ,
\end{equation}
where $z_{*2}$ is the turning point of the outer branch of $\Gamma_O$.
Note that by plugging in $\tilde{z}_*(z_1)$, we can express the second intersection location $z_2$ as a function of the first, $z_1$.
Finally, we have to impose that the second intersection $z_2$ satisfies
\begin{equation}
 \frac{\sqrt{\pi} \, \Gamma \left( \frac{3d-2}{2(d-1)} \right)}{d \, \Gamma \left( \frac{2d-1}{2(d-1)} \right)} \tilde{z}_* + \frac{l_B}{2} + s_x + \chi_{-,\tilde{z}_*}(z_2) = \chi_{+,z_{*2}} (z_2) \, .
\end{equation}
In summary, $\Gamma_{\tilde{A}}$ is described by $(z_1,s_x)$ that satisfy the above conditions.
These uniquely fix the solution with minimal area
\begin{equation}
\label{eq:area_Gamma_Ctilde}
\mathcal{A}(\Gamma_{\tilde{A}}) = \left. L^{d-1} \ell^{d-2} \frac{z^{2-d}}{2-d} \, {}_2 F_1 \left( \frac{1}{2}, \frac{2-d}{2(d-1)}; \frac{d}{2(d-1)}; \left( \frac{z}{\tilde{z}_*(z_1)} \right)^{2(d-1)} \right) \right|_{z_1}^{z_2 (z_1)} \, .
\end{equation}
Note that $\Gamma_{\tilde{A}} \cup \Gamma_{\tilde{C}}$ is also a candidate for $\Gamma_{\tilde{B}}$. However, we have checked numerically that $\mathcal{A}(\Gamma_{\tilde{B}})$ in eq.~\eqref{eq:area_Gamma_Btilde} is less than twice $\mathcal{A}(\Gamma_{\tilde{A}})$ in eq.~\eqref{eq:area_Gamma_Ctilde}, so $V_{A:B:C} \neq 0$ when $\Gamma_O$ is in the fully connected phase.

\subsection{Computation for definition $2$}
\label{app:monotonicity_def2}
We consider the configuration in the right panel of Fig.~\ref{fig:EWP-Poincared-mixed}.
By symmetry, $\Gamma_{\ti{B}}$ is centered at $x=0$ and $\Gamma_{\ti{A}}$ meets $\Gamma_{\ti{C}}$ on the external branch of $\Gamma_O$ at $x=0$ and at its turning point
\begin{equation}
    z_{*O} = \frac{d \, \Gamma \left( \frac{2d-1}{2(d-1)} \right)}{\sqrt{\pi} \, \Gamma \left(  \frac{3d-2}{2(d-1)}\right)} \, \frac{2l_A+ l_B + 2s}{2} \, .
\end{equation}
Additionally, the intersection point between the $+$ branch of $\Gamma_{\ti{B}}$ and $\Gamma_{\ti{A}}$, that we denote by $\left( z_1, x_1 \right)$, is related to the intersection point between the $-$ branch of $\Gamma_{\ti{B}}$ and $\Gamma_{\ti{C}}$ by $\left( z_1, -x_1 \right)$.
The value of the turning point $z_{*\ti{B}}$ of $\Gamma_{\ti{B}}$ can be expressed as a function of $z_1$ by the equality
\begin{equation}
    \chi_{+,z_{*\ti{B}}}(z_1) = \frac{l_B +s}{2} + \chi_{-,z_{*s}}(z_1) \, ,
    \qquad
    z_{*s} = \frac{d \, \Gamma \left( \frac{2d-1}{2(d-1)} \right)}{\sqrt{\pi} \, \Gamma \left(  \frac{3d-2}{2(d-1)}\right)} \, \frac{s}{2} \, .
\end{equation}
Similarly, the value of the turning point $z_{*\ti{A}}$ of $\Gamma_{\ti{A}}$ can be given as a function of $z_1$ by
\begin{equation}
    \chi_{+,z_{*\ti{A}}}(z_1) -
    \chi_{+,z_{*\ti{A}}}(z_{*O})  = \frac{l_B +s}{2} + \chi_{-,z_{*s}}(z_1) \, .
\end{equation}
Given this, the total area $\mathcal{A}_{\rm tot}(z_1)=\mathcal{A}(\Gamma_{\ti{A}})+\mathcal{A}(\Gamma_{\ti{B}})/2$ is given by
\begin{align}
    \frac{\mathcal{A}_{\rm tot}(z_1)}{L^{d-1} \ell^{d-2}} &=
    \left. \frac{z^{2-d}}{2-d} \, {}_2 F_1 \left( \frac{1}{2}, \frac{2-d}{2(d-1)}; \frac{d}{2(d-1)}; \left( \frac{z}{z_{*\ti{B}}(z_1)} \right)^{2(d-1)} \right) \right|_{z_1}^{z_{*\ti{B}}(z_1)} \nonumber \\
    &+ \left.\frac{z^{2-d}}{2-d} \, {}_2 F_1 \left( \frac{1}{2}, \frac{2-d}{2(d-1)}; \frac{d}{2(d-1)}; \left( \frac{z}{z_{*\ti{A}}(z_1)} \right)^{2(d-1)} \right) \right|_{z_1}^{z_{*O}} \, .
\end{align}
By minimizing the area with respect to $z_1$, we obtain $z_1^{\rm min}$.
This fixes the location of the intersection point $(z_1^{\rm min},x_1^{\rm min})$.
The volume subtended by $\Gamma_{\ti{A}}$ and the boundary, between $x=0$ and $x=x_1$, can be obtained as
\begin{align}
    \frac{\mathcal{V}_{\ti{A}}(z_1)}{L^d \ell^{d-1}} &= \int_{z_1}^{z_{*O}} dz \frac{x_+(z)}{z^d} 
    = \left. \frac{x_+(z)}{(1-d) z^{d-1}} \right|_{z_1}^{z_{*O}} - \int_{z_1}^{z_{*O}} dz \frac{x_+'(z)}{(1-d) z^{d-1}} \nonumber \\
    &= \left. \frac{x_+(z)}{(1-d) z^{d-1}} \right|_{z_1}^{z_{*O}} - 
    \left. \mathrm{V}_{z_{*\ti{A}}}(z) \right|_{z_1}^{z_{*O}} \, ,
\end{align}
\begin{equation}
    \mathrm{V}_{z_{*\ti{A}}}(z) \equiv \frac{z}{(d-1) z_{*\ti{A}}^{d-1}} \, {}_2 F_1 \left( \frac{1}{2}, \frac{1}{2(d-1)}; \frac{2d-1}{2(d-1)}; \left( \frac{z}{z_{*\ti{A}}} \right)^{2(d-1)} \right) \, ,
\end{equation}
where $x_+(z)$ is the profile of $\Gamma_{\ti{A}}$ given in eq.~\eqref{eq:GammaA-Poincare-d}. Note that $x_+(z_{*O})=0$. In the first line we have integrated by parts, while in the second line we have used
\begin{equation}
    x_+'(z) = - \frac{z^{d-1}}{z_{*\ti{A}}^{d-1}} \frac{1}{\sqrt{1- \left( \frac{z}{z_{*\ti{A}}} \right)^{2(d-1)}}} \, .
\end{equation}
In a similar way, we can obtain the volume subtended by $\Gamma_{\ti{B}}$ and the boundary, between $x=0$ and $x=x_1$, that we denote by $\mathcal{V}_{\ti{B}}(z_1)$.
Finally, the volume of the EWP reads
\begin{align}
    \frac{V^{(2)}_{A:B:C}}{L^d \ell^{d-2}} &=
    2 \frac{\mathcal{V}_{\ti{A}}(z_1^{\rm min}) + \mathcal{V}_{\ti{B}} (z_1^{\rm min})}{L^d \ell^{d-2}} \nonumber \\
    &=2 \left( \mathrm{V}_{z_{*\ti{B}}}(z_{*\ti{B}}) - \mathrm{V}_{z_{*\ti{B}}}(z_1^{\rm min}) - \left( \mathrm{V}_{z_{*\ti{A}}}(z_{*O}) - \mathrm{V}_{z_{*\ti{A}}}(z_1^{\rm min}) \right) \right) \, .
\end{align}
The value of $z_1^{\rm min}$ has been obtained numerically.

\bibliographystyle{JHEP}
\bibliography{EWPolygon}


\end{document}